\documentclass[universe,accept,moreauthors,pdftex,10pt,a4paper]{mdpi}
\firstpage{1}
\articlenumber{104}
\doinum{10.3390/universe7040104}
\pubvolume{7}
\pubyear{2021}
\copyrightyear{2021}
\externaleditor{Academic Editor: name}
\history{Received: date; Accepted: date; Published: date}
\pdfoutput=1

\Title{Electron-positron  vacuum instability in strong electric
fields. Relativistic semiclassical approach.}

\Author{D. N. Voskresensky $^{1,2}$}
\AuthorNames{D. N. Voskresensky}

\address{%
	$^{1}$ \quad Joint Institute for Nuclear Research, Joliot-Curie 6,  Dubna 141980,  Russia \\
$^{2}$ \quad
National Research Nuclear University (MEPhI),
Kashirskoe sh. 31,  Moscow 115409, Russia}
\corres{Correspondence: d.voskresen@gmail.com}

\abstract{Instability of electron-positron  vacuum in strong electric fields is studied. First, falling to the Coulomb center is discussed at  $Z>137/2$ for a spinless boson and at $Z>137$ for electron. Then,   focus is concentrated on description of deep electron levels  and  spontaneous positron production in the field of a finite-size nucleus  with the charge $Z>Z_{\rm cr}\simeq 170$. Next, these effects are studied in application to the low-energy heavy-ion collisions. Then, we consider phenomenon of  ``electron condensation'' on levels of upper continuum crossed the boundary of the lower continuum $\epsilon =-m$ in the field of a supercharged nucleus with  $Z\gg Z_{\rm cr}$. Finally, attention is focused on many-particle problems of polarization of the QED vacuum and electron condensation at ultra-short distances from a  source of  charge. We argue for a principal difference of  cases, when the size of the source is larger than the pole size $r_{\rm pole}$, at which the dielectric permittivity of the vacuum reaches  zero, and smaller $r_{\rm pole}$. Some arguments are presented in favor of the logical consistency of QED. All problems are considered within the same relativistic semiclassical approach.}


\newcommand{\be}{\begin{equation}}
\newcommand{\ee}{\end{equation}}

\usepackage{amssymb}

\newcommand{\lsim}{\stackrel{\scriptstyle <}{\phantom{}_{\sim}}}
\newcommand{\gsim}{\stackrel{\scriptstyle >}{\phantom{}_{\sim}}}

\begin{document}
\section{Introduction}
I
 dedicate this review to the blessed memory of Vladimir Stepanovich Popov, who recently  left us as the result of a many-year hard illness, which prevented him working actively in his last years.  The problem of the electron-positron pair  production when the ground-state electron level dives below the energy $-mc^2$ ($m$ is the electron mass, $c$ is the speed of light)
was of his  interest starting from the end of 1960-th. Especially he contributed to this problem during the
 1970s. V. S. Popov was awarded the I. Y. Pomeranchuk Prize in 2019 for his outstanding contributions to the theory of ionization of atoms and ions in the field of intense laser radiation and the theory of the creation of electron-positron pairs in the presence of superstrong external fields.

We worked together with Vladimir Stepanovich on problems of supercritical atoms with the charge $Z>Z_{\rm cr}=170-173$ during 1976--1978 when we developed semiclassical treatment of  this problem. These works, cf. \cite{Migdal:1976wx,Migdal:1977rn,Eletskii:1977na,Mur:1978nb,Mur:1978ke,Popov:1978kk,Popov:1979gq} became a part of my PhD thesis \cite{Voskresensky1977} that was defended in 1977  under the guidance of Arkadi Benediktovich Migdal.

As follows from the Dirac equation in the Coulomb field of a point-like nucleus with  $Z>1/e^2$ (in units $\hbar =c =1$, which will be used in this paper, $e^2\simeq 1/137$),  the electron that occupied  the ground-state level should fall to the center.
Following the  idea of I. Pomeranchuk and Ya. Smorodinsky   \cite{PomeranchukSmorodinsky1945}, the
 solution of the problem  of the falling of the electron to the center can be found while taking into account the fact that the real nuclei have a finite radius. With increasing $Z$, the energy of the ground state level decreases and, at $Z>Z_{cr}$, crosses the boundary of the lower continuum $\epsilon=-m$.
 The problem received a new push in the end of the 1960s
The important role of the Pauli principle was emphasized in \cite{GershteinZeldovich1970}.    However the authors erroneously assumed delocalization of the electron state with $\epsilon \simeq -m$.
 Independently, W. Pieper and W. Greiner \cite{Pieper1969} (in numerical analysis)  and  \mbox{V. S. Popov \cite{Popov1970ZhETF,Popov1970YadFiz,Popov1971ZhETF32,Popov1971ZhETF33,ZeldovichPopov1972}} (in analytical and numerical  studies) correctly evaluated the value of the critical charge to be   $Z_{\rm cr}\simeq 169-173$, depending on assumptions regarding the charge distribution inside the nucleus and the ratio $Z/A$. It was argued that  two positrons
with the energies $>m$ go off to infinity and electrons with $\epsilon <-m$ screen the  field of the nucleus by the charge  $-2e$.
 The typical distance characterizing electrons of the vacuum $K$ shell  is $\sim 1/(3m)\gg R_{\rm nucl}$, cf. \cite{Popov:1979gq}. Subsequently, there appeared an idea to observe positron production in heavy-ion collisions, where the supercritical atom is formed for a short time \cite{Muller1972,PopovJETP1974}. As the reviews of these problems, I can recommend \cite{Popov2001,Rafelski1916,Voskresensky1990}.

In 1976, with the inauguration of the UNI-LAC accelerator in GSI, Darmstadt, it became possible to accelerate heavy ions up to uranium below and above the Coulomb barrier. Instead of a positron line that is associated with the spontaneous decay of the electron-positron vacuum, mysterious line structures were observed, which, in spite of many attempts, did not get a reasonable theoretical interpretation. The experimental results on the mentioned positron lines proved to be erroneous. New experiments were conducted during 1993--1995, cf. \cite{Ahmad1995,Ganz1996,Leinberger1997}. The presence of the line structures  was not observed. Events, which could be interpreted as the effect of the decay of the QED vacuum with the spontaneous production of the electron-positron pair,  were not selected. In spite of  the effect of the spontaneous production of positrons in the  electric field of the supercharged nucleus being predicted many decades ago, it  has not yet been observed experimentally in heavy-ion collisions.

One also studied a possibility of a nuclear sticking in the process of the heavy-ion collisions \cite{Greiner1985,ZagrebaevGreiner}. Although these expectations   did not find a  support in further investigations, extra arguments were given for a possibility of the observation of the spontaneous positron production in the heavy-ion collisions, cf.  \cite{Maltsev2019}. Especially, the usage of  transuranium ions looks very promising \cite{Zagrebaev2006}.  Besides a spontaneous production of positrons, a more intensive induced production of pairs  occurs due to an excitation of nuclear levels, cf. \cite{Rafelski1916}.
Therefore, the key question is how to distinguish  spontaneous production of positrons that originated  in the decay of the electron-positron vacuum from the induced production  and other competing  processes.

New studies of low-energy heavy-ion collisions at the supercritical regime are anticipated at the upcoming accelerator facilities
in Germany, Russia, and China \cite{Gumberidze2009,Ter2015,Ma2017}.
This possibility renewed theoretical interest to the problem \cite{Godunov2017,Popov2018,Maltsev2019,Popov:2020xmd}. As one can  see from the numerical results reported in \cite{Popov:2020xmd},  these results  support those that were obtained in earlier works, although a comparison with the analytical results derived in  \cite{Migdal:1976wx,Migdal:1977rn,Eletskii:1977na,Mur:1978nb,Mur:1978ke,Popov:1978kk,Popov:1979gq} was not performed.
Additionally, it should be noted that there recently appeared statements that the spontaneous production of positrons should not occur in the problem under consideration. I see no serious grounds for these revisions and, thereby, will not review these works.

{\subsection{ A General Picture}}

\begin{figure}[H]
\includegraphics[width=6.5cm,clip]{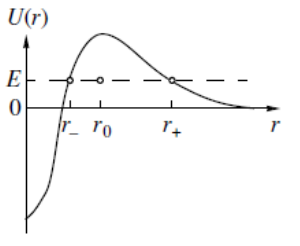}
\caption{Typical
 dependence of effective Schr\"odinger potential $U$ on $r$ for  a charged particle in an electric central-symmetric potential well, $r_{\pm}$ are turning points, and $r_0$ corresponds to maximum of the effective potential $U$. The dashed line describes  the quasistationary level with $\epsilon <-m$.}\label{UeffSchrod}
\end{figure}

States with $|\epsilon|<m$ correspond to the energy $E=(\epsilon^2 -m^2)/2m<0$ and effective potential $U$, see Figure \ref{UeffSchrod}. In terms of the  Schr\'odinger equation   these are ordinary bound states.
Let the ground state level be empty and we are able to adiabatically increase the charge of the nucleus $Z$.
The latter  means that  the time $\tau_Z$ characterizing the increase of $Z$ is much larger when compared to $1/|\epsilon_0 -\epsilon_{njm}|$, where $\epsilon_{njm}$ are the energies of other bound states in the potential well, and  $\tau_Z >1/m$ for the case of transitions from the ground-state level, $\epsilon_0$, to the continues spectrum. The empty level with $\epsilon <-m$ becomes quasistationary, see Figure \ref{UeffSchrod}. When penetrating the barrier between continua, see Figure \ref{ContinuaBoundaries} below,  two electrons (with opposite spins)  are produced, which occupy this level, whereas two positrons of the opposite energy  go off through the barrier to infinity.
In the standard interpretation, \mbox{cf. \cite{ZeldovichPopov1972},}  the electron states, $\psi\propto e^{-i\epsilon t}$,  with $\epsilon =\epsilon_0 +i\Gamma (\epsilon_0)/2$ for $\epsilon_0 < -m$,
$\Gamma >0$, cf. \mbox{Equations (3.5) and (3.6) in} \cite{MurPopovFerm}, are occupied due to the redistribution of the charge of the vacuum. The vacuum  gets the  charge $2e<0$ distributed in the region of the supercritical ion. Two positrons with $\epsilon_{e^+} =-\epsilon_0 -i\Gamma (\epsilon_0)/2$ go off to infinity after passage of a time $\sim \tau_0 e^{\Gamma t}$, $\tau_0\sim R$, where $R$ is the size of the potential well for $R\gsim 1/m$,  as it occurs for any decaying quasistationary state, producing a diverging spherical wave $\psi\propto e^{ikr}$, $k=\sqrt{\epsilon_{e^+}^2 -m^2}$ for the positron. For far-distant potentials, the situation is similar to that for the charged bosons, cf. \cite{Voskresensky1988}.
For the case $V=-Ze^2/r$ for $r>R_{\rm nucl}$, one obtains $\Gamma (-m)=0$.

For $Z<Z_{\rm cr}$ electrons of the lower continuum (with $\epsilon <-m$), fill all energy levels according to the Dirac picture of the electron-positron vacuum. They are spatially distributed at large distances. For $Z>Z_{\rm cr}$ the process of the tunneling of the electron of the lower continuum to the empty (localized) state that was prepared in the upper continuum with $\epsilon <-m$ can be  treated as the tunneling of the virtual positron (electron hole)  with $\epsilon_{e^+} =-\epsilon_0 -i\Gamma/2$ from the region of the potential well to infinity, where it already can be observed. If one scatters an external real positron with a resonance energy $\epsilon_{e^+} \simeq -\epsilon_0 >m$ on such a potential, this positron, for a short time, forms a resonance quasistationary state in the effective potential, which, after passage of a time $\sim 1/\Gamma$, is decayed. As the result,  the positron  goes back to infinity. After that, during  a time of the same order of magnitude, two positrons, being  produced in a fluctuation together with two electrons, go off to infinity and those two electrons fill the stationary negative-energy state, as was explained.

If the ground state level was initially occupied by two electrons of opposite spins, then, at  adiabatic change of the potential (in the sense clarified above), they remain on this  level $\epsilon =\epsilon_0$. At the adiabatic change of the potential, electrons have no energy to escape anywhere from this level.
The production of pairs does not occur, since the level is occupied by electrons. During a time $\sim 1/\Gamma$, their charge $2e<0$ is redistributed over the range of energies $\epsilon_0-\Gamma(\epsilon_0)/2\lsim \epsilon\lsim \epsilon_0+\Gamma(\epsilon_0)/2$. This charge is localized at distances  ($\sim 1/(3m)$ that are typical for the ground state in  the Coulomb field \cite{Popov:1979gq}). In this sense, one formally requires a many-particle description of the stationary  electron with $\mbox{Re}\epsilon_0 <-m$ at $\Gamma \neq 0$. However, neglecting a tiny  $\Gamma$ correction, for the finding of $\epsilon (Z)$, one may continue to employ the one-particle description. If the experimenter scatters an external positron with $\epsilon_{e^+} \simeq -\epsilon_0>m$ on such a potential, the positron  annihilates with one of the two electrons have occupied the ground-state  level. After the passage of a time $\sim 1/\Gamma$, there occurs spontaneous production of the one new pair, the electron fills empty state  (after that, again, two electrons occupy  the ground-state level) and the positron goes to infinity.
\\

{\subsection{ Semiclassical Approximation}}
Semiclassical approximation is one of the most important approximate methods of quantum mechanics  \cite{Migdal1975}. Classical and semiclassical ideas are widely used in quantum field theory in problems dealing with the spontaneous vacuum symmetry breaking for bosons, cf. \cite{ZuberItzykson1989,Voskresensky1990,Voskresensky1988},  in condensed matter physics, cf. \cite{LL5,LL9,Voskresensky:1993ux}, and in physics of nuclear \linebreak  matter \cite{Migdal1978,Migdal:1990vm}.

As a consequence of the instability  of the boson vacuum in a strong external field, there appears a reconstruction of the ground state  and there arises a condensate of the classical boson field \cite{Migdal1971,Migdal1978RevModPhys}. Many-particle repulsion of  particles in the condensate provides the stability of the ground state.
After that, excitations prove to be stable, cf. \cite{Migdal1978,Migdal:1990vm}. They  are also successfully described using semiclassical methods, e.g., such as the loop expansion \cite{ColemanWeiberg,Voskresensky1988}.

For fermions, there exist two possibilities. In the first situation, fermions heaving attractive interaction, being rather close to each other, may form Cooper pairs, cf. \cite{LL9}.  In the second situation, which we focus on here, electron-positron pairs, being produced in a strong static  electric field, are well separated from each other by the potential barrier. Consequently, the   electric potential attracts particles of one sign of the charge and repels antiparticles. Because of the Pauli principle, each unstable single-particle state is occupied by only one fermion. Therefore, it is natural to prolong a single-particle description in a overcritical region (until there appeared still not too many dangerous states). Classical approximation does not work for fermions, but semiclassical methods prove to be working.   As is known, the semiclassical approach yields correct results for the values  of the energy  levels with big quantum numbers and in the case of spatially smooth potentials, when $d\tilde{\lambda}/dx\ll 1$, where $\tilde{\lambda} =1/p(x)$ is the reduced electron De Broglie length, $p(x)$ is the momentum, and $x$ is the coordinate. For the Coulomb field for the ground-state level, a rough estimate yields $d\tilde{\lambda}/dr\sim 1/(Ze^2)$ for $r\to 0$. However, even for $d\tilde{\lambda}/dx \sim 1$, semiclassical approximation continues to work not bad in calculation of the energy levels, with an error not larger than 10\% due to the presence of a numerically small parameter $\sim 1/\pi^2$, cf. \cite{Migdal1975}.

{\bf Instability of the vacuum near a nucleus heaving a supercritical charge. 
} It proves to be that the semiclassical approximation is applicable with an appropriate accuracy for the description of the electron energy levels in the supercritical field of a nucleus with the supercritical charge $Z>(170-173)$. Semiclassical approximation allows for finding rather simple expressions for the critical value of the charge, cf. Refs. \cite{Krainov1971,Marinov1974,Voskresensky1977}, for energies of deep levels as a function of $Z$ and for the probabilities of the penetration of the barrier between continua, cf. \cite{Eletskii:1977na,Mur:1978nb,Mur:1978ke,Popov:1978kk,Popov:1979gq}.

{\bf The spontaneous positron production in low-energy heavy-ion collisions.}
A comparison of the theory and experiment should check the application of QED in the region of strong fields outside the applicability of the perturbation theory.  The description of the spontaneous production of positrons in heavy-ion collisions  needs a solution of the two-center problem for the Dirac equation. Because variables are not separated in this case, the problem does not allow for the analytical treatment and numerical calculations are cumbersome. However, the use of the semiclassical approximation  results in simple analytical expressions for the energies of the electron levels, cf. \cite{Popov:1978kk,Popov:1979gq},  valid with error less than few \%. Thereby, this is one more example of the efficiency of the semiclassical approach.

{\bf Electron condensation in a field of a supercharged nucleus.}  In supercritical fields, many energy levels cross the boundary of the lower continuum and the problem of the finding of the vacuum charge density becomes of purely many-particle origin. It  can be considered within the relativistic Thomas-Fermi method, cf. \cite{Migdal:1977rn}. All of the initially empty states, which  crossed the boundary $\epsilon =-m$, are filled after a while. In this sense, one may speak about "electron condensate''.

{\bf Vacuum polarization and electron condensation at super-short distances from  Coulomb center.}
In spite of the successes in explanation of all purely electrodynamical phenomena, QED is a principally unsatisfactory theory, since relations between the bare mass and charge and observable ones contain divergent integrals \cite{LL4,Bogolyubov1984}. As the result, as one thinks, there is no not contradictive manner to pass from super-short to long distances. In spite of this, as is well known, it is possible to remove divergencies from all observable quantities with the help of the renormalization procedure.

The problem of the so-called ``zero charge'' or Moscow zero, cf. \cite{LandauPomeranchuk1955,Fradkin1955}, is one of central problems related to renormalization of the charge. When considering the square of the charge of electron $e^2(r)$ as a function of the radius $r$ and assuming finite value of the bare charge $e^2(r_0)=e_0^2>0$ for the source-size $r_0\to 0$, one derives $e^2 (r\to \infty)\to 0$ instead of an expected value $e^2 (r\to \infty)\to e^2=1/137$. The same problem appears, when one considers the screening of the central source with the charge density $n_{\rm ext}=Z_0 \delta (r-r_0)$ for $r_0\to 0$, \mbox{cf. \cite{Eletskii:1977na}. }The problem of a distribution of the charge near an external source of the charge with the radius $R\ll 1/m$, as well as  the problem of the distribution of the charge of the electron at distances  $r\ll 1/m$ are the key principal problems of QED.
The semiclassical approach proves to be very promising in the calculation of the vacuum dielectric permittivity in strong inhomogeneous electric fields \cite{Migdal1972}. The density of the polarized charge is supplemented by the density from the electron condensation \cite{Eletskii:1977na,Migdal1978}. The problem proves to be specific and it depends on whether the radius of the external source of the charge is larger than a distance $r_{\rm pole}$, where the dielectric permittivity decreases to zero, or smaller $r_{\rm pole}$, cf. \cite{V1992,Voskresensky:1993uy}.
 References \cite{V1992,Voskresensky:1993uy} argued for the  condensation of electron states  in the upper continuum at distances larger than $r_{\rm pole}$ for $r_0>r_{\rm pole}$ and for the condensation of electron states originated in  the lower continuum at distances smaller than $r_{\rm pole}$ (for $r_0<r_{\rm pole}$), at which the dielectric permittivity proves to be negative and $e_0^2<0$. The semiclassical consideration of this problem allows for presenting arguments in favor of a logical  consistency of QED.

{\bf  Similar effects in semimetals and in stack of graphene
layers.} The existence of the Weyl semimetals, i.e., materials with the
points in Brillouin zone, where the completely filled valence and
completely empty conduction bands meet with  a  linear dispersion
law, $\epsilon =v_{\rm F}p$, where the Fermi velocity is $v_{\rm F} \sim
10^{-2} $, has been predicted in \cite{AB70}. Systems with the relativistic dispersion law are likely to be realized in some doped silver chalcogenides, pyrochlore iridates,
 and in topological insulator multilayer structures.
Weyl semimetals are three-dimensional analogs of  graphene \cite{Novoselov}, where the energy of excitations is also approximately presented by the linear function
of the momentum, but the electron subsystem is a two-dimensional one,
whereas the photon subsystem remains three-dimensional. Even
though the mass of excitations $m=0$ for ideal graphene and Weyl
semimetals without interactions, a non-zero mass, $m\neq 0$, can be induced in many
ways \cite{Kotov}, resulting in
a dispersion relation characterized by a gap, i.e.
$\epsilon^2 =p^2 v_{\rm F}^2 +m^2 v_{\rm F}^4\,.$
 In difference with a small  value
of the fine structure constant in QED, $e^2=1/137$, the effective coupling in Weyl semimetals and in graphene is $\alpha_{\rm ef} =e^2/v_{\rm F}\varepsilon_0 $, where
$\varepsilon_0$ is the dielectric permittivity of the substance. The coupling constant $\alpha_{\rm ef}$ can be as  $\ll 1$ as $\gsim 1$, depending on the substance, and both
weak and strong coupling regimes are experimentally accessible.
 Thus, Weyl semimetals and an infinite stack of graphene layers make it possible to experimentally study various effects
have been considered in 3+1 quantum electrodynamics (QED) for weak
and effectively strong couplings,  cf. \cite{Kolomeisky:2013vwa,Voskresensky2012}.

Not concerning spontaneous production of positrons of our interest here, the electron-positron production in heavy-ion collisions was studied in many papers, cf. \cite{Bertulani:1987tz,Baur:1990za,Ivanov1991,Baur:2007zz}.

Additionally, electron-positron pair production   from the vacuum can be triggered   by the laser  electromagnetic fields, e.g., see  \cite{Popov1971imaginary,Bulanov2006,Blaschke2006,Han2010,Piazza2012,
Narozhny2015,Popov:2016ebl,Yakimenko:2018kih}. However it seems unlikely to realize such a possibility at least in the nearest future, cf. \cite{Palffy:2019scn} and the references therein.

Electric fields with the strength $E\gg m^2$ may exist
in astrophysical environments, e.g., they may occur at  phase transitions  in neutron and hybrid stars \cite{Migdal:1990vm,Voskresensky:2002hu} and in  neutron star mergers
\cite{Paschalidis:2017qmb},  and they also exist at surfaces of hypothetical    nuclearites and abnormal superheavy nuclei \cite{Migdal1972,Voskresensky:1977mz,Witten1984,Migdal:1990vm,Gani:2018mey}.

 Various radiative corrections to the deeply bound electron levels should certainly be taken into account, e.g., cf. \cite{Flambaum2005,Dyall2007,Khetselius2009} and the references therein. These higher-order corrections will not be considered in the given paper.

Below, attention is focused on a semiclassical description. I  describe the instabilities of the boson and fermion vacua in static  potentials, in particular in the Coulomb field. Afterwards, focus is concentrated  on the description of the spontaneous positron production in low-energy heavy-ion collisions. Next, a many-particle semiclassical description of the electron condensation is considered. Finally, modification of the Coulomb field at super-short distances due to the vacuum polarization and electron condensation is studied.

The paper is organized, as follows.
Section \ref{boson} starts with a brief discussion  of instability for the charged bosons in static electric fields, in particular  in the  Coulomb field of a point-like nucleus with the charge  $Z>Z_{\rm cr}=1/(2e^2)$. The behavior of deeply bound electrons obeying the Dirac equation in the strong static electric fields is considered in Section \ref{DiracCentral}. First, I consider the case of a one-dimensional field and then of a spherically symmetric field. The Dirac equation is transformed to  equivalent Schr\"odinger form  in an effective potential and the interpretation of the solutions is discussed. Subsequently, in Section \ref{CoulombF}, I demonstrate exact solution of the problem of bound states in the strong Coulomb field of a point-like center. The focus is made on the problem of the falling of the electron to the center for a nucleus with the charge $Z\geq 1/e^2$. Section \ref{Finite} describes how the problem is resolved while taking into account that nuclei have a finite size. In Section \ref{transformed}, I  introduce a semiclassical approach to the Dirac equation, being  transformed to the  second-order differential equation. Electron levels crossed the boundary of the lower continuum are considered. The mean radius of the K-electron shell  and the  critical charge of the nucleus are  found for $\epsilon =-m$, as well as the number of levels that crossed the boundary of the lower continuum and  their energies. The critical charge of the nucleus for the muon is also found. A comparison of semiclassical expressions with much more cumbersome exact expressions permits understanding the merits of the semiclassical approach. In Section \ref{Diraclinear}, a semiclassical approximation is developed for the system of linear Dirac equations. Semiclassical wave functions in classically allowed and forbidden regions are introduced, and the Bohr--Sommerfeld quantization rule is formulated.
Next, the probability of the positron production is calculated. Subsequently, semiclassical approximation is applied to non-central potentials. In Section \ref{HIC}, focus is concentrated  on problems of the spontaneous positron production in low-energy  collisions of heavy ions.  The energies of  deep levels as a function of the distance between colliding nuclei and the angular distribution of the positron production are found while employing semiclassical approach. Subsequently, I consider a screening of the charge at collisions of not fully striped nuclei. Semiclassical approximation (imaginary time method) is adequate for describing dynamics of the tunneling of electrons from the lower continuum to the upper one.  In such a way, a correction on non-adiabaticity to the probability of the production of positrons is found.
The electron condensation  in the field of a supercharged nucleus  is considered in Section \ref{RTF}. Section \ref{Polarization} presents the effects that are associated with the polarization of the electron-positron vacuum in weak and strong fields.
Subsequently, in Section \ref{DistrVacCh}, I focus on the description of the charge distribution at super-short distances from the charge source. The effects of polarization of the vacuum and the electron condensation in the upper and lower continua will be considered. Section \ref{Conclusion} contains a conclusion.
\vspace{12pt}
\section{Relativistic Spinless Charged Particle in Static Field \boldmath{$A^\mu =(A^0,\vec{0})$}}\label{boson}

\subsection{Reduction of Klein-Gordon-Fock Equation to  Schr\"odinger Equation}

Consider a spinless negatively charged boson placed in a stationary attractive potential well $V$.
The Klein--Gordon--Fock equation renders
\be
\Delta \phi +[(\epsilon -V)^2 -m^2 ]\phi =0\,.\label{KGFV}
\ee

With the help of notations
\be
E=\frac{\epsilon^2 -m^2}{2m}\,,\quad U_{\rm ef}=-\frac{V^2-2\epsilon V}{2m}\,,\label{relBos}
\ee
we may rewrite Equation (\ref{KGFV}) in the form of the Schr\"odinger equation,
\be
\Delta \phi +2m (E-U_{\rm ef})\phi =0\,.
\ee

As we see from Equation (\ref{relBos}), for relativistic particles  there appears to be an attractive term in the effective potential $-V^2/ (2mc^2)$, even for a purely repulsive potential $V$. In the limit case $E\ll m$ and $|V|\ll m$, we have $\epsilon \simeq m +E$  and $U_{\rm ef}\simeq V$, and  we recover the Schr\"odinger equation for a nonrelativistic particle. For $|\epsilon|<m$ the "nonrelativistic'' energy is $E<0$, which corresponds to bound states in the interval of energies $-m <\epsilon <m$. For a sufficiently deep potential well, the energy of the ground state level  may cross the boundary $\epsilon =-m$. In a deeper potential, other levels cross this boundary. For $\epsilon <-m$, here $\mbox{Re}E>0$,  the levels become quasistationary,  see Figure \ref{UeffSchrod}.

A comment is in order (D. N. Voskresensky 1974, see comment in  \cite{MurPopov1976bos}). For a spinless particle under consideration, the ground-state single-particle level only crosses the boundary $\epsilon =-m$ for far-distant potentials, when $-V(r\to \infty)>C_{\rm cr}/r^2$, for a constant $C_{\rm cr}>0$. For potentials obeying condition $-V(r\to \infty)<C_{\rm cr}/r^2$,  there appears to be a bound state for  the antiparticle. In both cases for a broad potential well of a typical radius  $R\gg 1/m$ the vacuum instability occurs at $|V|\simeq |V|_{\rm cr}\simeq 2m (1 \pm O(1/(m^2R^2))$  either at  $\epsilon_{\rm cr}=-m$ or at  $\epsilon_{\rm cr}\simeq -m(1 -O(1 /(m^2 R^2))$. In the case of a broad potential well, solutions of many-particle problems in both cases are almost the same, cf.   \cite{Voskresensky1988}. For $-V>-V_{\rm cr}$  there appears production of  pairs. Positively charged antiparticles go to infinity and  negatively charged particles form a condensate, see \cite{Migdal1978,Voskresensky1988}.

\begin{figure}
\includegraphics[width=6cm,clip]{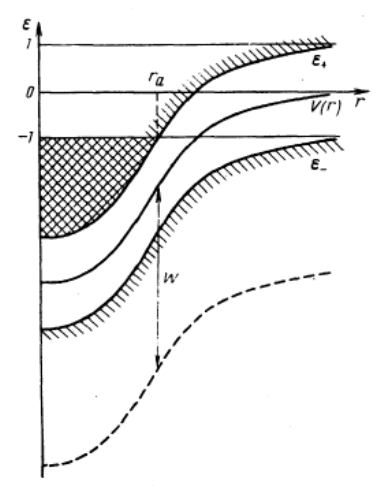}
\caption{Illustration
of the deformation of the upper and lower continua in a
strong external electric field (the boundaries of the continua are
shaded). Electrons that belong to the vacuum shell in the upper continuum  fill the cross-hatched region. The states
below the curve $\epsilon_{-}/m = V ( r )/m- 1$ form the unobservable Dirac
sea. The quantity $W$ shows an artificial  cutoff energy.
}\label{ContinuaBoundaries}
\end{figure}

Let us illustrate how the deformation of boundaries of upper and lower continua occurs  in a static  electric field forming a broad potential well for a negatively charged particle, \mbox{cf. \cite{Migdal:1977rn}.} To be specific, consider a spherically symmetric field. Boundaries of continua, $\epsilon_{\pm}$, are determined by
\be
\vec{p}^{\,2}(r)=(\epsilon_{\pm}-V)^2 -m^2 =0\,.
\ee
They are shown in Figure \ref{ContinuaBoundaries}. In upper and lower  continua $p^2 (r)>0$, these are classically allowed regions. In the gap between continua $p^2(r)<0$. This is a classically forbidden region. For $V<V_{\rm cr}=-2m-O(1/(m^2R^2))$, there arises a region of the overlapping of the continua that means that the negatively charged particle may penetrate from the lower continuum (from the exterior of the potential well) to the upper one (to the interior\mbox{ of the well).}

With an exponential accuracy, the probability of a passage of the one-dimensional  barrier is determined by
\be
W\sim e^{-2\mbox{Im} S}\sim e^{-2\int_{x_{1}}^{x_{2}} |p|dx}\,,\label{Wsemicl}
\ee
where $x_1$ and $x_2$ are the turning points at which $p(x)=0$. This expression is applicable \mbox{for $W\ll 1$.}

 As example, consider a uniform static electric field $eE=-\nabla V =const$, $|eE|\ll m^2$.
 Then we have $p\simeq \sqrt{(\epsilon +eEx)^2 -m^2}$ . From Equation (\ref{Wsemicl}), we immediately obtain
\be
W\sim e^{-\pi E_0/E}\,,\quad E_0=m^2\,.
\ee
This expression coincides with the first term of the infinite series  solution \cite{Schwinger}.

A question arises as to whether it is possible to observe a process of the production of pairs already in a weak attractive electric field with the strength $|E|\ll m^2$ at $-\delta V>2m$?
The critical difference $-\delta V\simeq -2m$ can be easily reached in the field of the capacitor, where $\nabla A_0 =const$,  at the increase of the distance $d$ between plates. Employing $|\nabla A_0|=|\vec{E}|\sim 10^{4}$ V$/$cm, the value, which is easily produced in electrical engineering, we estimate $|\delta V|> 2m_\pi$ already for $d\gsim 10^3$ cm. Here, $m_\pi \simeq 140$ MeV is the mass of the lightest charged boson,  the pion.
However the probability of the production of the pairs
$W\sim e^{-2\mbox{Im}S}$, $\mbox{Im}S=\int_{x_{1}}^{x_{2}} |p| dx$,  is negligibly small at these conditions.
Indeed, for $V=-eEx$, we  get
$\mbox{Im}S=\int_{x_{1}}^{x_{2}} |p| dx=\frac{\pi}{2}\frac{E_0}{E}\,.$
For pions $E_0\simeq  10^{21}$ V$/$cm. For electrons $E_0\simeq 1.3\cdot 10^{16}{\rm V}/{\rm cm}$.

\subsection{Relativistic Spinless Charged Particle in Coulomb Field of Point-Like  Center}

 In the case of the Coulomb field of a point-like nucleus, $V=-Ze^2/r$, with the help of the replacement  $\phi (\vec{r})=R(r)Y_{lm}$, we obtain equation for the radial wave function $R(r)$ \mbox{in the form}
\be
\Delta_r R +{2m}\left[E+\frac{(Ze^2)^2}{2m  r^2}-\frac{l(l+1)}{2m  r^2}+\frac{\epsilon Ze^2}{m r}\right]R=0\,,\quad \Delta_r =\frac{1}{r}\frac{\partial^2 (rR)}{\partial r^2}\,,\label{bosR}
\ee
where $E=\frac{\epsilon^2 -m^2}{2m}$ is the effective nonrelativistic  Schr\"odinger
energy of the particle,
\be
U_{\rm ef}(r)=-\frac{(Ze^2)^2}{2m  r^2}+\frac{l(l+1)}{2m  r^2}-\frac{\epsilon Ze^2}{mr} \label{UefBos}
\ee
is the effective potential, now, depending on $l$. Equation (\ref{bosR}) and the ordinary Schr\"odinger equation  for the radial function in the effective potential coincide
after  undertaking replacements
\be
l(l+1) -(Ze^2)^2 =\lambda(\lambda +1)\,,\quad \epsilon Ze^2/m \to Z^\prime e^2\label{lambda}
 \ee
 in the former one. Thus, instead of the expression for the energy of the
Schr\"odinger particle in the Coulomb field, we derive
\be
{E}_{n_r,l}=-\frac{(Z^\prime e^2)^2 m}{2  (n_r+\lambda +1)^2}\,.
\ee
Here, $n_r+\lambda +1=n+\lambda -l $, $n_r =0,1,...$ is the radial quantum number. \textls[-40]{
Solving \linebreak Equation (\ref{lambda}) and retaining solution with positive-sign  square root, } $\lambda =-\frac{1}{2}+\sqrt{(l+\frac{1}{2})^2 -
{(Ze^2)^2}}$, because, for $Z=0$, $l=0$, one should have $\lambda =0$, we find the Sommerfeld formula for a spinless particle,
\be
\epsilon^2_{n_r,l} =\frac{m^2 }{1+\frac{Z^2 e^4}{ (n -l-1/2+\sqrt{(l+1/2)^2 -(Ze^2)^2}\,\,)^2}}\,.\label{epsilonCoul}
\ee

There are two square-root solutions of this equation. Solution, which yields $\epsilon \to m$ for $Ze^2\ll 1$, $n=1$, $l=0$, describes a negatively charged particle in the attractive Coulomb field ($Z>0)$.
Solution, which yields $\epsilon \to -m$ for $Ze^2\ll 1$, $n=1$, $l=0$, $Z>0$, after a change of $\epsilon \to -\epsilon$ describes the positively charged particle of the same mass in the field $Z<0$,  since Equation (\ref{KGFV}) does not change under simultaneous replacement $\epsilon\to -\epsilon$ \mbox{and $Z\to -Z$.
}

In the limit $Ze^2\ll 1$, Equation (\ref{epsilonCoul}) for a negatively charged spinless boson, in the ground state ($n=1$, $l=0$), produces $\epsilon \to m -\frac{(Ze^2)^2 m}{2 n^2}$ in accordance with the result for the Schr\"odinger particle.

For $Z>Z_{cr}=1/(2e^2)$ the particle, being  in the ground state ($n=1$), falls down to the center. Let $Ze^2 =1/2 +\delta$ for $0<\delta \ll 1$. Subsequenty, choosing positive-sign square root of solution (\ref{epsilonCoul}) we have for $Ze^2 =1/2 +\delta$,
$\epsilon \simeq \frac{m(1+i\delta)}{\sqrt{2}}$ and the wave function
$$\phi \propto e^{-i\epsilon t}\propto e^{m \delta t/\sqrt{2}}\to \infty \quad
{\rm for} \quad t\to \infty\,,$$
is not normalized, reflecting the fact of the falling of the negatively charged  particle to  the Coulomb center with $Z>0$ and the falling of the positively charged  particle to the Coulomb center at  $Z<0$. We dropped the negative-root solution of Equation (\ref{epsilonCoul}) as not physical one, since it arises  at $\epsilon \simeq -m$
already for small $Z>0$. However, note that the negative-root solution of Equation (\ref{epsilonCoul}),  $-\frac{m (1+i\delta)}{\sqrt{2}}$, for the negatively charged  particle  near the Coulomb center
for $Z>Z_{\rm cr}=1/(2e^2)$ yields  $\phi\propto e^{-m \delta t/\sqrt{2}}$, i.e.,  decreasing at $t\to \infty$.
This implies a possibility of a multi-particle  interpretation of the $\epsilon<0$ solution for the negatively charged particle in the field $Z>0$. We return to this question in Section \ref{DistrChsupershort}.

The value $Z_{cr}=68.5$. It means that the Mendeleev table would be closed on element with $Z_{cr}=68$, if the nuclei were point-like. As we have mentioned, the lightest spinless meson is the pion. The radius of the real nucleus with atomic number $A$ is found from the condition $4\pi \rho_0 R^3/3=A$, where
$\rho_0\simeq 0.16$ fm$^{-3}\simeq 0.48 m_\pi^3$. For a symmetric nucleus $A\simeq 2Z$ we estimate $R>a_{1\rm B}^{\pi}=1/(m_\pi Ze^2)$ (radius of the ground-state orbit for the pion) already for $Z>40$. Subsequently, the lowest pion orbit enters inside the nucleus and approximation of a point-like nucleus becomes invalid.

Note that, for $Z=Z_{\rm cr}$, $\epsilon_{\rm part}+\epsilon_{\rm a.part}=m \sqrt{2}>0$, and thereby pairs are not produced at such conditions. This peculiarity appears only for the case of the point-like Coulomb field. For a field, being cut at   $R\neq 0$ ($R\ll 1/m_\pi $, such that $
V=-Ze^2/r$ for $r>R$ and $V=-Ze^2/R$, the  model I,  or for $V=-\frac{Ze^2}{R} (\frac{3}{2}-\frac{r^2}{R^2})$, the model II at   $r<R$,
the ground state particle level continues to decrease with increasing $Z$ and decreasing $R$ and for  $Z=Z_{\rm cr}(R)>Z_{\rm cr}$,
it reaches $\epsilon =-m $. At $Z=Z_{\rm cr}(R)$, the sum  $\epsilon_{\rm part}+\epsilon_{\rm a.part}$ is  zero, corresponding to the spontaneous production of the  pairs for $Z\geq Z_{\rm cr}(R)$, at $R<R_{\rm cr}$.

{\bf Sommerfeld formula for electron.}
Electron has spin $1/2$. In the absence of the magnetic field spin and orbital spaces 
are orthogonal.  Thus one may expect that expression (\ref{epsilonCoul}) continues to hold also for electron after replacement $l+1/2\to |\vec{J}|+1/2=|\kappa|$, where $\kappa =-1,0,1...$ is integer number, since axial vectors of angular momentum  and spin are summed up,  $\vec{L}\to \vec{J}=\vec{L}+\vec{s}$. Subsequently, we have
\be
\epsilon^2_{n_r ,\kappa} =\frac{m^2}{1+\frac{Z^2 e^4}{ (n_r +\sqrt{\kappa^2 -(Ze^2)^2}\,\,)^2}}\,,\label{epsilonCoule}
\ee
where $n_r=n -|\kappa|=0,1,...$ is a radial quantum number. Now, falling to the center appears when  the ground state level reaches the value $\epsilon =0$. It occurs for $Z=Z_{\rm cr} =1/e^2 =137$. For a field cutted at $R\neq 0$, e.g., for the case $V=-Ze^2/r$ for $r>R$ and $V=-Ze^2/R$  for $r<R$, the ground state level continues to decrease with increasing $Z$ and for  $Z=Z_{\rm cr}(R)>Z_{\rm cr}$ it reaches $\epsilon =-m$. After that, the sum  $\epsilon_{\rm part}+\epsilon_{\rm a.part}$ reaches zero, corresponding to the  spontaneous production of the electron-positron pairs. Two electrons occupy the ground-state level and two positrons with $-\epsilon >m$ move  to infinity. 

Note that the same expression (\ref{epsilonCoule}) is derived from the exact solution of the Dirac equation in the Coulomb field, as we will see in Section \ref{CoulombF}.

\section{Dirac Equation for  Particle in  Static Electric Field, \boldmath{$A^\mu =(A^0,\vec{0})$}} \label{DiracCentral}
We are now at the position to focus on the problem of our main interest in this paper, i.e., to describe the behavior of  electrons in a strong static \mbox{electric field.}

Interaction with $4$-vector field $A^\mu =(A^0,\vec{A})$ is constructed with the help \linebreak of minimal coupling
\be
(\gamma^\mu\hat{p}_\mu -e\gamma^\mu {A}_\mu -m)\Psi =0\,, \label{DirField}
\ee
$p_\mu =i\partial_\mu$, $\gamma^\mu$ are ordinary Dirac matrices.

\subsection{ Dirac System in Case of  One-Dimensional Electric Field}

In the case of a static one-dimensional electric field ($\vec{A}=0$) using
replacement
\be\Psi =e^{-i\epsilon t}\tilde{\psi} (x)\ee
 we rewrite
Equation (\ref{DirField}) as
\be
(\epsilon -V +i\gamma^0\vec{\gamma}\frac{d}{dx} -\gamma^0 m  +V(x))\tilde{\psi} (x)=0\,.\label{Dirone}
\ee

We may rewrite Equation (\ref{Dirone}) as
\begin{eqnarray}
\label{psipr}
\tilde{\psi}^{\prime}=\hbar^{-1} \hat{D}\tilde{\psi}\,,\quad
\hat{D} =
\left(\begin{array}{ccc}
0 && m +\epsilon -V\\[3mm]
m-\epsilon +V&&
0
\end{array}\right),\quad \tilde{\psi} =
\left(\begin{array}{ccc}
G \\[3mm]
F
\end{array}\right)\,.
\end{eqnarray}

For further convenience, here we retained dependence on $\hbar$.

\subsection{ Dirac System in Central-Symmetric Field}
Introducing
\begin{eqnarray}
\label{GFsp1}
{\psi}_{jlm}=
\frac{1}{r}\left(\begin{array}{ccc}
G(r)\Omega_{jlM}(\vec{n}) \\[3mm]
iF(r)\Omega_{jl^{\prime}M}(\vec{n}^{\,\prime})
\end{array}\right),\quad \Omega_{jl^{\prime}M}=-\vec{\sigma}\vec{n} \Omega_{jlM}(\vec{n})\,,
\end{eqnarray}
where $\Omega_{jlM}$ is the spherical spinor, $j,M$ are full angular momentum and its projection, $j=l\pm 1/2$, $l$  is the orbital angular momentum, $l+l^{\,\prime}=2j$, $\vec{n}=\vec{r}/r$, $|\kappa|=j+1/2$.

After the separation of angular and spin variables, the Dirac system becomes
\begin{eqnarray}
\label{psiprsph}
\psi^{\prime}=\hbar^{-1} \hat{D}\psi\,,\quad
\hat{D} =
\left(\begin{array}{ccc}
-\tilde{\kappa}/r && m +\epsilon -V\\[3mm]
m-\epsilon +V&&
\tilde{\kappa}/r
\end{array}\right),\quad \psi =
\left(\begin{array}{ccc}
G \\[3mm]
F
\end{array}\right)\,,
\end{eqnarray}
$\tilde{\kappa}=\hbar \kappa$\,, $|\kappa|=j+1/2$. The ground state corresponds to $\kappa =-1$.
The one-dimensional result, see (\ref{psipr}),  follows from (\ref{psiprsph}) provided one puts $\kappa =0$ and replaces $d/dr\to d/dx$.

\subsection{Reduction of Dirac System to  Schr\"odinger Equation}\label{Schrod}

With the help of the replacement
\be
\phi =(m +\epsilon -V)^{-1/2} G\,,\label{phiG}
\ee
Equation (\ref{psiprsph}) is reduced to the  equation of the second-order in $r$-derivative, similar to the Schr\"odinger equation,
\be
\phi^{\,\prime\prime} +p^2(r)\phi =0\,,\quad p^2 =2m (E-U_{\rm ef}(r))\,,\label{phiGsecond}
\ee
where
\be
E=\frac{\epsilon^2 -m^2}{2m}\,,
\quad U_{\rm ef}(r)= \frac{\epsilon V}{m}-\frac{V^2}{2m}+\frac{\kappa (1+\kappa)}{2r^2 m}+U_s\,,\label{phiGsecondU}
\ee
\be
U_s = \frac{1}{4m}\left[\frac{V^{\,\prime\prime}}{m +\epsilon -V}+\frac{3}{2}\left(\frac{V^{\,\prime}}{m +\epsilon-V}\right)^2-\frac{2\kappa V^{\,\prime}}{r(m +\epsilon -V)}\right]\label{spinterm}
\ee
is the term appeared due to the spin. If $U_s$ were zero, after the replacement $\kappa\to l$ we would recover the Klein--Gordon--Fock equation for a spinless particle.

 At $r\to 0$, for $V=-Ze^2/r$, we have $U_s \to -\frac{1+4\kappa}{8mr^2}$. For 1 s level $\kappa =-1$, $U_s \to \frac{3}{8mr^2}$. In the latter case
\be
U_{\rm ef}(r)\to -\frac{(Ze^2)^2}{2m r^2} +\frac{3}{8mr^2}\,
\ee
for $r\to 0$.
The falling to the center in such a Schr\"odinger potential occurs  when \linebreak  $U_{\rm ef}(r)<-1/(8mr^2)$, cf. \cite{LL3}, which corresponds to  $Ze^2 >1$.

\subsection{Interpretation of Bound States in a Weak Field }\label{Interpretation}

The Dirac equation describes the electron and positron simultaneously. Therefore at appearance of the bound state in a potential well there arises a question regarding whether it relates to the electron or to the positron.
As example, consider the case of a weak external static  central-symmetric electric field produced by  a static source of a  positive charge distributed in a range $r$. Subsequently,
$V=-\zeta v(r)<0$ for the electron, where $\zeta >0$ is a parameter proportional to the depth of the potential well. As is known, for sufficiently small $\zeta$, the Dirac equation, as the Klein--Gordon--Fock equation,  can be transformed to the Schr\"odinger equation for a nonrelativistic particle.  The bound state for the electron appears first at a certain value of $\zeta$. At decreasing $\zeta$, this state is diluted in the continues spectrum with $\epsilon \geq m$.

The system of Dirac equations (\ref{psiprsph}) is symmetric in respect to replacements $\epsilon \to -\epsilon$, $V\to -V$, $\kappa\to -\kappa$, $G\to F$. Equation describing energy levels does not depend on $G$ and $F$. Thereby, it is symmetric, respectively, replacements $\epsilon \to -\epsilon$, $V\to -V$, $\kappa\to -\kappa$. In the case of the source of a positive charge, the electron undergoes attraction.
In the field of the  opposite-sign charge ($V\to -V$), the  electron undergoes repulsion. Because, in the attractive field, there appears the electron energy level going from the upper continuum, in the repulsive field there appears the electron energy level originating from the lower continuum. However, because the Dirac equation simultaneously describes electron and positron, if the electron moves in a repulsive field, then the positron moves in an attractive one. Thereby, the  electron level moving in a repulsive field from the lower continuum can be interpreted as  the positron  level ($\epsilon\to -\epsilon$, $\kappa\to -\kappa$) going  from the upper continuum (now in the field of attraction to the positron). It is natural to think that in a weak repulsive field for the electron for a small $\zeta <0$  a deeply bound level with $\epsilon\simeq -m$  should not exist. Because such a state nevertheless exists in the full set of solutions of the Dirac equation, after the replacement $\epsilon\to -\epsilon$, $\kappa\to -\kappa$, it should be interpreted as the positron state. This interpretation is  confirmed experimentally. In the field of a proton, there are electron bound states lying near the boundary of the upper continuum but there are no positron states with $\epsilon \simeq -m$. Vise versa, in the field of an antiproton, there exist positron levels with $\epsilon \simeq m$, but there are no electron levels with $\epsilon \simeq -m$. This picture is also established by the minimization of the energy in the mentioned cases. Namely, in the field of a positive charge, the  presence of the bound electron  is more energetically favorable when compared to the presence of the positron.

Statements done above seem obvious except the case, which I shall consider below in Section \ref{DistrChsupershort}, when polarization of the vacuum may result in a negative dielectric permittivity and attraction is  replaced by repulsion.

\subsection{Exact Solution for Electron in Coulomb Field of Point-Like Center}\label{CoulombF}

Consider the discrete spectrum $\epsilon <m$ of the Dirac equation in the potential $V=-Ze^2/r$.
We search $G$ and $F$ in Equation (\ref{psiprsph}) as
\be G=\sqrt{m +\epsilon}\, e^{-\tilde{r}/2}\tilde{r}^{g} (Q_1+Q_2)\,,\quad F=-\sqrt{m-\epsilon} \, e^{-\tilde{r}/2}\tilde{r}^{g} (Q_1-Q_2)\,,\label{GFrho}
\ee
where
\be
\tilde{r} =
2r\sqrt{m^2 -\epsilon^2}\,,\quad g =\sqrt{\kappa^2 -(Ze^2)^2}\,.\ee

This form of the solution, cf. \cite{LL4}, follows from asymptotic behavior of $G,F\sim r^{\pm g}$ at $r\to 0$ and
$G,F\sim e^{-\tilde{r}/2}$ at $r\to \infty$. Solutions $G,F\sim C_{1,2}r^{-g}$ are dropped (i.e., we put $C_{1,2}=0)$ due to the divergence of their contribution to the probability ($\int |\psi|^2 dr\to \infty$).

Setting (\ref{GFrho}) in  Equation (\ref{psiprsph}), we obtain a system of equations
\begin{eqnarray}
&\tilde{r} Q_1^{\,\prime}+\left(g -\frac{Ze^2 \epsilon}{\sqrt{m^2-\epsilon^2}}\right)Q_1 +\left(\kappa -\frac{Ze^2 m}{\sqrt{m^2-\epsilon^2}}\right) Q_2 =0\,,\nonumber\\
&\tilde{r} Q_2^{\,\prime}+\left(g +\frac{Ze^2 \epsilon}{\sqrt{m^2-\epsilon^2}}-\tilde{r}\right)Q_2 +\left(\kappa +\frac{Ze^2 m}{\sqrt{m^2-\epsilon^2}}\right) Q_1 =0\,.\label{Q1Q2a}
\end{eqnarray}

These equations are reduced to
\begin{eqnarray}
&\tilde{r} Q_1^{\,\prime\prime}+\left(2g +1-\tilde{r}\right)Q_1^{\,\prime} -\left(g -\frac{Ze^2 \epsilon}{\sqrt{m^2-\epsilon^2}}\right) Q_1 =0\,,\nonumber\\
&\tilde{r} Q_2^{\,\prime\prime}+\left(2g +1-\tilde{r}\right)Q_2^{\,\prime} -\left(g +1-\frac{Ze^2 \epsilon}{\sqrt{m^2-\epsilon^2}}\right) Q_2 =0\,.\label{Q1Q2}
\end{eqnarray}

As is seen, equations (\ref{Q1Q2}) are  
 symmetric under simultaneous replacement $\epsilon\to -\epsilon$
and $Ze^2\to -Ze^2$.

The finite solution for $\tilde{r}\to 0$ gets the form
\be
Q_1 =AF\left(g -\frac{Ze^2 \epsilon}{\sqrt{m^2-\epsilon^2}}\,, 2g +1,\tilde{r}\right)\,,\quad
Q_2  =B F\left(g +1 -\frac{Ze^2 \epsilon}{\sqrt{m^2-\epsilon^2}}\,, 2g +1,\tilde{r}\right)\,,\label{Q1Q21}
  \ee
where $F(\alpha,\beta,z)$ is the degenerate hypergeometric function. Setting $\tilde{r} =0$ in one of \mbox{Equation (\ref{Q1Q2a})}, we find relation
\be
B=-\frac{g -\frac{Ze^2 \epsilon}{\sqrt{m^2-\epsilon^2}}}{\kappa -\frac{Ze^2 m}{\sqrt{m^2-\epsilon^2}}}A\,.\label{BA}
\ee

Both of the hypergeometrical functions in (\ref{Q1Q21}) are reduced to polynomials, otherwise they would grow as $e^{\tilde{r}}$ for $\tilde{r}\to \infty$, which results in the divergence of the probability. From this requirement follows that $\alpha$ in $F(\alpha,\beta,z)$ equals a non-positive integer number, i.e.,
\be
g -\frac{Ze^2 \epsilon}{\sqrt{m^2-\epsilon^2}}=-n_r\,,\quad n_r =1,2,...\label{gamnr}
\ee

For $n_r =0$, only one of two functions is reduced to a polynomial. Subsequenty, $g =\frac{Ze^2 \epsilon}{\sqrt{m^2-\epsilon^2}}$ and $\frac{Ze^2 m}{\sqrt{m^2-\epsilon^2}}=|\kappa|$. If $\kappa <0$, then $B=0$ in Equation (\ref{BA}) and $Q_2 =0$, and the required condition is fulfilled. If $\kappa >0$,  then $B=-A$ and $Q_2$ is a divergent function at $n_r =0$. Thereby, permitted states are $n_r = 0, 1,...$ for $\kappa<0$ and $n_r = 1,2,...$ for $\kappa >0$. From (\ref{gamnr}), it also follows the solution for the negatively charged particle with $\epsilon <0$ for $Z<0$. In a single particle problem under consideration, one should drop such a solution, since it describes a strongly bound particle  already in  a weak field.
However, such a solution can be appropriately treated within a many-particle picture with taking the vacuum polarization and the electron condensation that originated in the lower continuum into account, as we argue below \mbox{in Section \ref{DistrChsupershort}.}

From (\ref{gamnr}), we obtain the Sommerfeld expression
\be
\epsilon =\pm m\left[1+ \frac{(Ze^2)^2}{(\sqrt{\kappa^2 -(Ze^2)^2}
+n_r)^2} \right]^{-1/2} \,,\label{Some}
\ee
 cf. Equation (\ref{epsilonCoule}). Note that, for $Z>0$, $Z<1/e^2$, only solution $\epsilon >0$ follows from (\ref{gamnr}), since $n_r+g >0$.
Thereby, ``$+$'' sign solutions (\ref{Some}) correspond to particles (electrons) in the field of the positively charged Coulomb center (or to antiparticles (positrons) in the field of the negatively charged Coulomb center).
The "$-$'' sign solutions (\ref{Some}), after replacements $\epsilon\to-\epsilon$, $\kappa\to -\kappa$ (after that "$-$'' sign  branch  coincides with "$+$'' sign branch) describe antiparticles with $\epsilon >0$ in the field of negatively charged Coulomb center ($Z<0$).

The ground state 1 s-level of the electron in the field of the positively charged Coulomb center ($Z>0$) corresponds to $\kappa =-1$, $n_r =0$. Its energy is
\be
\epsilon_{0}=m g_0\,,\quad g_0=\sqrt{1-(Ze^2)^2}\,.\label{enZ137}
\ee

At $Ze^2\geq 1$, there occurs falling of the electron to the center. Indeed, for $r\to 0$ following \mbox{(\ref{GFrho}), (\ref{Q1Q2})} we get
\be
G=a_1 r^g +a_2 r^{-g}\,,\quad F=b_1 r^{g}+b_2 r^{-g}\,.
\ee

For $Ze^2=1+\delta >1$, the value $g =i\sqrt{2\delta}$ becomes imaginary and solutions oscillate as
\be
C_1\cos({|g|}\ln r) +C_2\sin ({|g|}\ln r)\,,
\ee
that corresponds to not normalized probability $\int_{0}^{\infty}|\psi|^2 dr$.
At $Ze^2 =1+\delta$, $0<\delta \ll 1$, solution of Equation (\ref{enZ137}) yields $\epsilon =+i m\sqrt{2\delta}$ and the electron wave function
grows as $\Psi \propto e^{+m\sqrt{2\delta}t}$, indicating the falling of the electron to the center. The solution of opposite sign  (see Equation (\ref{Some})) arises from the lower continuum at $V\to 0$. In the single-particle problem a negative-energy solution should be dropped. Note that at $Ze^2 =1+\delta$, it yields $\epsilon =-i m\sqrt{2\delta}$ and  $\Psi\to 0$ at $t\to \infty$ that may suggest an interpretation.
However, an appropriate interpretation proves to be possible only beyond the single-particle problem, as will be shown in Section \ref{DistrChsupershort}.

Solutions  (\ref{Some}) and (\ref{enZ137}) hold formally for the positron in the Coulomb potential of the nucleus with the charge  $Z<0$. Within the single-particle problem under consideration, appropriate interpretation  again exists for  the solution, where energy originates from the upper continuum  decreasing with increasing $-Z$, rather than the negative-energy solution, similarly to that happened for the electron at $Z>0$.

For $Z>0$, only two  electrons (due to Pauli principle),  if they have  occupied the ground state, undergo falling to the Coulomb center
for $Ze^2 =1$. For levels with the quantum number $n_r >0$, we have $\epsilon_{n_r,\kappa}> 0$ for $Z=1/e^2$.
Now, assume that  the ground-state level was empty and we adiabatically increase $Z$. There is no appropriate solution of the single-particle problem for the point-like nucleus with $Z>1/e^2$ in this case.

{\bf Avoiding  problem of falling to the center.}
 A reasonable interpretation may appear, only if one assumes that the  nucleus has a size  $R\neq 0$, and then we may safely decrease $R$. First assume that $R\ll r_\Lambda =1/m$. In the limit $\Lambda =\ln (r_\Lambda/R)\gg 1$ for the ground-state level of the electron, one gets  \cite{Popov1971ZhETF33,ZeldovichPopov1972}
\be
\epsilon_0 (\zeta <1)=mg_0/\mbox{th}(\Lambda g_0)\,,\quad {\rm for}\quad \zeta=Ze^2 <1\,.\label{gamma0a}
\ee

For $\zeta <1$, $\Lambda g_0 \gg 1$, the value $\mbox{th}(\Lambda g_0)\simeq 1-2e^{-2\Lambda g_0}$  rapidly tends to unity and Equation (\ref{gamma0a}) coincides with (\ref{enZ137}). For $R\neq 0$, the point $\zeta =1$ is already not a singular point for the function $\epsilon_0 (\zeta)$.
Equation (\ref{gamma0a})  is analytically continued in the region $\zeta >1$. For $\zeta$ close to unity, we have
\be
\epsilon_0 (\zeta >1)=m\tilde{g}_0/\mbox{tg}(\Lambda \tilde{g}_0)\,,\quad {\rm for}\quad \zeta =Ze^2 >1\,,\label{gamma01}
\ee
where $\tilde{g}_0=\sqrt{\zeta^2 -1}$.
At any $R\neq 0$ the curve $\epsilon_0 (\zeta >1)$ continues to decrease with increasing $\zeta$ and reaches the boundary of the lower continuum. It occurs at $\zeta_{\rm cr}=1+\pi^2/(2\Lambda^2)+O(\Lambda^{-3})$.

 A comment is in order. The single-particle solution for $R\to 0$ should be modified.
Indeed,  for $R$ as small as $R\sim r_{\rm L}\simeq r_{\Lambda}e^{-3\pi /(2e^2)}$,  the multi-particle effects of the polarization of the vacuum should be included, and the problem goes beyond the single-particle one, see the below consideration  in Section \ref{Polarization}.
\\

\subsection{Avoiding Problem of Falling to  Center in Realistic Treatment. Spherical Nucleus of Finite Size}\label{Finite}

For  
the Coulomb field with the charge $Z<1/e^2$, the electron in the ground state is typically situated at distances
$\sim a_{1\rm B}=1/(Z_{\rm obs} e^2 m)> 1/m$ and distribution of the charge $Z(r)$ at distances $r\sim R_{\rm nucl}\ll a_{1\rm B}$ almost does not affect the electron motion. In the realistic problem, the  nucleus has a finite size, $R_{\rm nucl}\simeq r_N A^{1/3}\ll a_{1\rm B}$, where $A$ is the atomic number, $r_N\simeq  1.2$ fm, and, thereby, the potential is smoothen at $r<R_{\rm nucl}$.
The falling to the centrum does not occur, as it has been mentioned. Even for $Z\gg 1/e^2$, the electron density remains to be distributed at finite distances.

Taking into account of the distribution of the charge inside the nucleus, we have
\be
V(r)=-\zeta f(r/R_{\rm nucl})/R_{\rm nucl} \quad {\rm for}\quad 0<r<R_{\rm nucl}, \quad V=-\zeta/r\,,\quad {\rm for}\quad r>R_{\rm nucl}\,.\label{modelI}
\ee

Two models have been employed in the literature:
model I, when $f(x<1)=1$, that corresponds to the surface distribution of the charge, and model II, when  $f(x<1)=(3-x^2)/2$, which describes distribution of protons with the constant volume density.

The energy shift of the electron level can be found with the help of the perturbation theory that is applied to  the Dirac system (\ref{psiprsph}). Following \cite{ZeldovichPopov1972},
\be
\beta =\frac{\partial\epsilon}{\partial \zeta}=\int V(r)(G^2 +F^2)dr/\zeta <0\,,\label{betaZP}
\ee
i.e., the curve $\epsilon (\zeta)$ decreases monotonically with increasing $\zeta$ and crosses the boundary of the lower continuum with a finite value $\beta$. After that, $\epsilon (\zeta)$ acquires an exponentially small imaginary part.

Because the exact solution of the Coulomb problem for $r>R$ looks rather cumbersome and for $r<R$ is impossible for a realistic cut of the potential, it is natural to use approximate methods. Most economical is a semiclassical approach. Here, we should notice that the replacement (\ref{phiG}) becomes singular for $\epsilon <-m$ in the point $V(r_1)=m+\epsilon <0$. Because to this, the effective potential
\be
U_{\rm ef} (r,\epsilon)=\frac{3}{8m} (r-r_1)^{-2}+...\to \infty\,, \quad {\rm for}\quad r\to r_1\,,
\ee
and semiclassical expressions loose their sense due to the divergency of the integral \linebreak $\int^r 2m (E-U_{\rm ef} (r,\epsilon))^{1/2} dr$. However, this is only a formal problem, since the initial Dirac \mbox{system (\ref{psiprsph})} has no singularity  at $r\to r_1$. To avoid the problem one should bypass the singular point in the complex plane, as one usually does bypassing turning points,  or one may apply the semiclassical consideration straight to the linear Dirac equations. Note that, in the one-dimensional case corresponding to $\kappa =0$, see Equation (\ref{psipr}), the mentioned singularity occurs in the turning points, and one may use standard semiclassical methods.

The probability of the spontaneous production of positrons is determined by the width of the corresponding electron level, $\mbox{Im}\,\epsilon$, for $\mbox{Re}\,\epsilon <-m$. Thus the width is found from the solution of the Dirac equation. The value $\Gamma$, which determines probability of the positron production, $W\sim e^{\Gamma t}$, can be expressed directly through components of the Dirac bispinor ($G$ and $F$).
It yields the flux of particles going to infinity (at normalization on one particle):
\be
\Gamma =\int \psi^{\dagger}\gamma^0\vec{\gamma}\psi d\vec{f}=2\mbox{Im}(FG^*)|_{r\to  \infty}\,.\label{gammaflux}
\ee

\section{Semiclassical Approach to Dirac Equation Transformed to Second-Order Differential Equation}\label{transformed}

\subsection{Accuracy of Calculation of  Energy Levels in Semiclassical Approximation}
 Substituting $\psi =Ae^{iS/\hbar}$, where $A$ and $S$ are real quantities,  in equation
\be
\hbar^2\Delta\psi +p^2 (r)\psi =0\,
\ee
we find two equations
\be
\hbar^2\Delta A+p^2 A=A(\nabla S)^2\,,\quad i\hbar(2\nabla A\nabla S+A\Delta S)=0\,.
\ee

For a convenience, the dependence on $\hbar$ is recovered here.
The Hamilton--Jacobi equation for the action $(\nabla S)^2 =p^2$ is obtained provided
\be
\hbar^2 \frac{A^{\,\prime\prime}}{p^2 A}\sim \frac{\hbar^2}{(pl)^2}\sim \left(\frac{d\tilde{\lambda}}{dr}\right)^2\ll 1\,,\quad \tilde{\lambda} =\frac{\hbar}{ p}\,,\label{semiclaccuracy}
\ee
where $l$ is the typical size of the potential $V$. For the Coulomb potential at typical distances $r\sim 1/(2m)$ characterizing ground-state electron with $\epsilon \simeq -m$ we have $p\sim \tilde{g}/r$. From estimate (\ref{semiclaccuracy}), we see that the semiclassical approximation for the wave function for such distances is accurate up to terms $1/\tilde{g}^2$, $\tilde{g} =\sqrt{\zeta^2 -\kappa^2}$ for $\zeta >|\kappa|$.

Using the Bohr--Sommerfeld quantization rule, we have
\be
\frac{\hbar^2}{(pl)^2}\sim \frac{\hbar^2}{(\int_{r_0}^{r_{-}}pdr)^2}\sim \frac{1}{\pi^2(n_r +\gamma)^2}\,,
\ee
where the phase $\gamma \sim 1$, $n_r=0,1,...$, $r_0$, and $r_{-}$ are the turning points separating the classically allowed region. Thus even in calculation of the energy  of the levels with small quantum numbers one may consider on the error not larger that 10\% .

Finally, let us notice that the transition  from the Dirac equation in the external field to the corresponding more simple Hamilton--Jacobi equation has been used in many investigations, cf. \cite{Pauli1980,Rubinow1963,Stachel1977}. The case of the deep electron levels, with the energy $\epsilon \lsim -m$, was studied in \cite{Eletskii:1977na,Mur:1978nb,Mur:1978ke,Popov:1978kk,Popov:1979gq}.

\subsection{Semiclassical Approximation to  Coulomb Field of Point-Like Nucleus}
In the field $V=-\zeta/r$, for $\zeta <|\kappa|$, the semiclassical method results in exact expression for the energy spectrum. Let us show this.
For that, we do replacements
\be
G=\sqrt{\frac{m+\epsilon}{r}}\,(\chi_1 +\chi_2)\,,\quad F=\sqrt{\frac{m-\epsilon}{r}}\,(\chi_1 -\chi_2)\,.
\ee

Subsequently, the system of two Dirac equation (\ref{psiprsph}) reduces to equations
\be
\chi_i^{\,\prime\prime}+p_i^2(r)\chi_i =0\,,\quad i=1,2\,,
\ee
with
\be
p_i (r)=\left[\epsilon^2 -m^2 -\frac{2\epsilon \zeta \pm \sqrt{m^2-\epsilon^2}}{r}+\frac{\zeta^2-\kappa^2+1/4}{r^2}\right]^{1/2}\,.
\ee

Adding the Langer correction to the effective potential results in replacements $p_i\to p_i^*$, we find

\be
p_i^* (r)=\sqrt{-a+2b/r -g^2/r^2}\,,\quad a=m^2-\epsilon^2\,,\quad b=\epsilon\zeta\pm \frac{1}{2}\sqrt{m^2 -\epsilon^2}\,,\quad g^2=\kappa^2-\zeta^2\,.\label{abc}
\ee

Subsequently, applying the Bohr--Sommerfeld quantization rule, we have
\be
\int_{r_0}^{r_{-}}p_i^* dr= (a^{-1/2} b -g)\pi =(n_r +1/2)\pi\,.\label{BZabc}
\ee

From here, we recover the  exact result (\ref{Some}). To get  (\ref{Some}) from an exact solution of the Dirac equations, we have performed a cumbersome analysis of hypergeometric functions, whereas the semiclassical approach needs taking only one simple integral.

After replacements $b\to\epsilon \zeta$, $g^2\to (l+1/2)^2 -\zeta^2$,
Equation (\ref{abc}) is also valid for spinless bosons. Performing integration  leads us to the  exact expression (\ref{epsilonCoul}).

\subsection{Finite Nucleus. Semiclassical Wave Functions and Quantization Rule}

Certainly 
, it is also possible to apply semiclassical approach to Equation (\ref{phiGsecond}) with effective potential in the form (\ref{phiGsecondU}), (\ref{spinterm}). In the range, where the parameter of
applicability of semiclassical approximation is $|d\tilde{\lambda}/dr| \sim 1$,  the usage of  Dirac equations presented in different  forms leads to slightly different results. For instance, applying (\ref{phiGsecond}) to the Coulomb field does not yield the exact result for the energy of the levels, although the accuracy of the approximation proves to be appropriate. For the electron energy  $\epsilon <-m$ the variable replacement (\ref{phiG}) leads to the  singularity  in the point $r_1$,
where $V(r_1)=m+\epsilon <0$ . Near this point, semiclassical expressions become invalid due to divergence of the contribution to the action $\int[2m(E-U_{\rm ef})]^{1/2} dr$. However, as it was mentioned, this circumstance is not reflected on the calculation of the energy levels, since $r_1$ is situated under the barrier, where wave functions prove to be exponentially small.

The electron energy levels can be found with the help of the Bohr--Sommerfeld quantization rule \cite{Eletskii:1977na} applied to the Dirac equation presented in the form (\ref{phiGsecond}) with effective potential in the form (\ref{phiGsecondU}), (\ref{spinterm}). We have
\be
\int_{r_0}^{r_{-}}p^* dr =(n_r +\gamma^{\,\prime})\pi\,.\label{quantruleef}
\ee

Value $p^*$ is obtained from expression (\ref{phiGsecond}) after taking the Langer correction into account, i.e., after  doing the replacement $\kappa (1+\kappa)/r^2 \to (\kappa+1/2)^2/r^2$ in the expression for the effective potential. The value of the phase $\gamma^{\,\prime}$ depends on   whether the turning point is inside the nucleus or outside it.
In the latter case, the potential is $V=-\zeta/r$ and $\gamma^{\,\prime} =3/4$ for $\kappa =-1$ and
$\gamma^{\,\prime} =1/2$ for $\kappa \neq -1$.

The contribution to the normalization of the semiclassical wave function from the classically forbidden region is usually dropped. In order to understand accuracy of this approximation consider the probability of the presence of the electron in sub-barrier region $r_{-}<r<r_{+}$:
\be
W_0 =\int_{r_{-}}^{r_{+}}(G^2+F^2)dr\,.
\ee

To be specific, let us put $\epsilon =-m$ and consider $\zeta \gg |\kappa|$. The wave function in the classically allowed region is \cite{Migdal1975}:
\be\chi =(c_0/\sqrt{p^*})\sin (\int_{r_0}^r p^* dr +\pi/4)dr\,.\label{chinorm}
\ee

Constant $c_0$ is found from the normalization condition \cite{Migdal:1977rn},
$$2\int_{r_0}^{r_{-}}(\epsilon -V)\chi^2 dr/m \simeq 1\,.$$

 Subsequently, we expand the effective potential (\ref{phiGsecondU}) near the turning point. For $V=-\zeta/r$, \mbox{we obtain}
\be
U_{\rm ef}=\zeta/r -\tilde{g}^2/(2r^2 m)=U(r_{-})+4m^2(r-r_{-})/\zeta +..., \quad \tilde{g}=\sqrt{\zeta^2-\kappa^2}\,,\label{Unearturning}
\ee
for $r-r_{-}\ll r_{-}\sim \zeta$. The solution of Equation (\ref{phiGsecond}) in potential (\ref{Unearturning}) is expressed through the Airy function
\be
\chi (r)=(-V)^{-1/2}G=c_0 Ai (2\zeta^{-1/3}(r-r_{-}))\,.\label{Airy}
\ee

 The probability of finding the particle in the sub-barrier region is
\be
W=-\int_{r_{-}}^{\infty}(V/m +1)\chi^2 dr \simeq c_0^2 \int_{0}^{\infty}Ai^2 (2x\zeta^{-1/3})dx =c_1 \zeta^{-1/3}\,,\label{AiryExp}
\ee
where $c_1 =3^{4/3}\Gamma^2 (2/3)/16\pi \simeq 0.158$.

Thus, the probability of a penetration of the electron in classically forbidden region is  numerically small  for $\zeta \sim 1$, and it falls down with increasing $\zeta$. This justifies  that we neglected the contribution of the region $r>r_{-}$ at the normalization of the wave functions  (taking $r_0 <r<r_{-}$).
Note that the quantization rule remains applicable with a larger accuracy, $1/\zeta^2$, since, at its derivation, it was not used how wave functions are normalized. Strictly speaking, in the case of quasistationary levels, the quantization rule is slightly modified, due to  $\mbox{Im}\epsilon \neq 0$, cf. \cite{Sergeev1991}. However, changes of the energy levels are exponentially small, due to the exponential smallness of the penetrability of the barrier.

With the semiclassical  $\chi$ function, we obtain an expression for the averages $\overline{r^\lambda}$.
For $\epsilon =-m$ and $\zeta \gg |\kappa|$, one has  \cite{Eletskii:1977na},
\be
\overline{r^\lambda}=\zeta^{\lambda}\frac{3\pi^{1/2}(\lambda +2)\Gamma (\lambda +1)}{m^\lambda 2^{\lambda +3}\Gamma (\lambda +5/2)}\left[1-\frac{(\lambda +3/2)\kappa^2}{(\lambda +2)\zeta^2}+...\right]\,.\label{rlambda1}
\ee

$\Gamma (x)$ is the Euler $\Gamma$-function. For $\zeta \sim 1$, the accuracy of this expression is not as good, but it increases appreciably with increasing $\zeta$.

The quantity $\overline{r}$ characterizes the mean radius of the bound state at
$\epsilon =-m$, values $\overline{r^{\lambda}}$ at $\lambda =1/2,3/2,2$ are met  in the problem of the modification of the value $Z_{\rm cr}$  due to a screening of the charge by other electrons of the ion (if they are), see below in Section \ref{screenK}. A comparison of the semiclassical expressions with  the exact solutions numerically found shows an appropriate accuracy of the semiclassical results, even for $\zeta\sim |\kappa|\sim 1$. For $\zeta\gg \kappa \gg 1$, the result (\ref{rlambda1}) coincides with the corresponding asymptotic of the \mbox{exact solution.}

\subsection{Critical Charge of the Nucleus}

Let us calculate the critical charge of the nucleus  (when the electron level with quantum numbers $n,\kappa$  reaches $\epsilon =-m$). Using the Bohr--Sommerfeld quantization rule in the\mbox{ form (\ref{quantruleef}),}
one obtains, cf. \cite{Marinov1974},
\be
mR_{\rm nucl}=\tilde{g}^2/(2\zeta\mbox{ch}^2 y)\,,\label{Rcrch}
\ee
where $y$ is positive root of the equation
\be
y-\mbox{th} y =\frac{(n_r +\gamma_1)\pi -\tilde{\gamma}}{2\tilde{g}}\,,\quad \tilde{\gamma}=\mbox{arcctg} (\Xi/\tilde{g})\,,\label{Rcr}
\ee
$n_r =0,1,...$ radial quantum number, $\gamma_1 =3/4$ for $ns$ levels and $\gamma_1 =1/2$ for $\kappa \neq -1$.

 In Ref. \cite{Marinov1974}, quantity $\Xi$ was found from matching of the exact solution inside the nucleus and semiclassical one outside the nucleus. As was shown in \cite{Voskresensky1977}, usage of the  semiclassical solutions both inside and outside the nucleus does not spoil the accuracy of the result. Therefore we further follow  consideration of  \cite{Voskresensky1977}.

For the model I, the semiclassical solution inside the nucleus coincides with the exact one and we find
\be
\Xi =\beta\mbox{ctg}\beta\,\quad \beta =\sqrt{\zeta(\zeta -2R_{\rm nucl}m)}\,.\label{xiI}
\ee

Here, note that a first estimate of $R_{\rm cr}$ in this model was performed in \cite{Krainov1971}, where it was taken $\tilde{\gamma}=\zeta$, that differs from that follows from (\ref{Rcr}), (\ref{xiI}).

For the model II, an analytical expression can be found expanding $p(r<R)$  in the parameter $\zeta$,
\be
\tilde{\gamma}=\mbox{arcctg}[((p^*(R_{\rm nucl})/(\tilde{g}m))\mbox{ctg}\int_0^{R_{\rm nucl}}p^* dr]\,,
\ee

\begin{eqnarray}
&\int_0^{R_{\rm nucl}}p^* dr =\int_0^1 dx[\zeta^2 f^2(x)-2\zeta R_{\rm nucl}m f(x)-9/(4f^2(x))]^{1/2}\nonumber\\
&=\frac{4}{3}\zeta\left[1-\frac{c_2}{\zeta^2}-\frac{3}{4}\frac{R_{\rm nucl}m}{\zeta}+O\left(\frac{1}{\zeta^4}, \frac{R_{\rm nucl}^2m^2}{\zeta^2}\right)\right]\,,\quad c_2=\frac{9}{32}\left(1+\frac{1}{\sqrt{3}}\mbox{arth}\frac{1}{\sqrt{3}}\right)\,,\quad \kappa =-1\,,
\end{eqnarray}
where $f(x)$ follows Equation (\ref{modelI}), here for the model II.
Although the parameter of applicability of semiclassical expressions to the Coulomb field is $\tilde{g}\gg 1$, the difference of the above obtained expression with the result of the exact calculation is less than few percents, even at \mbox{ $\zeta =\zeta_{\rm cr}\simeq 1.24$.}

For $\zeta \sim 1$, expanding  (\ref{Rcr}) in $1/y$ and dropping numerically small term $e^{-2y}$, from
Equation (\ref{Rcrch}), we finally  find
\be
R_{\rm nucl}\simeq \frac{2\tilde{g}_{\rm cr}^2}{\zeta_{\rm cr}m}\left[\mbox{exp}\left(\frac{\pi(n_r+\gamma_1)-\tilde{\gamma}}{\tilde{g}_{\rm cr}}+2\right)+2\right]^{-1}\,,\label{Rcre}
\ee
from where we find $Z_{\rm cr}(R_{\rm nucl})$.

\subsection{Number of Levels Which Crossed Boundary of Lower Continuum}

Now, let us find   the number of levels $n_{\kappa}$ with fixed quantum number $\kappa$ and  the total number of levels $N$, which have crossed the boundary $\epsilon =-m$. For this aim \cite{Mur:1978ke}, we need to use the Bohr--Sommerfeld quantization rule at $\epsilon =-m$.
For $\tilde{g}\gg 1$, we have $d\tilde{\lambda}/dr\ll 1$. For $\zeta\gg 1$, this means that $\zeta -|\kappa|\gg \zeta^{-1}$, i.e., semiclassical approximation can only be violated for states with the  momenta at which $\zeta -|\kappa|\lsim \zeta^{-1}$. The accuracy of the semiclassical expressions for the wave function is $\sim 1/\zeta^2$, cf. \cite{Migdal:1977rn}. Taking these approximations into account, employing the Bohr--Sommerfeld quantization rule, we obtain
\be
n_\kappa =\frac{1}{\pi}\int (V^2 +2Vm -\kappa^2/r^2)^{1/2} dr \,.\label{nkappa}
\ee

For the potential that is given by Equation (\ref{modelI}), for $R_{\rm nucl}m\ll 1$,
 we obtain
\be
n_\kappa =\frac{g}{\pi}[2(\mbox{Arth}\sqrt{1-\eta}-\sqrt{1-\eta})+h(\rho)\,,
\ee
where $\rho =|\kappa|/\zeta$\,, $\eta =R_{\rm nucl}/r_{-}=2R_{\rm nucl}m/(\zeta(1-\rho^2))$, $r_{-}$ is the turning point in the effective potential, $h(\rho)$ takes into account integral over the interior region of the nucleus
$0<r<R_{\rm nucl}$,
\be
h(\rho)=(1-\rho^2)^{-1/2}\int_{x_0}^1 [f^2 (x)-\rho^2 x^{-2}]^{1/2}dx\,,\label{hrho}
\ee
where $x_0=x_0(\rho)$ is the root of equation $xf(x)=\rho$.

For $Ze^3\ll 1$ (at this condition distribution of  electrons, which fill the vacuum shell, only slightly modifies the bare potential, as we shall see below),
Equation (\ref{nkappa}) correctly determines the distribution of electrons with $\epsilon <-m$ of the supercritical atom over the momenta $j=|\kappa|-1/2$. The maximum value of $j$ corresponds to $r_{-}=R_{\rm nucl}$, $\eta =1$,
\be
\kappa_{\rm max}=\zeta -R_{\rm nucl}m+O(R_{\rm nucl}^2m^2/\zeta)\,.
\ee

The total number of levels with $\epsilon <-m$,
\be
N=\sum_{\kappa} n_{\kappa}\,,
\ee
can be found by replacing the summation by the integration. We should take into account that, in the Dirac equation, $|\kappa|\geq 1$. Thereby, we still should subtract spurious term $\kappa =0$. Thus,
\begin{eqnarray}
&N=\int dr[\frac{1}{2} (V^2 +2Vm)r -\frac{1}{\pi}(V^2+2Vm)^{1/2}]\nonumber\\
&=A_1 \zeta^2 \ln \frac{\zeta}{R_{\rm nucl}m}+A_2 \zeta^2 +A_3 \zeta\ln\frac{\zeta}{R_{\rm nucl}m}+A_3\zeta+A_4+...,\label{Nsem}
\end{eqnarray}
where $A_1 =1/2$, $A_2 =\int_0^1 f^2(x)xdx-\ln 2 -1$\,, $A_3 =-1/\pi$, \linebreak $A_4 =-\frac{1}{\pi}(\int_0^1 f(x) dx +\ln 2 -2)$.

\begin{figure}[H]
\includegraphics[width=5.8cm,clip]{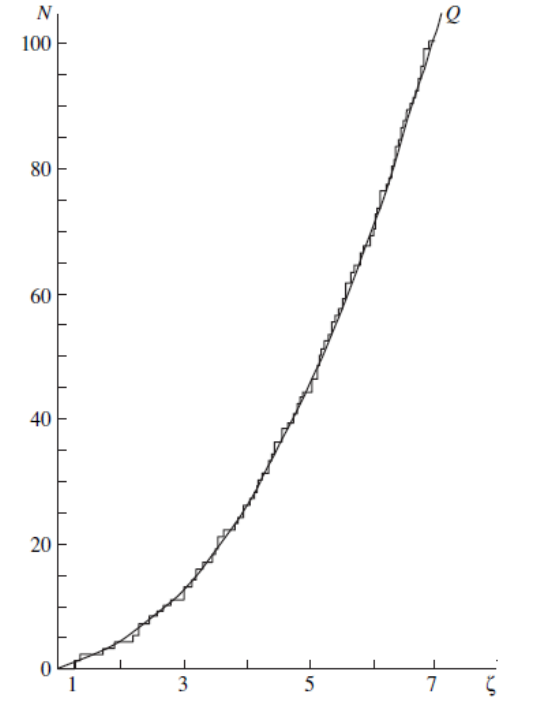}
\caption{The number of levels with $\epsilon <-m$ for the potential of the model II, cf. \cite{Mur:1978ke}. The stepwise broken line represents a numerical solution of the Dirac equation, while the curve $Q$ was computed according to the semiclassical Equation (\ref{Nsem}).}\label{NumberLevels}
\end{figure}

For the model II, the result of this calculation is shown in Figure \ref{NumberLevels}. Again, we observe an excellent accuracy of the semiclassical result, even for $\zeta\sim 1$.

\subsection{Energy of Single-Particle Levels at $\epsilon <-m$}

\subsubsection{Energy Spectrum for $|\epsilon|-m\ll m$}

Expand the effective potential in $m+\epsilon$, cf. \cite{Eletskii:1977na}:
\be
U_{\rm ef}(r,\epsilon)=\sum_{n=0}^{\infty} (m+\epsilon)^n u_n (r)\,,
\ee
where $U_{\rm ef}(r,\epsilon)$ can be taken following  Equation (\ref{phiGsecondU}).
Here, $u_0 (r)=U_{\rm ef}(r,\epsilon =-m)$. \mbox{For $n\geq 1$,}
\be
u_n = \frac{V}{m^n}\delta_{n1}+\frac{1}{mV^n}\left[-\frac{V^{\,\prime\prime}}{4V}+
\frac{3(n+1)}{8}\left(\frac{V^{\,\prime}}{V}\right)^2+\frac{\kappa V^{\,\prime}}{2rV}\right]\,,
\ee
where $\delta_{n1}$ is the Kronecker symbol.

The energy of the levels is found from the Bohr--Sommerfeld quantization condition
\be
\int_0^{r_{-}}\sqrt{-2mu_0} dr+(m+\epsilon)\int_0^{r_{-}}\sqrt{-2 u_1} dr+O((1+\epsilon/m)^{3/2})=
(n_r +\gamma^{\prime})\pi\,.\label{u0u1}
\ee
As before, $\gamma =3/4$ for levels with $\kappa =-1$ and $\gamma =1/2$ for $\kappa \neq -1$.
With the help of (\ref{u0u1}) we find
\be
\epsilon =-m+\beta (\zeta_{\rm cr}-\zeta)+...\,,
\ee
\be
\beta =f_2/f_1\,,\quad f_1 =\int_0^{r_{-}} \sqrt{-2m^2 u_1}\,dr\,,
\quad f_2 =\int_0^{r_{-}}\frac{\zeta f^2/(mR_{\rm nucl})-f}{\sqrt{-2u_0/m}\,R_{\rm nucl}}dr\,.\nonumber
\ee

A comparison of numerical calculation done following these expressions with that for the
exact Dirac equation again shows a good agreement. Note that the value $\beta$ determines the threshold behavior of the probability of the production of positrons.

 \subsubsection{Energy Spectrum for $|\epsilon|\ll -m$}

 This spectrum has been found in \cite{Mur:1978ke}. For $\zeta\gg \zeta_{\rm cr}$, many levels have energies $|\epsilon|\ll -m$. In this case, as follows from
 Equation (\ref{phiGsecondU}) and (\ref{spinterm}), the  terms $\propto \kappa$ in the centrifugal potential and in the spin term cancel each other. Approximately, we have
\be
p^*(r)\simeq [(\epsilon -V)^2-\kappa^2/r^2]^{1/2}\,.\label{pdeep}
\ee

For $k=\sqrt{\epsilon^2-m^2}<\zeta/R_{\rm nucl}$, the turning point $r_{-}$ lies outside the nucleus, $r_{-}>R_{\rm nucl}$. Employing the Bohr--Sommerfeld quantization condition, we get
\begin{eqnarray}
&k_n \simeq |\epsilon_n| =c_0\zeta R_{\rm nucl}^{-1}e^{-n\pi/\zeta}=\zeta R_{\rm nucl}^{-1} e^{-(n-n_{*})\pi/\zeta}\,,\quad n>n_{*}\,,\label{kne}\\
&c_0 =\mbox{exp}(\int_0^1 f(x)dx -1)\,,\quad n_{*}=\zeta \pi^{-1}(\int_0^1 f(x)dx -1)\,.\nonumber
\end{eqnarray}

For deeper levels, $k>\zeta R_{\rm nucl}^{-1}$, classically permitted region $r_0<r<r_{-}$  is completely inside the nucleus. Thereby, the spectrum is entirely determined by the expression for $f(x)$:
\be
k_n=\zeta R_{\rm nucl}^{-1}f(\Xi_n)\,,\quad 1\ll n\ll n_{*}\,,
\ee
where $\Xi_n$ is the root of equation
\be
\int_0^\Xi f(x)dx -\Xi f(\Xi)=n\pi/\zeta\,.
\ee

For example, for the model II at  $1\ll n\ll n_{*}$, we have
\be
k_n =\frac{\zeta}{2R_{\rm nucl}}\left[3-(n/n_{*})^{2/3}\right]\,,\quad n_{*}=\frac{\zeta}{3\pi}\,.
\ee

From these expressions, it is easy to find expression for the level density $dn/d\epsilon$. For model II, we find
\be
dn/d\epsilon =C y^{-1}\,,\quad {\rm for}\quad 0<y<1\,
\ee
and
 \be
dn/d\epsilon =C (3-2y)^{1/2}\,,\quad {\rm for}\,,\quad  1<y<3/2\,
\ee
for $y=kR_{\rm nucl}/\zeta$, $C=const$. From here, we see the accumulation of levels toward the boundary $\epsilon =-m$ ($k\to 0$).

For levels with arbitrary angular momenta the ``Coulomb'' part of the spectrum gets the form
\be
\epsilon_{n\kappa}=-\zeta R_{\rm nucl}^{-1} c(\rho) \mbox{exp}(-n\pi/\tilde{g})\,,\label{encrho}
\ee
where $\rho =|\kappa|/\zeta$, $0<\rho <1$. Pre-exponential factor
\be
c(\rho)=\mbox{exp}\left[\ln (2 (e\rho)^{-1}(1-\rho^2))-(1-\rho^2)^{-1/2} \mbox{Arth} (1-\rho^2)^{1/2}+h(\rho)\right]\,,\label{crho}
\ee
where $h(\rho)$ that is given by Equation (\ref{hrho})  depends on the $f(x)$, $e=2.718...$ is the Euler number. The function $c(\rho)$ monotonically decreases with increase of $\rho$ from 1 for model I and from $\simeq 1.4$ for model II at $\rho =0$ up to zero  for $\rho =1$ in both models.

Equation (\ref{encrho}) is obtained at the condition that the turning point $r_{-}$ lies inside the nucleus. The condition of applicability of Equation (\ref{encrho}) is $\tilde{g}/\pi\ll n< n_{\kappa}$. Because $n_\kappa \simeq (\tilde{g}/\pi)\ln (\zeta/R_{\rm nucl})$, then, due to large values of the logarithm, this equation describes most of the levels  crossed the boundary  $\epsilon <-m$.

The exponential dependence of $\epsilon_n$ on $n$ and the accumulation of levels near $\epsilon =-m$, as follows from Equations (\ref{kne}) and (\ref{encrho}), are related to the fact that $U_{\rm ef}\simeq -\tilde{g}^2/r^2$ for $r\to 0$. If  $R$ was zero, the electrons would collapse to the center. The spectrum of the Schr\"odinger equation in such a potential behaves as \cite{MorsFeshbach},
\be
E_n=E_0 e^{-2\pi n/\tilde{g}}\,,
\ee
where $E_0$ is the energy of the lowest level. In our case, $E\simeq \epsilon^2/2m$, and thereby we recover Equation (\ref{encrho}) for $c(\rho)=1$.

\subsection{Exponential Estimate of Probability of Spontaneous Production of Positrons}\label{Exponent}

Because, following Dirac the process of the production of $e^-e^+$  pairs can be treated as the penetration of electrons of the lower continuum into the upper continuum through the classically forbidden region ($p^2 <0$), the probability of this process is, as in case of spinless particles, determined by Equation (\ref{Wsemicl}). Equivalently, one can find the coefficient of transmission of the barrier in the effective potential or find semiclassical asymptotic of the functions $G$ and $F$ for $r\to \infty$. This single-particle picture is distorted with a deepening of the level and with the increase of the number of levels crossed the boundary $\epsilon =-m$. We may use Equations (\ref{phiGsecond})--(\ref{spinterm}) while taking the Langer correction into account, which improves the application of semiclassical expressions.

In the threshold region of positron energies setting $\epsilon \simeq -m$ in the expression for the spin term $U_s$, we obtain
\be
p^{*\,2} (r)\simeq (\epsilon -V)^2 -m^2 -\kappa^2/r^2\,,\label{pspher}
\ee
cf. with Equation (\ref{pdeep}) we have used for a description of the very deep levels. In case of the Coulomb field $V=-\zeta/r$, replacing (\ref{pspher}) in (\ref{Wsemicl}), we obtain

\be
W\sim \mbox{exp}\left[-2\pi\zeta \left(\frac{(m^2 +k^2)^{1/2}}{k}-(1-\rho^2)^{1/2}\right)\right]\,,\quad \rho =\kappa/\zeta\,, \quad k=\sqrt{\epsilon^2 -m^2}\ll m\,,
\ee
that coincides with the asymptotic of the exact solution of the Coulomb problem.

\subsection{Critical Charge of Nucleus for Muon}

For the electron, one has $R_{\rm nucl}\ll 1/m$, since $1/m \simeq 386$\,fm and $R_{\rm nucl}\simeq r_0 A^{1/3}\simeq A^{1/3}/m_{\pi}$, $m_\pi \simeq 280m$. For muon $R_{\rm nucl}\gg 1/m_\mu$, $m_\mu \simeq 207 m_e$.

In order to find the critical charge for the muon, $\zeta_{\rm cr}^\mu$, when $\mu^{-}$ level reaches $\epsilon =-m_\mu$,  we continue to apply the semiclassical approximation. For the model I, the turning point lies outside the nucleus.
Let us expand $U_{\rm ef}(r,\epsilon)$ near the turning point. Using \mbox{Equation (\ref{Airy}),} after the replacement $r_{-}\to r_0$, and matching solutions $G^{\,\prime}/G$  at $r=R_{\rm nucl}$, we find \cite{Eletskii:1977na}:
\be
\frac{\mbox{Ai}^{\,\prime}(0)}{\mbox{Ai}(0)}\frac{2}{\zeta^{1/3}}\simeq \frac{\beta \mbox{ctg}\beta}{m_\mu R_{\rm nucl}}\,,\quad \beta =\sqrt{\zeta (\zeta -2R_{\rm nucl}m_\mu)}
\ee
for the $ns$ level. From here follows
\be
\zeta_{\rm cr}^\mu \simeq 2Rm_\mu +\frac{(n\pi)^2}{2R_{\rm nucl}m_\mu}(1+a(R_{\rm nucl}m_\mu)^{-2/3}+...)\,,\quad a=-2^{4/3}3^{-5/6}\pi \Gamma^2(2/3)\,,
\ee
that coincides with expression, which follows from the direct solution of the Dirac equation at $\epsilon =-m_\mu$.
In the model II we obtain  $\zeta_{\rm cr}^\mu \simeq 16.7$ that corresponds to $Z_{\rm cr}^\mu \simeq 2300$, and in the  model I, respectively $Z_{\rm cr}^\mu \simeq 3700$.

\section{Semiclassical Approximation to System of Linear Dirac Equations}\label{Diraclinear}

\subsection{Semiclassical Wave Functions}

Let us apply semiclassical expansion to Equation (\ref{psiprsph}), cf. \cite{Popov:1979gq}. The parameter of expansion $\tilde{\lambda} /l$ is $\propto \hbar $, where $l$ is the typical length for the change of the potential. We present
\be
\psi =\phi e^{\int^r y dr}\,,
\ee
\be
y(r)=\frac{1}{\hbar}y_{-1}(r)+y_0 (r)+...\,,\quad \phi =\sum_{n=0}^{\infty}\hbar^n \phi^{(n)}\,,
\ee
and arrive at the chain of equations for $y_n$ and $\phi^{(n)}$:
\be
(\hat{D}-y_{-1})\phi^{(0)}=0\,,\quad (\hat{D}-y_{-1})\phi^{(1)}=\phi^{(0)\,\prime}_r +y_0\phi^{(0)}\,,...\label{uniformsyst}
\ee

One usually restricts expansion by consideration of first two terms. Because semiclassical series is an asymptotic one, retaining of too many terms may worsen the convergence of the series to the exact solution.

In order the system of homogeneous Equations (\ref{uniformsyst}) to have
nontrivial solution,    $y_{-1}(r)$  should be an eigenvalue and $\phi^{(0)}\equiv \phi_i$, $i=1,2$, the  eigenfunction of one of two-component eigenvectors of the matrix $\hat{D}(r)$. From the condition
$\mbox{det} \hat{D}=0$, we get
\be
y_{-1}\equiv \lambda_i =\pm i\sqrt{(\epsilon -V)^2 -m^2 -\tilde{\kappa}^2/r^2}\equiv \pm q\,.
\ee

Replacing $y_{-1}$ back to Equations (\ref{uniformsyst}), we obtain
\begin{eqnarray}
\label{psiprsphphi}
\phi_i =A
\left(\begin{array}{ccc}
m+\epsilon -V \\[3mm]
\lambda_i +\kappa/r
\end{array}\right)\,=
A_1
\left(\begin{array}{ccc}
\lambda_i -\kappa/r  \\[3mm]
m-\epsilon +V
\end{array}\right)\,,
\end{eqnarray}
where $A$ and $A_1$ are normalization constants.

Because the matrix $\hat{D}$ is not symmetrical,
besides the right-hand eigenvectors $\phi_i$, we should introduce  the left-hand
eigenvectors $\tilde{\phi}_i$:
\begin{eqnarray}
&(\hat{D}-\lambda_i)\phi_i =\tilde{\phi}_i (\hat{D}-\lambda_i)=0\,,\label{systDlamb}\\
&\tilde{\phi}_i =A (m-\epsilon +V, \lambda_i +\kappa/r)=A_1(\lambda_i -\kappa/r , m-\epsilon -V)\,.\nonumber
\end{eqnarray}

Note  that the left eigenvectors do not coincide with transposed right eigenvectors \mbox{($\tilde{\phi}_i\neq \phi_i^{T}$)} and the left-hand and right-hand vectors are mutually orthogonal,
\be
(\tilde{\phi}_i , \phi_j)=\sum_{\alpha =1}^2 (\tilde{\phi}_i)_\alpha (\phi_j)_\alpha \sim \delta_{ij}\,.
\ee

 To determine $y_0$, let us   put  $\phi^{(0)}=\phi_i$ in Equation (\ref{uniformsyst})
and multiply both sides of equation from the left by $\tilde{\phi}_i$. As follows from the first Equation (\ref{systDlamb}), the term with $\phi^{(1)}$ vanishes, and we obtain
\be
y_0 =-(\tilde{\phi}_i ,\phi_i^{\,\prime})/(\tilde{\phi}_i ,\phi_i)\,.
\ee

Further calculations entail no difficulty, cf. \cite{Popov:1979gq,PopovEletskyMur1976}. The resulting wave functions of the
quasistationary state with energy $\epsilon <-m$ in the region of classically permitted motion $r_0<r<r_{-}$ to have the form:
\begin{eqnarray}
&G=C_1 \left[\frac{\epsilon +m-V}{p}\right]^{1/2} \mbox{sin} \theta_1\,,\quad F=\mbox{sgn} \,\kappa\,\cdot
C_1 \left[\frac{\epsilon -m-V}{p}\right]^{1/2} \mbox{sin} \theta_2\,,\label{GFcl}
\\
&p=-iq=\sqrt{(\epsilon -V)^2 -m^2 -\frac{\kappa^2}{r^2}}\,,\quad \theta_1=\int_r{_0}^{r} (p+\frac{\kappa w}{pr})dr +\pi/4\,, \nonumber\\
&\theta_2 =\int_{r_{0}}^{r} (p+\frac{\kappa \tilde{w}}{pr})dr +\pi/4\,,\quad w=\frac{1}{2}\left(\frac{V^{\,\prime}}{m+\epsilon -V}-\frac{1}{r}\right)\,,\quad \tilde{w}=\frac{1}{2}\left(\frac{V^{\,\prime}}{m-\epsilon +V}+\frac{1}{r}\right)\,.\nonumber
\end{eqnarray}

Here, $C_1$ is  normalization constant. As it was discussed, semiclassical wave functions can be normalized neglecting  penetration of the particle into the classically forbidden
regions $r<r_0$ and  $r>r_{-}$, i.e., $\int_{r_0}^{r_{-}} (G^2+F^2)dr =1$. Thus, we find
\be
C_1 =\left[\int_{r_0}^{r_{-}} \frac{\epsilon -V}{p}dr\right]^{-1/2}=\left(\frac{2}{T}\right)^{1/2}\,,\label{C1T}
\ee
where $T$ is the period of the particle motion in the classically allowed region.

In the sub-barrier region $r_{-}<r<r_{+}$, where $p^2<0, p=iq$ and $q$, $y_{-1}$ and $y_0$ are real, wave functions attenuate exponentially with increasing $r$. The resulting expressions have different forms in dependence on the sign of $\kappa$. For $\kappa <0$, i.e., for $\kappa =-1$, we have
\begin{eqnarray}
\label{psiprsphphiunder}
\psi =
\left(\begin{array}{ccc}
G\\[3mm]
F
\end{array}\right)\,=
C_{2-} (Qq)^{-1/2} \,\mbox{exp}\left[-\int_{r_{-}}^r \left(q-\frac{V^{\,\prime}m}{2Qq}\right)dr\right]
\left(\begin{array}{ccc}
m+\epsilon -V   \\[3mm]
-Q
\end{array}\right)
\end{eqnarray}
with $Q=q-\kappa/r$.

For $\kappa >0$, we have
\begin{eqnarray}
\label{psiprsphphiunderpol}
\psi =
C_{2+} (Qq)^{-1/2} \,\mbox{exp}\left[-\int_{r_{-}}^r \left(q+\frac{V^{\,\prime}m}{2Qq}\right)dr\right]
\left(\begin{array}{ccc}
-Q   \\[3mm]
m-\epsilon +V
\end{array}\right)
\end{eqnarray}
with $Q=q+\kappa/r$, $C_{2\pm}$ are normalization constants.

In the region $r>r_{+}$, the quasistationary state
describes outgoing positron and represents a diverging wave. For $\kappa <0$:
\begin{eqnarray}
\label{psiprsphphiunderpolrneg}
\psi =
iC_{3-} (Pp)^{-1/2} \,\mbox{exp}\left[-\int_{r_{+}}^r \left(ip-\frac{V^{\,\prime}m}{2Pp}\right)dr\right]
\left(\begin{array}{ccc}
m+\epsilon -V  \\[3mm]
iP
\end{array}\right)
\end{eqnarray}
with $P=p-i\kappa/r$. The flux of particles moving to infinity 
is then given by $\Gamma =\mbox{lim}\,\,\mbox{Im} (F^* G)$ at $r\to \infty$.

For $\kappa >0$:
\begin{eqnarray}
\label{psiprsphphiunderpolrnol}
\psi =
iC_{3+} (Pp)^{-1/2} \,\mbox{exp}\left[-\int_{r_{+}}^r \left(ip+\frac{V^{\,\prime}m}{2Pp}\right)dr\right]
\left(\begin{array}{ccc}
 iP \\[3mm]
m-\epsilon +V
\end{array}\right)
\end{eqnarray}
with $P=p+i\kappa/r$. $C_{2\pm}$ are normalization constants.

The obtained formulas  are valid for all $r$, except  regions  $\delta r\propto 1/\zeta^{2/3}$   near the turning points.
The usual procedure is employed to match semiclassical solutions. The solution is either expressed in terms of
an Airy function or one may use the Zwaan's method.  Consequently, we have
\begin{eqnarray}
&C_{2\pm}& =-iC_{3\pm} \\
&=&-\frac{\mbox{sgn}\,\kappa}{2}C_1 m \left[\frac{|\kappa|}{mr_{-}+(\kappa^2 +r_{-}^2 m^2)^{1/2}}\right]^{-\mbox{sgn}\,\kappa/2}\mbox{exp}\left[-\int_{r_{-}}^{r_{+}} \left(q+\mbox{sgn}\,\kappa \frac{V^{\,\prime}m}{2Qq}\right)dr\right]\,.\nonumber
\end{eqnarray}

Note that the effective potential, which we have used in (\ref{phiGsecond}), can be presented while employing function $w$ that appeared in (\ref{GFcl}):
\be
U_{\rm ef}=-\frac{V^2}{2m}+\frac{\epsilon V}{m}+\frac{\kappa(\kappa+1)}{2r^2 m}-\frac{\kappa}{rm}w+\frac{1}{2m}(w^{\,\prime}+w^2 +\frac{w}{r})\,.\label{Uefw}
\ee

The terms in Equation (\ref{Uefw}), which contain the function $w$, are due to the electron spin. For $|V|\gg m$, they
are small compared to the first three terms. Subsequently, the expression for the effective potential takes the same form as for a scalar particle. At the turning points $r_{-}$ and $r_{+}$, the effective potential is not singular.

 The action becomes
\be
S=\int^r dr \sqrt{2m(E-U_{\rm ef})}=\int^r dr \left[p+\frac{2\kappa w}{pr} - m^{-1}(w^{\,\prime}+w^2 +\frac{w}{r})\right]^{1/2}\,.
\ee

Expanding $S$ in $1/\zeta \ll 1$, we obtain
\be
S=\int^r dr \left[p+\frac{\kappa w}{pr} +O(m/\zeta^2)\right]\,,
\ee
that coincides with Equations (\ref{GFcl})--(\ref{psiprsphphiunderpolrnol}), which we have derived in this section.

Using Equations (\ref{GFcl})--(\ref{psiprsphphiunderpolrnol}) for $\epsilon =-m$, we obtain
\be
\overline{r}=\frac{3(\zeta^2 -\kappa^2+1/4)(\zeta^2+2\kappa^2/3 -5\kappa/3+1)}{10\zeta (\zeta^2+(\kappa^2-3\kappa/2+1/2)/2)}\,.\label{rlambda}
\ee

This expression yields $\overline{r}=0.301/m$ for the ground-state level, whereas the exact result gives $0.303/m$. For  $\zeta\gg |\kappa|\gg 1$, result (\ref{rlambda}) coincides with that follows from (\ref{rlambda1}).

\subsection{Nonrelativistic Limit}

To be specific consider case $\kappa <0$ and the classically allowed region. Introducing a nonrelativistic
energy $\tilde{\epsilon}=\epsilon -m$ and the variable  $\tilde{q}=(q^2+\kappa/r^2)^{1/2}\simeq q+\frac{\kappa}{2qr^2}$,
let us transform the factor in exponent  (\ref{psiprsphphiunder}) as
\be
\int_{r_{-}}^{r} dr \left(q - \frac{V^{\,\prime}m}{2Qq}\right)=\int_{r_{-}}^{r} dr
\left[\tilde{q} - \frac{1}{2q} (V^{\,\prime}/Q+\kappa/r^2)\right]=\int_{r_{-}}^{r} dr \left[\tilde{q} -\frac{1}{2}(\ln Q)^{\,\prime}\right]\,,
\ee
where $Q=q-\kappa/r$. The latter term in the integral cancels with the pre-exponential factor $Q^{-1/2}$. Now let us take into account that $\kappa(1+\kappa)=l(1+l)$.  Subsequently, we have
\be
\tilde{q}(r)=\left[2m (-\tilde{\epsilon}+V(r)-l(l+1)/(2mr^2)\right]^{1/2}\,,\quad G(r)=\frac{C}{\tilde{q}^{1/2}(r)}\mbox{exp}(-\int_{r_{-}}^{r} dr \tilde{q}(r)\,,
\ee
where $C=const$ that reproduces the Schr\"dinger wave function in this region. Note that $\tilde{q}(r)$ enters not $\kappa =\mp (j+1/2)$, but orbital moment $l$.  We formally considered case $\kappa <0$ just to be specific. Case $\kappa >0$ is considered similarly. Additionally, note that, for $\kappa\neq -1$, one should  add to $\tilde{q}(r)$ the Langer correction.

\subsection{Bohr-Sommerfeld Quantization Rule}

From (\ref{psiprsphphiunder}), (\ref{psiprsphphiunderpol}) we derive \cite{Popov:1979gq},
\be
\int_{r_{0}}^{r_{-}} dr (p+\frac{\kappa w}{pr})=(n+\gamma^{\,\prime})\,.\label{BohrSomDir}
\ee

As we have mentioned, value $\gamma^{\,\prime}$ depends on the fact does $r_0$ lie inside the nucleus or outside it. In the latter case, $\gamma^{\,\prime}=1/2$ for $\kappa \neq -1$ and $\gamma^{\,\prime}=3/4$ for $\kappa = -1$.

Equation (\ref{BohrSomDir}) determines the real part of the energy  $\epsilon_{n\kappa}$. It differs from the ordinary Bohr--Sommerfeld rule used in nonrelativistic quantum mechanics by expression for relativistic momentum $p(r)$ and by the term $\propto w$ appeared due to the spin--orbital interaction. Taking into account of the term $\propto w$ is legitimate within semiclassical scheme. Let us show it on an example of the
Coulomb field $V=-\zeta/r$. Subsequently, $w(r)=-\frac{m+\epsilon}{2(\zeta + (m+\epsilon)r)}$ and $p(r)$ is determined by Equation (\ref{GFcl}). For $r_0<r<r_{-}$, the momentum $p(r)\sim \tilde{g}/r$ and the ratio $|\frac{\kappa w}{p^2 r}|\sim |\kappa \tilde{g}^{-2} rw|\sim |\kappa|/\zeta^2$ for deep levels. Because semiclassical approximation for wave functions is valid up to $1/\zeta^2$,  the second term in the integral (\ref{BohrSomDir}) should be retained in the case of deep levels $|\epsilon|\gg m$ for $|\kappa|\gg 1$, but it can be dropped for $|\kappa|\sim 1$. For $\epsilon =-m$, we \mbox{have $w=0$.}

Note that the results of calculations performed with the help of the  quantization rules (\ref{quantruleef}) and  (\ref{BohrSomDir})  differ only in  correction terms. For instance, from
(\ref{BohrSomDir}), we derive exactly the same electron energy spectrum as that given by Equations  (\ref{encrho}) and (\ref{crho}), with the help of the quantization rule in the form  (\ref{quantruleef}).

\subsection{Probability of Spontaneous Production of Positrons}

Let us calculate the probability of spontaneous production of positrons, $\Gamma =-2\mbox{Im}\epsilon$. Replacing  (\ref{psiprsphphiunderpolrneg}), (\ref{psiprsphphiunderpolrnol}) in (\ref{gammaflux}), we find
\be
\Gamma =\Gamma_0 e^{-2\int_{r_{-}}^{r_{+}}q(r)dr}\,,\quad \Gamma_0 ={T}^{-1} e^{2\kappa {\rm Pr} \int_{r_{-}}^{r_{+}} w dr/(qr)}\,.\label{gamzero}
\ee
The last integral  is understood in the sense
of the principal value, being denoted as ${\rm Pr}$, due to singularity at the point
where $V(r)=m+\epsilon$.

In the nonrelativistic limit, the value $\Gamma_0 =1/T$ has the meaning of the
number  of impacts per unit time of the particle (localized inside the
region $r_0 <r<r_{-}$)  against the potential barrier at $r=r_{-}$, and the exponential is the probability of the penetration of  the barrier in each impact. The allowance for the relativistic effects and  the spin change the expression for the period of the
oscillations and add to (\ref{gamzero}) a factor depending
 on the sign of $\kappa$.

While taking into account  that  in the region of the barrier $V$  is the purely
Coulomb field, for $w=0$ all of the integrals are calculated
exactly:
\begin{eqnarray}
&\Gamma =\Gamma_0 \mbox{exp}\left[-2\pi\zeta\left((m^2+k^2)^{1/2}/k -(1-\rho^2)^{1/2}\right)\right]\,,\label{GammaGamma0}\\
&1/\Gamma_0 = \frac{2\zeta }{k^2}\left[(1-\rho^2)^{1/2}(m^2 +k^2)^{1/2}-\frac{m^2}{k}\mbox{Arth}\left(k\left(\frac{1-\rho^2}{m^2 +k^2}\right)^{1/2}\right)\right]\,.\nonumber
\end{eqnarray}

For the positron momentum $k=\sqrt{\epsilon^2 -m^2}\to 0$, we have $\Gamma_0 =c_1=3/[2\zeta (1-\rho^2)^{1/2}(2+\rho^2)]$, and, for $k\to \infty$, we have  $\Gamma_0 =c_2 k =k/[2\zeta(1-\rho^2)^{1/2}]$.
For  $k\ll \zeta^{1/2}m$ the width $\Gamma$ is exponentially small for any $\kappa$. For $|\kappa|\gg (\zeta/\pi)^{1/2}$ expression simplifies as
\be
\Gamma\simeq k[2\zeta (1-\rho^2)]^{-1/2}\mbox{exp}[-2\pi\zeta (1-(1-\rho^2)^{1/2})]\,.
\ee

For  $|\kappa|\lsim \kappa_0=(\zeta/\pi)^{1/2}$,
the exponential factor in $\Gamma$ becomes of the order of unity, and
the semiclassical approximation becomes invalid.
Note that $\kappa_0/\kappa_{max}=1/\sqrt{\pi\zeta}$. Therefore, a number of levels diffused in the continuum, for which $\Gamma$ is not exponentially small,  is tiny for $\zeta\gg 1$.

\subsection{Semiclassical Method for Noncentral Potentials Obeying System of Linear Dirac Equations}

We described the spectrum of the quasistationary levels in the lower continuum  for a spherical  nucleus with the charge $Z>Z_{\rm cr}$. The results can be generalized to the case, when the
potential does not obey  spherical symmetry \cite{Popov:1979gq}. Let us present the Dirac equation as
\be
-i(\hat{\vec{\alpha}}\nabla)\psi =\hbar^{-1}\hat{D}\psi\,,\quad \hat{D}=\epsilon -m\hat{\beta}-V(\vec{r})\,,\label{Dnoncentr}
\ee
where  $\hat{\vec{\alpha}}=\gamma^0\vec{\gamma}$, $\hat{\beta}=\gamma^0$ are Dirac matrices, and we recovered dependence on $\hbar$. Let us present bispinor  $\psi$ as $\psi =\phi e^{i\sigma}$
and expand real quantities $\phi$ and $\sigma$ in the parameter that is proportional to $\hbar$:
\be
\sigma =\hbar^{-1}\sigma_{-1}+\sigma_0 +...,\quad \phi =\phi^{(0)}+\hbar \phi^{(1)}+...
\ee

Replacing these series to Equation (\ref{Dnoncentr}), we obtain the chain of equations
\begin{eqnarray}
&[\hat{D}-(\hat{\vec{\alpha}}\nabla\sigma_{-1})]\phi^{(0)}=0\,,\label{systnonz}\\
&[\hat{D}-(\hat{\vec{\alpha}}\nabla\sigma_{-1})]\phi^{(1)}=\hat{\vec{\alpha}}\nabla)\phi^{(0)}
+(\hat{\vec{\alpha}}\nabla\sigma_{0})\phi^{(0)}\,,...\nonumber
\end{eqnarray}

The condition of existence of a nontrivial solution $\phi^{(0)}$,
\be
\mbox{det}[\hat{D}-(\hat{\vec{\alpha}}\nabla\sigma_{-1})]=0\,,
\ee
results in the Hamilton--Jacobi equation
\be
(\nabla \sigma_{-1})^2 =(\epsilon -V)^2 -m^2\,.
\ee

In difference with the spherically-symmetric case, the matrix
\be
\hat{D}-\hat{\vec{\alpha}}\nabla\sigma_{-1}=\epsilon -V(\vec{r})-m\hat{\beta}-\hat{\vec{\alpha}}\nabla S
\ee
is Hermitian; therefore, its left-hand, $\tilde{\phi}_i$,  and
right-hand, ${\phi}_i$, eigenvectors are Hermitian conjugates, $\tilde{\phi}_i =\phi_i^{\dagger}$, and
\be
(\hat{D}-\hat{\vec{\alpha}}\nabla S)\phi_i = \phi_i^{\dagger}(\hat{D}-\hat{\vec{\alpha}}\nabla S)=0\,,\quad i=1,2,3,4\,.
\ee

With the help of this equation, from (\ref{systnonz}), we find a system of equations for $\sigma_0$,
\be
\phi_i^{\dagger}(\hat{\vec{\alpha}}\nabla \sigma_0)\phi_j = -\phi_i^{\dagger}\hat{\vec{\alpha}}\nabla \phi_j\,.\label{rightsp}
\ee

Bispinors $\phi_i$ are found by diagonalizing the matrix $\hat{D}-\hat{\vec{\alpha}}\nabla S$, so that the right-hand side of Equation (\ref{rightsp}) contains known quantities. Determining from this equation $\sigma_0$, we obtain the quasiclassical solution of the Dirac equation
\be
\psi =\phi_i \mbox{exp}(\hbar^{-1}\sigma_{-1}+\sigma_0)\,.
\ee

In practice, the calculation of the functions $\sigma_{-1}$ and $\sigma_{0}$ for
noncentral potentials  is a complicated
mathematical problem requiring  the solution of first-order
 differential equations in partial derivatives. In contrast
to the case when $V$ is spherically symmetric, in general case
the result is not expressed in quadratures. If a parameter of a ``non-sphericity'' is small, then one may develop a perturbation theory.

\section{Spontaneous Production of Positrons in Heavy-Ion Collisions}\label{HIC}

\subsection{Approach to the Problem}
The minimal distance between colliding nuclei with charges $Z_1$ and $Z_2$ is as follows \cite{LL1,PopovJETP1974},
$$R_{\rm min}=(Z_1+Z_2)^2 e^2/(2E_{\rm c.m.})+\sqrt{(Z_1+Z_2)^2 e^2/(2E_{\rm c.m.})^2+b^2}\,,$$
where $E_{\rm c.m.}$ is the kinetic energy of colliding nuclei in c.m. reference frame, $b$ is the impact parameter.
In order the energy of the electron, $\epsilon_{\rm 1s}$, in the quasi-molecule would become  $<-m$ the colliding heavy nuclei  should reach distances $|\vec{r}_1 -\vec{r}_2|=R<R_{\rm cr}$, where $R_{\rm cr}\simeq 33$ fm for central U$+$U collisions, see below. Thus, $R_{\rm cr}$ is approximately twice larger than $ 2R_{\rm nucl}$,  where $R_{\rm nucl}\simeq 1.2 A^{1/3}$ fm is the radius of the single nucleus $\simeq 7$ fm. On the other hand,
$R_{\rm cr}\ll \overline{r}_K\simeq 0.3 \zeta_{\rm cr}$,  where $\overline{r}$ is estimated using Equation (\ref{rlambda}). For U$+$U collisions, $R_{\rm cr}/\overline{r}_K \sim 0.2$.
Nuclei move with the velocity $v_A\sim (0.025-0.07)$, cf. \cite{PopovJETP1974}, whereas the electron of the K-shell has a typical velocity $v_e\simeq 1$. Thereby, one may use adiabatic approximation, i.e., we may use $\epsilon (R(t))$.  Because $R_{\rm cr}/\overline{r}_K \sim 0.2\ll 1$,
the  anisotropy  of the potential is not as large, and we may present
\be
V({r})=-\left(\frac{Z_1e^2}{r_1}+ \frac{Z_2e^2}{r_2}\right)=-\frac{Ze^2}{r}\left(1+\frac{R^2}{(2r)^2}P_2 (\cos \theta)+...\right)\,,\label{VrR}
\ee
where $Z=Z_1+Z_2$,  $\vec{r}_{1,2}=|\vec{r}\pm \vec{R}/2|$, $P_2$ is the second Legendre polinomical,
$R(t)$ is the distance between centers of nuclei. In the second equation and, further we for simplicity, consider the case $Z_1=Z_2$. Otherwise, odd-power terms appear in the expansion. In inclusive experiments, this anisotropy disappears due to the averaging. However for event-by-event collisions such  terms may lead to the forward-backward anisotropy reflecting in some observable effects. In the first approximation in $(R/2\overline{r}_K)^2$, the problem is reduced to that we have considered above for the spherical nucleus with the charge $Z=Z_1+Z_2$. The effective nucleus radius now is $2R_{\rm nucl}$.

The  process of the spontaneous production of positrons can also be  described in adiabatic approximation, since, as we have argued, we may use that $\epsilon(R(t))$ and, since $1/\Gamma (\epsilon(R(t)))\gg \tau_{\rm col}\gsim 2R_{\rm cr}/v_A$.  The most serious experimental problem is to separate  spontaneous production of positrons  in the tunneling process from the frequency dependent  processes also resulting in a production of positrons.
For example, the parameter $2R_{{\rm nucl}}/R_{{\rm{ cr}}}\sim (1/2 - 1/3)$ is not as small.
Therefore, a serious competing time-dependent process is associated with an induced production of positrons occurring due to excitation of the nuclear levels,  cf.  \cite{Oberackner1976,Rafelski1916} and the references  therein. However, the difference between characteristics of the induced and  spontaneous production of positrons is significant. The induced positron production exists in both subcritical and supercritical regimes. When the electron level crosses the boundary $\epsilon =-m$, there appears a narrow energy-line in the positron spectrum owing to the switching on of the spontaneous positron production occurring in the tunneling process. Thus, there is a principal difference between the subcritical and the supercritical regimes that may help in the experimental identification of the spontaneous positron production.

Another effect is associated with the presence of a magnetic component of the field. First, an  indication on presence of strong magnetic fields in heavy ion collisions was performed in \cite{VA1980}. For peripheral collisions of heavy ions at collision energies $\lsim$ GeV$\cdot A$  it yields $h\sim H_\pi (Ze^6)^{1/3}$ for $R\simeq A^{1/3}/m_{\pi}$, $v_{A}\sim 1$, $H_{\pi}=m_{\pi}^2/e$.
More generally, replacing $1\to v_A\gamma$, $\gamma =1/\sqrt{1-v_A^2}$, we have
\be
eh \sim Ze^2 v_A\gamma /R^2\,.
\ee

For collisions with low energies $E\sim (5-10)$MeV$\cdot A$ of our interest here, it follows that
$h\sim 10^{15}$G, for $R\sim R_{cr}\simeq (30-50)$ fm, and $v_{A}\sim 10^{-1}$, cf. also \cite{RGS1987}.

In the presence of a “weak” homogeneous magnetic field, the
reduction of $Z_{\rm cr}$ in the case of the supercritical atom has been  found by using the perturbation theory \cite{Oraevsky1977},
\be
\zeta_{\rm cr}(h)=\zeta_{\rm cr}(0)-\frac{5\pi^2\mu}{6\ln (1/R^3_{\rm nucl})}\frac{h}{H_0}\,,
\ee
$H_0=m^2/e\simeq 4.4\cdot 10^{13}$G, $\mu\simeq 1/3$ for $\zeta=\zeta_{\rm cr}$.

For strong fields, numerical evaluations
\cite{Oraevsky1977}, see also \cite{Popov2001}, yielded $Z_{\rm cr} = 165$ for $h=H_0$, and
$Z_{\rm cr} = 96$ at $h=10^2 H_0$. For $h=10^{18}$G, one gets $Z_{\rm cr} = 41$. This effect appears  because of the exact compensation of the diamagnetic and paramagnetic
contributions to the ground state for the electron. Although these estimates are performed for the case of purely uniform static magnetic field, they show that a magnetic effect also should be carefully studied for the case of realistic time-space configuration of the field.

Below, I only focus on the description of the spontaneous production of positrons and, simplifying this consideration, I also ignore the mentioned  magnetic effects.

\subsection{Electron Energy as a Function of Distance between Nuclei}

Usage of the Bohr--Sommerfeld quantization rule allows for considering the problem  analytically \cite{Voskresensky1977}, cf. \cite{Popov:1979gq}.
From (\ref{phiG})--(\ref{phiGsecondU}), taking into account of the Langer correction resulting in the replacement $p\to p^*$, we have
\be
p^*(r)=\frac{F(r,\epsilon)}{r}\,,\quad F(r,\epsilon)=\left[(\epsilon^2 -m^2)r^2 +2\epsilon \zeta r +\frac{\kappa +1}{\tilde{a}}-\frac{3}{4\tilde{a}^2}-(\kappa +{1}/{2})^2 +\zeta^2\right]^{1/2}\,,\label{pst}
\ee
where $\tilde{a}=1+r(m+\epsilon)/\zeta$, $\zeta =Ze^2$, $Z=Z_1+Z_2$.
Applying the quantization rule (\ref{quantruleef}), first for $\epsilon \neq -m$ and then for $\epsilon = -m$, and subtracting one result from the other, we obtain
\be
\int_{R/2}^{r_\epsilon}dr F(r,\epsilon)/r=\int_{R_{\rm cr}/2}^{r_{-m}}dr F(r,\epsilon =-m)/r\,.
\ee

Here, $r_\epsilon$ is the turning point for the given $\epsilon$ and $r_{-m}$ is the turning point for $\epsilon =-m$.
I used that in integration over the regions  $r< R/2$, $r< R_{\rm cr}/2$ dependence on $\epsilon$  can be dropped, since at $|\epsilon|\sim m$ of our interest, we have $|V|\gg |\epsilon|$. Thereby, the specifics of the behavior $V(r)$ in the region $r<R_{\rm cr}/2$ almost does not affect the result. To be specific, we may use $V=const$ for $r<R_{\rm cr}/2$. Integrals undergo logarithmic diverge at the lower limit. After their regularization, the dependence on $R$ and $R_{\rm cr}$ is
separated in the explicit form:
\be
\int_{0}^{r_\epsilon}dr [F(r,\epsilon)- F(r,\epsilon =-m)]/r +\int_{r_{-m}}^{r_\epsilon}dr F(r,\epsilon =-m)/r =\tilde{g}\ln \frac{R}{R_{\rm cr}}\,.\label{Fgln}
\ee

Integrals in (\ref{Fgln}) are calculated numerically. A comparison with the exact solution of two-center Dirac problem shows that the error of the semiclassical result does not exceed $0.1\%$.
We can proceed further using that $r|m+\epsilon|/\zeta <r_{\epsilon}|m+\epsilon|/\zeta\ll 1$ at least for $|\epsilon|\sim m$ of our interest. Thereby, we expand $\tilde{a}$ in Equation (\ref{pst}) in the series of $r$. As the result, we find
\begin{eqnarray}
&F(r,\epsilon)=(\tilde{g}^2+br+cr^2)^{1/2}\,,\label{Fabc}\\
 &b=2\epsilon \zeta -(\kappa -1/2)(m+\epsilon)/\zeta\,,
\quad c =\epsilon^2-m^2 +(\kappa -5/4)(\epsilon +m)^2/\zeta^2\,.\nonumber
\end{eqnarray}

From (\ref{Fgln}) and (\ref{Fabc}), we obtain
\be
\frac{R}{R_{\rm cr}}=-\frac{2\zeta}{b}\left(1+\frac{\tilde{g}^2 c}{3b^2}+O(c^2)\right)\,.\label{RRcr}
\ee

For $|\epsilon +m|\ll m$, we find
\be
\epsilon =-m-\beta m (R-R_{\rm cr})/R_{\rm cr}\,,\quad \beta =\left(1-\frac{\kappa -1/2}{2\zeta^2}-\frac{\tilde{g}^2}{3\zeta^2}\right)^{-1}\,.
\ee

For U$+U$ collisions for the ground-state level, we find $\zeta\simeq 1.343$ and $\beta\simeq 0.79$. The slope-parameter $\beta$ determines the probability of the production of positrons for $|\epsilon +m|\ll m$.
 The semiclassical approximation reproduces the $Z$ dependence of $\beta$ correctly, the difference with exact calculation done within solution of the two-center problem for the Dirac equation \cite{Lisin1980} is approximately (3--4)\%.

Setting $c=0$ in Equation $(\ref{RRcr})$, we obtain a very simple and accurate result \cite{Voskresensky1977,Popov:1978kk,Popov:1979gq}:
\be
\epsilon (R)=\epsilon (R/R_{\rm cr})=-m\frac{R_{\rm cr}/R -(\kappa -1/2)/(2\zeta^2)}{1-(\kappa -1/2)/(2\zeta^2)}\,.\label{epsUU}
\ee

The difference of this simple expression with exact solution of the two-center Dirac \mbox{Equation \cite{Lisin1980}} is less than (1--2)\% \, already for $\zeta\to 1$ when the parameter of applicability of the semiclassical approximation is $~1$. Such an accuracy is sufficient; therefore, here I do not present a more accurate semiclassical expression \cite{Popov:1979gq} obtained without using expansion in $c^2$, which has still higher accuracy.  It may be curious to notice that, when in 1976 I showed the result  (\ref{epsUU}) to Vladimir Stepanovich Popov, he did not believe in it, saying that one of  his collaborators   during   a year  is trying to solve the  Dirac equation for the two-center problem numerically on ITEP big computer and, yet, only obtained the result for $\zeta  =1$. He took the slide rule (that time there were no PCs) and confirmed  that for $\zeta  =1$ the whole \mbox{curve (\ref{epsUU})} fully coincides with the result of the exact numerical calculation. Because the criterion of applicability of the semiclassical approximation for the ground state is $\tilde{g}_0\gg 1$, it became clear that, for $\zeta  >1$, the accuracy of approximate solution  (\ref{epsUU}) should at least not be worse than in case $\zeta =1$.

\begin{figure}[H]
\includegraphics[width=7cm,clip]{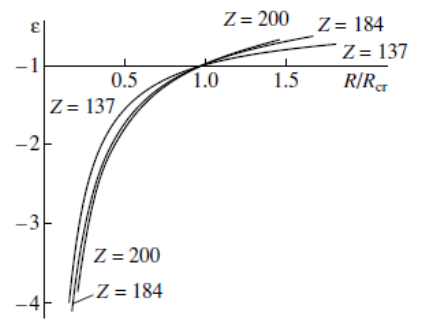}
\caption{Solution
$\epsilon (R/R_{\rm cr})$ of Equation (\ref{epsUU}) for various values of the parameter $\zeta$.}\label{EnergyLoverContUU}
\end{figure}

Subsequently, the result (\ref{epsUU}) was reflected in  our publications \cite{Popov:1978kk,Popov:1979gq}. Result  (\ref{epsUU})  is shown in Figure \ref{EnergyLoverContUU}. For $\zeta=1.343$, $\kappa =-1$, we get $-\epsilon (R/R_{\rm cr})=0.705 (R_{\rm cr}/R)+0.295$.

The expression for the critical distance between nuclei, $R_{\rm cr}$, can be found from \linebreak Equation (\ref{Rcre}) for a spherical nucleus after replacement of the nucleus radius $R_{\rm nucl}$ by $R/2$, where, now, $R$ is the distance between nuclei and $Z\to Z_1+Z_2$. Consequently, we find
\be
R_{\rm cr}=\frac{4\tilde{g}^2}{\zeta m}\left[\mbox{exp}\left(\frac{\pi (n+\gamma_1)-\tilde{\gamma}}{\tilde{g}}+2\right)+2\right]^{-1}\,.
\ee

For the case of U$+$U collisions, in the model I that we obtain $R_{\rm cr}\simeq 33$ fm, whereas exact solution of the Dirac equation \cite{Lisin1980} yields $R_{\rm cr}\simeq 34.3$ fm.

\subsection{Tunneling in the Two-Center Problem. Angular Distribution of Positrons}

The potential of the system of
two nuclei (\ref{VrR}) contains, at $r\gg R$, a quadrupole correction. In the sub-barrier region, the correction is $\lsim (R_{\rm cr}/(2r_{-}))^2\lsim 10^{-2}$. Therefore, the problem is reduced to the  calculation of the penetrability of a three-dimensional barrier that only differs little from a spherically symmetrical one.  Thus, we may use expansion
\be
V=V_0 +m^2 R^2 V_1\,,\quad S=S_0+m^2R^2 S_1\,.
\ee

We substitute these expressions to the Hamilton--Jacobi equation and  obtain
\begin{eqnarray}
&(\nabla S_0)^2 =2m (E-U_0)\,,\quad \nabla S_0\nabla S_1 =-U_1\,,\\
& U_0 (r)=-\left(\frac{\zeta^2}{2r^2 m}+\frac{\zeta\epsilon}{rm}\right)\,,\quad U_1 (r)=-\frac{\zeta}{4r^3 m^2}\left(\epsilon +\frac{\zeta}{r}\right)P_2 (\cos \theta)\,.\nonumber
\end{eqnarray}

The first equation is easily integrated, resulting in
\be
S_0 (r,\theta)=\int^r pdr +\kappa\theta\,.\label{S0theta}
\ee
Taking the first term into account leads to exponential term in Equation (\ref{GammaGamma0}). Second term in
(\ref{S0theta}) is due to anisotropy of the potential.

Equation for $S_1$ in the under-barrier region $r_{-}<r<r_{+}$ gets the form
\be
iq\frac{\partial S_1}{\partial r}+\frac{\kappa}{r^2} \frac{\partial S_1}{\partial \theta}=-U_1 (r,\theta)\,, \quad p=iq\,,
\ee
and it is solved by the method of separation of the variables. Supposing
\be
r^2 U_1 (r,\theta)=u(r)(\frac{3}{4}\cos  (2\theta)+\frac{1}{4})
\ee
and taking into account the boundary condition $\mbox{Im}S_1 (r_{-},\theta)=0$, for $r=r_{+}$ we obtain
\vspace{-14pt}
\begin{eqnarray}
&\mbox{Im}S_1 (r_{+},\theta)=aP_2 (\cos \theta)+a_1\,,\label{S1theta}\\
&a=\int_{r_{-}}^{r_{+}}dr \frac{u(r)m^2}{q(r)}\mbox{ch}\left(2\kappa \int_{r}^{r_{+}}\frac{dr^{\,\prime}}{q(r^{\,\prime})r^{\,\prime\,2}} \right)\,,\quad
a_1 =-\frac{1}{2}\int_{r_{-}}^{r_{+}}dr \frac{u(r)m^2}{q(r)}\mbox{sh}^2\left(\kappa \int_{r}^{r_{+}}\frac{dr^{\,\prime}}{q(r^{\,\prime})r^{\,\prime\,2}} \right)\,.\nonumber
\end{eqnarray}
For  the angular asymmetry of the positron production, the constant $a_1$ is immaterial.

A  remarkable fact is that the expression for $a$
acquires a hyperbolic cosine that enhances  the angular
anisotropy of the emitted particles when compared with
the anisotropy of the potential. The cause of this
effect is that the sub-barrier trajectory of a tunneling
particle with nonzero angular momentum is not a
straight line due to $\kappa \neq 0$.
 This  leads to
a substantial difference in the description  of the three-dimensional
and the one-dimensional tunneling of  particles.

For the Coulomb field integrals (\ref{S1theta})  can be calculated exactly.
However, the result looks cumbersome.  An estimate shows that $W(\theta)\simeq \mbox{exp}(-2\mbox{Im}S)=C \mbox{exp}(\alpha P_2 (\cos \theta))$, where $C$ is a constant, $\alpha\sim m^2R^2\eta^{-1}\mbox{sh}\eta\gg m^2 R^2$, $\eta =2\pi\kappa/\tilde{g}$.  For U$+$U collisions $\alpha \sim 1/3$, and we can expect a noticeable angular anisotropy.
The positrons are predominantly emitted  along the axis joining the nuclei at the instant of their closest approach. This question is worthy of experimental study.

Concluding, note that we needed the applicability of semiclassical approximation for both the radial motion and the angular motion. Strictly speaking, the latter takes place only for $|\kappa|\gg 1$. However, as it always occurs, even for $|\kappa|\sim 1$, one may expect good accuracy of  semiclassical expressions.

\subsection{Screening of K-Electron by Electron Cloud of Not Fully Stripped Quasi-Molecule}\label{screenK}

If the colliding nuclei
are not fully stripped, the quasi-molecule is surrounded by an
electron cloud.  Screening weakens the attraction of the $K$-electron to the nuclei in the quasi-molecule. Consequently, the  critical distance $R_{\rm cr}$, at which the K-electron level crosses the boundary  $\epsilon =-m$, is decreased.  This effect can be calculated using nonrelativistic many-particle semiclassical approximation (Thomas--Fermi method), cf. \cite{Voskresensky1977,Popov:1979gq}.  Let us \mbox{use that}
\be
R_{\rm cr}\ll \overline{r}_K\ll a_{\rm TF}=(9\pi^2/128)^{1/3}(Ze^6)^{-1/3}/m\simeq 30 \zeta^{-1/3}/m\,,
\ee
where $a_{\rm TF}$ is the mean radius of the Thomas--Fermi atom. The shift of the ground-state electron energy level can be found with the help of the perturbation theory. We have
\be
\Delta \epsilon_0 \simeq \overline{V(\vec{r})-V_0 (\vec{r})}\,,
\ee
where $V_0 (\vec{r})$ is the potential of the two striped  nuclei (\ref{VrR}) and $V (\vec{r})$ is the potential of the  two not fully striped ions. The typical size for the change of $\delta V$ is $a_{\rm TF}$. Therefore, with the accuracy $\sim (R_{\rm cr}/a_{\rm TF})^2\sim 10^{-5}$, the perturbation can be considered to be spherically symmetric. Thus,
\be
V(r)=V(r_i)-\frac{Ze^2\phi(r)}{r}\,,\quad V(r_i)=-\frac{Z_1 e^2}{r_i}\,,\label{VrVi}
\ee
$r_i =x_0 a_{\rm TF}$ is the radius of the ion, $\phi (r)$ is the solution of the Thomas--Fermi equation \cite{LL3},
\be
\phi^{\,\prime\prime}_x=x^{-1/2}\phi^{3/2}
\ee
with boundary conditions $\phi (0)=1$, $\phi (x_0)=0$, $x=r/a_{\rm TF}$, and $Z_1 =-Zx_0 \phi^{\,\prime}_x (x_0)$ is the observed charge of the two partially screened nuclei.

Expansion $\phi (x\to 0)$ yields \cite{LL3}:
\be
\phi (x)=1+ \phi^{\,\prime}_x (0)x +\frac{4}{3}x^{3/2}+...\label{phiexpan}
\ee
For the case of neutral atoms $\phi^{\,\prime}_x (0)=-1.588$.

From (\ref{VrVi}) and (\ref{phiexpan}) for the shift of the ground-state level, we obtain
\begin{eqnarray}
\Delta \epsilon_0 =V(r_i)+\phi^{\,\prime}_x (0) \frac{Ze^2}{a_{\rm TF}}=\frac{Ze^2}{a_{\rm TF}}[\phi^{\,\prime}_x (0)-\phi^{\,\prime}_x (x_0)]+\frac{4\zeta}{3a_{\rm TF}^{3/2}}\overline{r^{1/2}}+...
\end{eqnarray}

Values $\phi^{\,\prime}_x (0)$ and $\phi^{\,\prime}_x (x_0)$ are tabulated. We estimate $|\Delta R_{\rm cr}/R_{\rm cr}|\sim |\Delta \epsilon_0/\epsilon_0|\simeq 10\%$ for the ionization parameter $q=(Z_1+Z_2-N)/(Z_1+Z_2)\simeq 0.5$, and $\simeq 12\%$ for $q=0$,
where $N$ is the total number of electrons in the quasi-molecule.

\subsection{Calculation of Positron Production Employing the Imaginary-Time Method}

\subsubsection{General Description of the Method}

First, consider the problem of the one-dimensional motion of a relativistic  particle in the potential $V(x,t)$. The Lagrangian is as follows
\be
L=-m\sqrt{1-\dot{x}^2}-V(x,t)+V_0\,.
\ee

The constant is added to recover Lorentz invariance of the action
\be
S=\int_{t_{1}}^{t_{2}} Ldt\,,
\ee
since $t$ is not a scalar. At the initial time-moment particle was in the point $x_1(t_1)$ and, at the final moment, in $x_2(t_2)$.

In the semiclassical approximation, the wave function is
\be
\psi(x)\sim e^{iS(x_1,x)}=e^{i\mbox{Re}S(x_1,x)-\mbox{Im}S(x_1,x)}\,.\label{psiSim}
\ee
The action is found from the Hamilton--Jacobi equation.

In the imaginary-time method, the sub-barrier motion is formally considered at imaginary values of the time variable. Performing the variable replacement $\tau =it$, we arrive at the Euclidian action
\be
S_{\rm E} =\int_{\tau_{1}}^{\tau_{2}}[m\sqrt{1+(dx/d\tau)^2} +V(x,\tau)-V_0]d\tau\,.
\ee

The trajectory $x(\tau)$ in the under-barrier motion, where $S_{\rm E}$ is real,  is determined by the  condition $\delta S=0$. From here, one finds the equation of motion, which has a meaning of the Newton equation
\be
\frac{d\tilde{p}}{d\tau}=\frac{d}{d\tau}\frac{mdx/d\tau}{\sqrt{1+(dx/d\tau)^2}}=-\frac{\partial V_{\rm E} (x,\tau)}{\partial x}\,,\quad V_{\rm E} =-V\,.\label{NewtonImaginary}
\ee

With exponential accuracy, the probability to find the particle in the turning point of the exit from the barrier,  if it initially were in the point of the entrance of the barrier, is given by
\be
W(x_1,x_2)=e^{-2\mbox{Im}S (x_1,x_2)}=e^{-2S_{\rm E} (x(\tau_1),x(\tau_2))}\,.
\ee
This expression can be generalized to take the pre-exponential coefficient into account. However, we will restrict ourself by consideration of the exponential term.

It is essential that the sub-barrier trajectory satisfies the classical equation of motion, but now in the Euclidian time. To find it and to calculate $S$ and $W$, we may  formally use the known equations of the classical physics.

\subsubsection{Tunneling in Slowly Time-Dependent Potential}
The case of space-dependent and slowly time-dependent fields was considered in \cite{Popov:1979gq}, \mbox{cf. \cite{Han2010}.} For simplicity, consider a scalar  particle in a one-dimensional field.
Let the probability of the tunneling  in the static limit be known,
\be
W=e^{-2\int_{x_{1}}^{x_{2}}|p|dx}\,,
\ee
where $x_1$ and $x_2$ are the entrance and exit turning points, i.e., $p(x_1)=p(x_2)=0$.
Variation of the action due to a weak dependence of the potential on time $V(x,t)$ yields
\begin{eqnarray}
&\delta S =\delta \int_{t_{1}}^{t_{2}}[-m(1-\dot{x}^2)^{1/2}-V(x,t)]dt\\
&=\int_{t_{1}}^{t_{2}}[p\delta\dot{x}-
(\partial V/\partial x)\delta x -\delta V (t)]dt=-\int_{t_{1}}^{t_{2}}\delta V (x(t))dt\,.\nonumber
\end{eqnarray}
We used equation of motion and integration by parts. The last integral can be calculated while using imaginary-time method. Thus, we obtain
\be
\delta S_{\rm E} =\int_{\tau_1}^{\tau_2}\delta V_{\rm E} (x(\tau))d\tau\,.\label{deltaSE}
\ee

Dependence $x(\tau)$ is determined from (\ref{NewtonImaginary}) as
\be
\tau (x_1,x)=\int_{x_1}^{x_2}dx \frac{\sqrt{m^2-\tilde{p}^2}}{\tilde{p}}=\int_{x_1}^{x_2}dx \frac{V-\epsilon}{\sqrt{m^2-(\epsilon -V)^2}}\,,\label{taux12}
\ee
where we used  relation $\tilde{p}^2=m^2-(\epsilon -V)^2$ and that $\epsilon$ may only adiabatically change with time, i.e., it may depend on $\tau$ only via the dependence of one of the parameters.

\subsubsection{Correction on Non-Adiabaticity  to the Spontaneous Positron Production in Low-Energy Heavy-Ion Collisions}
As a specific example, consider the probability of the spontaneous positron production in low-energy heavy-ion collisions. Deriving Equations (\ref{GammaGamma0}) and (\ref{S1theta}), we assumed that, during a time of the tunneling ($(r_{+} -r_{-})\sqrt{m^2+k^2}/k$), the potential $V$ and $\epsilon$ did not have a time to change. Here, please do not mix  typical time, for which the particle passes the barrier, cf. \cite{Kolomeitsev:2013du},  and time $1/\Gamma$, with an inversed probability to observe the positron. As we see from this simple estimate, adiabatic approximation does not hold at least for $k\to 0$, i.e., in the vicinity of  the boundary of the continua, $|\epsilon|\simeq m$.

Let us find a correction to the penetrability of the Coulomb barrier due to finite speed of the colliding nuclei \cite{Popov:1979gq}. Following (\ref{VrR}), the $R(t)$ dependent correction to the static Coulomb potential is as follows
\be
\delta V=-\frac{\zeta}{4r^3}P_2 (\cos \theta) R^2 (t)\,.\label{deltaVR}
\ee

Further consider the case when positrons  are emitted  along the axis
that joins the nuclei, $P_2 (0)=P_2(\pi)=1$. Subsequently, the probability of their production is maximal.
Expanding $R(t)$ near the closest approach
point, we obtain
\be
R(t)=R_0+v^2t^2/(4R_0)\,.\label{Rt}
\ee

From (\ref{deltaVR}) and (\ref{Rt}), we have
\be
\delta V=-\frac{\zeta v^2}{8r^3} t^2\,.
\ee

The imaginary time $\tau =it$ is found from Equation (\ref{taux12}). Thus, we obtain
\be
\tau =\frac{\zeta}{k^3}[m^2\phi +(m^2+k^2)^{1/2}(m^2+\rho^2k^2)^{1/2}\sin \phi]\,,\label{tauphi}
\ee
where we introduced variable $\phi=2\mbox{arcsin}[(r_{+}-r)/(r-r_{-})]^{1/2}$, $0\leq \phi\leq \pi$,
$r=r_{+}\cos^2(\phi/2)+r_{-}\sin^2 (\phi/2)$, values $\tau =0$ and $\phi =0$ correspond to the
instant of emergence from under the barrier. The total imaginary tunneling time is $\tau_t =\pi\zeta m^2/k^3$, i.e.,
$\tau_t \to \infty$ for the electron energy  $\epsilon \to -m$, whereas, for deep electron levels, $\tau_t$ \mbox{strongly diminishes.}

The replacement of (\ref{tauphi}) in (\ref{deltaSE}) yields
\be
\delta S_{\rm E}=\delta\mbox{Im}S =-\frac{2Z}{Am_N}\frac{\zeta^2}{R_0 v_p^3}I(\epsilon_p,\eta)\,,
\ee
where $\epsilon_p=-\epsilon$, $v_p =(1-m^2/\epsilon_p^2)^{1/2}$ is the speed of the positron,
\begin{eqnarray}
&I(\epsilon_p,\eta)=-\frac{1}{8}\int_{0}^{\pi}d\phi \left[\frac{\sin\phi +(1-v_p^2)\eta}{\cos\phi +\eta}\right]^2 \frac{\cos\phi +(1-v_p^2)\eta}{\cos\phi +\eta}\,,\\
&\eta =[1-(1-\rho^2)v_p^2]^{1/2}\,.\nonumber
\end{eqnarray}

\begin{figure}[H]
\includegraphics[width=8cm,clip]{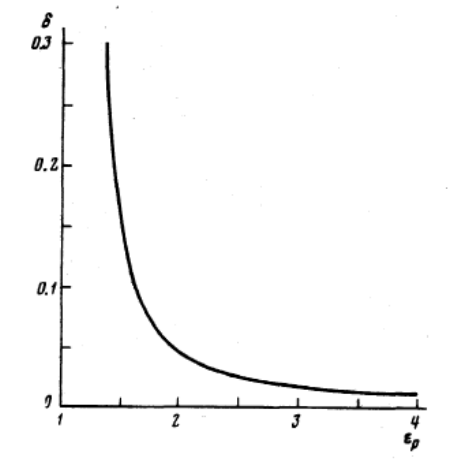}
\caption{ Correction on non-adiabaticity of the  motion of nuclei, $\delta$, cf. \cite{Popov:1979gq}, for  collisions U$+$U as a function of the positron energy $\epsilon_p$.}\label{AdiabaticCor}
\end{figure}

The ratio
\be
\delta =\frac{\mbox{Im}\delta S}{\mbox{Im}S_0}\,,
\ee
where $\quad \mbox{Im}S_0 =\pi\zeta[v_p^{-1}-(1-\rho^2)^{1/2}]\,,$
 for the collisions U$+$U ($\zeta =1.343$) is shown in \mbox{Figure \ref{AdiabaticCor}} as a function of the positron energy $\epsilon_p$. It is seen that $\delta <0.1$ for $\epsilon_p >1.65 m$. The adiabatic approximation in the problem of spontaneous production
of positrons becomes  invalid   near  $\epsilon_p=m$, where the positron production
cross section  is, in any case, tiny.

Numerical calculations \cite{PopovJETP1974,Marinov1974} have shown that $R_{\rm cr}$
rapidly increases with increasing  charge $Z=Z_1+Z_2$ of colliding nuclei. The cross section
of the spontaneous production of positrons increases in
this case $\propto R_{\rm cr}^{7/2}$, while the
correction for the non-adiabaticity of the tunneling decreases as $1/R_{\rm cr}$ at a fixed $\epsilon_p$. Therefore, it would be more convenient to perform
experiments with heavier nuclei, for which $R_{\rm cr}$ is larger.

\section{Many-Particle Semiclassical Approximation.  Electron Condensation in \mbox{Upper Continuum}}\label{RTF}
\subsection{Screening of a Source of Positive Charge in Presence of External Electrons}

In 
 a many-particle problem, most of the electrons in spherically symmetric  potential well, $V<0$, have angular momenta $l\gg 1$. Thereby, to find  distribution of the charge, we may deal with a more simple Klein--Gordon--Fock Equation (\ref{KGFV}) while assuming $j\simeq l$.  The value of the maximum  momentum,  at which the  electron placed in the positively charged ion where all levels with energies less than $\epsilon_{\rm bound}$ are already occupied  is  bound, satisfies the  condition
\be
p_{max}=\sqrt{(\epsilon_{\rm bound} -V)^2 -m^2}\label{KGFVmax}
\ee
with $\epsilon_{\rm bound}\geq -m$. If there is a sufficient amount of external electrons, the resulting system is charge-neutral. In this case, we should put $\epsilon_{\rm bound}=m$. Subsequently, $ p_{max}=\sqrt{ -2m V+V^2}\,,$ and taking into account that each cell of the phase space can only be occupied by two electrons of opposite spin, we have
\be
n_e =\frac{p_{max}^3}{3\pi^2}=\frac{(-2m V+V^2)^{3/2}}{3\pi^2}\,.
\ee

Thus, the relativistic Thomas--Fermi equation renders
\be \Delta V =4\pi e^2\left[n_{\rm nucl} -\frac{(-2m V+V^2)^{3/2}}{3\pi^2}\right]\,,\label{AtomRelTF}
\ee
$n_{\rm nucl}$ is the charged density of the nucleus.
It is curious to note that such an equation for neutral atom has been introduced long ago \cite{Vallarta1932}, but a relativistic term was then treated as a small correction  in nonrelativistic limit $|V|\ll m$.

\subsection{Filling of the Vacuum Shell by Electrons}

Note that, even in the absence of external electrons, which may fill the empty states, in case when the potential well  $V<-2mc^2$ electrons and positrons can be created already from the vacuum in the absence of any external electrons. Positrons go off to infinity, whereas electrons screen the initial positive charge of the source.
In this case, we should put $\epsilon_{\rm bound}=-m$. Subsequently, the relativistic Thomas--Fermi equation renders, cf. \cite{Rafelski1975,Migdal:1976wx,Migdal:1977rn},
\be \Delta V =4\pi e^2\left[n_{\rm nucl} -\frac{(2m V+V^2)^{3/2}}{3\pi^2}\theta(2m V+V^2)\right]\,,\label{TFvacShell}
\ee
where $\theta (x)$ is the step-function,
with the boundary conditions on the boarder of the ion
\be
V(r_i)=-2m =-Z_i e^2/r_i\,,\quad V^{\prime}(r_i)=Z_i e^2/r_i^2\,,\label{boundTF}
\ee
and with   $V(r)=-Z_i e^2/r$ for  $r>r_i$. Reference \cite{Rafelski1975} presented  numerical solutions.  The thorough analytical and numerical study of  the problem of the filling of the vacuum shell by many electrons was performed  in an independent study \cite{Migdal:1976wx,Migdal:1977rn}. This phenomenon was called "electron condensation", demonstrating that all of the vacuum levels are filled by electrons of the lower continuum, cf. \cite{Migdal1978}.

\subsection{A Detailed Derivation of Relativistic Thomas-Fermi Equation}

The electron density can be found by direct summation of the moduli squared of the  wave functions \cite{Migdal:1977rn}:
\be
n_e =-\sum_{n\kappa m}|\psi_{n\kappa m}|^2\,,\label{rhopsi}
\ee
where $\psi_{n\kappa m}$ are semiclassical wave functions presented in Equations (\ref{GFcl})--(\ref{psiprsphphiunderpolrnol}). Actually, we need wave functions in the classically allowed region given by (\ref{GFcl}).

Differentiating quantization rule (\ref{BohrSomDir}) over $n$, we obtain
\be
\frac{\partial\epsilon}{\partial n}\int_{r_{0}}^{r_{-}}\frac{\epsilon -V}{p}dr \simeq \pi\,,\label{difBS}
\ee
where we dropped the term $\frac{\partial}{\partial n} \frac{\kappa w}{pr}$, which only leads to a small correction $|w|/V^2 \sim 1/\zeta^2$, cf. \cite{Mur:1978ke}.

From (\ref{difBS}) and (\ref{C1T}), we obtain
\be
C_1 =\left(\frac{1}{\pi}\frac{\partial \epsilon}{\partial n}\right)^{1/2}\,.
\ee

Using that $\sum_{m=-j}^{j}|Y_{lm}|^2 =(2j+1)/(4\pi)$, where $Y_{lm}$ is the spherical function, from Equation (\ref{rhopsi}), we have
\be
n_e (r)=-\sum_{n\kappa}\frac{2j+1}{4\pi^2}\frac{\partial \epsilon}{\partial n}\frac{\epsilon -V}{pr^2}\,.
\ee

Here, we replaced $\sin^2\theta_1$ and $\sin^2\theta_2$ by $1/2$ due to multiple oscillations. Replacing summation in $n$ by integration, we find
\be
n_e =-\frac{1}{4\pi^2}\sum_{j}N_j =-\frac{1}{4\pi^2}\sum_{j}\frac{2(j+1/2)}{r^2}\sqrt{(\epsilon_{\rm bound} -V)^2 -m^2 -(j+1/2)^2/r^2}\,.\label{mj}
\ee

Doing further integration in $j$ with  $\epsilon_{\rm bound}=-m$, we recover (\ref{TFvacShell}).

Now, let us estimate the number of electrons in the vacuum shell, for which single-particle approximation fails, i.e., number of levels,  for which the width has no exponential smallness. Integrating (\ref{mj}) over the volume, we find the number of levels with momenta $j\leq \kappa -1/2$,
\be
\delta (\kappa)=\frac{1}{N_e}\sum_{j=1/2}^{\kappa -1/2}N_j =c\kappa^2\,,\label{deltakappa}
\ee
where
\be
c=3I_1/(2I_2)\,,\quad I_1=\int (V^2+2mV)^{1/2}dr\,,\quad I_2=\int (V^2+2mV)^{3/2} r^2dr\,,
\ee
$V^2\geq -2mV$.
In particular, for $V=-\zeta/r$ with logarithmic accuracy, we obtain
\be
I_1 =\zeta \ln (\zeta/R_{\rm nucl})\,,\quad I_2 =\zeta^3 \ln (\zeta/R_{\rm nucl})\,,\quad c=3/(2\zeta^2)\,.
\ee

For $\zeta \gg 1$ and $\kappa_0 =(\zeta/\pi)^{1/2}$ it follows that
\be
\delta (\kappa_0)=3/(2\pi\zeta)\ll 1\,.\label{deltakappa0}
\ee

Getting (\ref{deltakappa}), we counted all states with $|\kappa|<\kappa_0$, whereas not all of them have
exponentially  suppressed $\Gamma$. Taking a correction (\ref{deltakappa0}) into account leads to the appearance  of a numerical factor $\ln (\kappa_0/R_{\rm nucl})/\ln (\zeta/R_{\rm nucl})\simeq 1/4 $ for $\zeta\gg 1$ since $R\propto \zeta^{1/3}$. We estimate $\delta\simeq 0.1/\zeta$, i.e., $\delta \sim 1\%$ for $Z\sim 1/e^3$. The smallness of $\delta$ characterizes the accuracy of Equation (\ref{TFvacShell}).

Taking the exchange and correlation
corrections in the relativistic Thomas--Fermi equation into account is conveniently done by means of a variational
method analogously  to that is performed for the nonrelativistic
Thomas--Fermi equation \cite{Gambas1949}.  We arrive at
\be
n_e\simeq -\frac{1}{3\pi^2}[(V^2 +2mV)^{1/2} -\nu (V+m)]^3\theta (V^2 +2mV)\,,
\ee
$\nu \simeq e^2/\pi$. For $Ze^3\ll 1$ this correction can be safely dropped. For
$Ze^3\gg 1$ it can be taken into account  in Equation (\ref{TFvacShell}) by introducing  the renormalized coupling constant $e^2\to e^2 (1+3e^2/\pi)$, cf. \cite{Migdal:1977rn}.

Additionally, a correction appears due to that the dielectric permittivity of the vacuum, $\varepsilon(eE)$, differs from unity, $e\vec{E}=-\nabla V$. Thus, one should replace $\Delta V\to \nabla(\varepsilon(E)\nabla V)$ in \mbox{Equation (\ref{TFvacShell}).}
However, this correction, as the correlation correction, is tiny, since $\varepsilon(eE)=1-(e^2/(3\pi))\ln (eE/m^2)$, and at distances $r\gsim 1/(a_{Z})$ of our interest  $\varepsilon(eE)\simeq 1+O(e^2/(3\pi))$, cf. \cite{LL4} and Equation (\ref{epsfirstlog}), below.

\subsection{ Weak Screening, $ 1/e^2 \ll Z \ll 1/e^3$}
 Consider the screening of the positively charged nucleus of the initial proton number $Z$ and  the radius $R$
(typically $R_{\rm nucl}\simeq A^{1/3}/m_\pi$, $A\sim 2Z$). Assume that, inside the nucleus, the proton charge density is  $n_p^0=const$.
Introducing $\psi =-V/m-1$ in the region $V<-m$ ($\psi \geq 1$), where the electrons of the vacuum shell give some contribution to the screening of the charge $Z$, from Equation (\ref{TFvacShell}) we obtain
\be
\Delta (m\psi)=\frac{4e^2 m^3}{3\pi}(\psi^2 -1)^{3/2}\theta (\psi -1)-4\pi n_p^0\theta (R_{\rm nucl}-r)\,,\label{weakTF}
\ee
$\theta(x)$ is the step-function.
For $r>R_{\rm nucl}$, with the help of the replacement  $x=r/r_i$, we obtain

\be
\psi^{\,\prime\prime}_x +\frac{2\psi^{\,\prime}_x}{x} =\mu (\psi^2 -1)^{3/2}\,,\quad \psi (1)=1\,,\quad \psi^{\,\prime}(1)=-2\,,\quad \mu =\frac{4e^2 m^2 r_i^2}{3\pi}=\frac{(Z_{\rm obs}e^3)^2}{3\pi}\,.
\ee
Here, $Z_{\rm obs}$ is the charge seen at infinity. Because  $\mu\ll 1$, we may use expansion
\be
\psi (x,\mu)=\psi_0 (x)+\mu \psi_1 (x)+...
\ee

Subsequently, we have equations
\be
\Delta_x \psi_0 =0\,,\quad \psi_0(1)=1\,,\quad \psi_0^{\,\prime}(1)=2\,,
\ee
\be
\Delta_x \psi_1 =3(\psi_0^2-1)^{3/2}\psi_0\psi_1\,,\quad \psi_1(1)=0\,,\quad \psi_1^{\,\prime}(1)=0\,.
\ee

At the edge of the nucleus $x=2mR/\zeta \ll 1$.
At $x\ll 1$, we derive
\be
1+\psi (x,\mu)=2x^{-1}[1+4\mu (-\ln x +C_0)+O(x,\mu^2)]\,,\quad C_0 =2\ln 2 -11/3\,.
\ee

Inside the nucleus at the condition $Ze^3\ll 1$, the potential is close to the bare one. Setting
$\psi =\zeta y(\Xi)/(R_{\rm nucl}m)$, $\Xi =r/R$, we obtain
\be
y_{\Xi}^{\,\prime\prime}+\frac{2y_{\Xi}^{\,\prime}}{\Xi} = \frac{4e^2 R^3}{3\pi \zeta}
\left(\frac{\zeta^2 y^2}{R^2}-m^2\right)^{3/2} -\frac{4\pi R^3 e^2}{\zeta}n_p^0\,,\quad r<R_{\rm nucl}\,.
\ee

Using that inside the nucleus $|V|\sim \zeta/R_{\rm nucl}\sim Z^{2/3} m\gg m$ and $\frac{4\pi}{3}R_{\rm nucl}^3 n_p^0 =Z$,
\mbox{we get}
\be
y_{\Xi}^{\,\prime\prime}+\frac{2y_{\Xi}^{\,\prime}}{\Xi} =-3 +\nu y^3\,,\quad \nu = \frac{4(Ze^3)^2}{3\pi}\,.
\ee

Because $\nu \ll 1$, we expand
\be
y=y_0 (\Xi)+\nu y_1 (\Xi)+...
\ee
and get
\be
y_0 (\Xi)=\frac{1}{2} (3-\Xi^2)\,,\quad y_1 (\Xi)=C+\Xi^{-1}\int_0^{\Xi} y_0^3 (x) x (\Xi -x)dx\,.
\ee

Matching of $V$ and $V^{\,\prime}$ at the edge of the nucleus yields
\be
C=-1-\int_0^1 y_0^3 (x) x dx\,,
\ee
and
\be
Z_{\rm obs}=Z \left[1-\frac{4}{3\pi}(Ze^3)^2 (\ln \frac{\zeta}{R} +C_1)+...\right]\,,
\ee
\be
C_1 =\ln 2 -\frac{8}{3}+\int_0^1 y_0^3 (x) x^2 dx\simeq 1.38\,.
\ee

\subsection{Strong Screening, $Ze^3\gg 1$}
Continue to consider a nucleus with $Z\sim A/2$ and  $R_{\rm nucl}\simeq Z^{1/3}/m_\pi$.
Because $R$ grows with $Z$, one may expect that, for a sufficiently large $Z$, most of the electrons enter the nucleus and the  interior becomes   charge-neutral, as  infinite matter. For the bare nucleus, the energy that is associated with the electric field,
\be
{\cal{E}}_{\rm el}=\int\frac{(\nabla V)^2}{8\pi e^2}d^3 x\sim Z^2 e^2/R_{\rm nucl}\sim Z^{5/3}e^2m_\pi\,,
\label{EnPoi}\ee
increases with $Z$ more sharply when compared to the binding energy $\sim A\sim Z$, thereby
the volume-charged systems do not exist. The charge, if it exists, is repelled to the surface.

To approximately solve Equation (\ref{TFvacShell}), we now introduce variables $x=(r-R_{\rm nucl})/l$ and $V=-V_0 \chi (x)$.
Constant $V_0$ is found from the condition of the charge neutrality at $x\to -\infty$, i.e., $V_0^3/(3\pi^2)=n_p^0$ for $V_0\gg m$. Thus, in new variables, Equation (\ref{TFvacShell}) renders
\be
\chi_x^{\,\prime\prime}l^{-2}+2\chi_x^{\,\prime}l^{-2}/(x+R_{\rm nucl}/l)=\frac{4\pi e^2 n_p^0}{V_0}[\chi^3 -\theta (-x)]\,,\label{chiV}
\ee
with boundary conditions $\chi (-\infty)=1$, $\chi (\infty)=0$. The latter condition just means that typical decrease of the potential occurs already at $x\sim l$ near the nucleus boundary, whereas the transition to the Coulomb law occurs at $x\gg l$. The solution at  such large distances can only be found numerically.

Because, in dimensionless equation with dimensionless boundary conditions typical $|x|\sim 1$, for $R_{\rm nucl}\gg l$, which we assume, we can neglect the second term in l.h.s. of \mbox{Equation (\ref{chiV}).} In this case, geometry becomes one-dimensional and Equation (\ref{chiV}) \mbox{reduces to}
\be
\chi^{\,\prime\prime}=\chi^3 -\theta(-x)\,,\label{limitchi}
\ee
where we determined the length $l$, as
\be
l^{-2}=4\pi e^2 n_p^0/V_0=4e^2 (\pi/3)^{1/3}(n_p^0)^{2/3}\,.
\ee

Taking the boundary conditions into account, the first integral of \mbox{Equation (\ref{limitchi})} is as follows
\be
2\chi^{\,\prime\,2}=\chi^4 +(-4\chi +3)\theta(-x)\,,\label{firstint}
\ee
and the final solution is
\be
\chi(x)=1-3[1+2^{-1/2}\mbox{sh}(a-x/\sqrt{3})]^{-1}\,\quad x<0\,,\quad \mbox{sh} a=11\sqrt{2}\,,\label{shexact}
\ee
\be
\chi(x)=2^{1/2}(x+b)^{-1}\,,\quad x>0\,,\quad b=4\sqrt{2}/3\,.
\ee

Note that Equation (\ref{limitchi}) allows for very simple approximate solution for $x<0$.  To get it, we write $\chi =1+\psi$, $\psi \ll 1$ and, from (\ref{limitchi}), find
\be
\chi(x)\simeq 1-C^{\prime}e^{x\sqrt{3}}\,.
\ee
Using the boundary conditions at $x=0$, we find $C^{\prime}\simeq 0.24.$
This solution with an error less than 1.5$\%$ coincides with the exact solution.

The maximal strength of the electric field is reached at the edge of the nucleus,
$$E_{\rm max}=\frac{9\pi\sqrt{2}}{16}\left(\frac{3}{\pi}\right)^{1/6}(n_p^0)^{2/3}\simeq 8.2\cdot 10^{19}\,\,{\rm V}/{\rm cm}\,,$$
that $\simeq 6000$ times exceeds the electron QED unit $E_{\rm QED}=m^2 c^3/(e\hbar)\simeq 1.3\cdot 10^{16}\,\,{\rm V}/{\rm cm}$. Note that, to obtain this conclusion, we essentially used the relation $R_{\rm nucl}\sim Z^{1/3}/m_\pi$.

The energy of the system can be recovered  by the integration of Equation (\ref{TFvacShell}). For $|V|\gg m$, we have
\be
{\cal{E}}=\int\left[-\frac{(\nabla V)^2}{8\pi e^2}-\frac{V^4}{12\pi^2}-n_p^0\theta (R_{\rm nucl}-r) V\right]
d^3 x\,.\label{Enlim}
\ee
Expression (\ref{EnPoi}) is obtained, after one puts to zero the term $\frac{V^4}{12\pi^2}$ related to the electron condensation  and employs  the partial integration and Poisson equation.

In our case, $\nabla V=0$ inside the system for $R_{\rm nucl}\gg l$ and $V_0 =(3\pi^2 n_p^0)^{1/3}$.
With these values, Equation (\ref{Enlim}) yields
\be
{\cal{E}}=\frac{V_0^4}{4\pi^2}\cdot \frac{4\pi}{3}R_{\rm nucl}^3\,.
\ee
Accordingly, the energy is reduced to the kinetic energy of the degenerate relativistic electron gas filling all energy levels of the vacuum shell with  $\epsilon <-m$. One should add to it the energy that is associated with the strong interaction of nucleons resulting in the binding of the ordinary atomic nuclei. In such a way, we get 
transition to the description of infinite matter.
We see that, not taking into account a pion condensate or some other complex processes, we have  ${\cal{E}}>0$ and such a matter, without inclusion of the gravity,  is unstable, cf. \cite{Migdal:1977rn,Migdal:1990vm}.

If, instead of the usage that $A\sim 2Z$, we  assumed the validity of the $\beta$ equilibrium conditions, $n\leftrightarrow p+e+\bar{\nu}$, we would get $A\gg Z$, and taking into account the gravity and the filling of all electron levels up to $\epsilon=m$, we would recover the description of the ordinary neutron-star matter, cf.  \cite{Migdal:1990vm}.

\subsection{Falling to the Center in Relativistic Thomas-Fermi Equation}\label{fall}
For  $V=-Ze^2/r$, the number of electrons filling the vacuum shell is
\be
N_e \simeq \int^r \frac{|V|^3}{3\pi^2}d^3 x\sim \ln 1/(rm)\to \infty\ee
for $r\to 0$.

\textls[-23]{Now, consider  a formal solution of Equation (\ref{TFvacShell})  at $r<r_i$ with boundary \mbox{conditions (\ref{boundTF})}} corresponding to that for $r>r_i$, we deal with the Coulomb law with the  charge equal to the  observable charge $Z_{\rm obs}$. As we shall see, such a problem has a unique solution independently on the charge $Z_0$ put in the center, i.e., at $r\to 0$.
It proves to be that the exact solution of Equation (\ref{TFvacShell}) has the pole singularity already at a finite value $r=r_{\rm pole}(\mu)$. In a weak screening limit from Equation (\ref{weakTF}), for $r\to r_{\rm pole}(\mu)$,  in the dimensionless variable $x=r/r_i$, $x_{\rm pole}=r_{\rm pole}/r_i$, we  get \cite{Eletskii:1977na},
\be
\psi (x,\mu)=\frac{C}{x-x_{\rm pole}}\left[1+\frac{a_1(x-x_{\rm pole})}{x_{\rm pole}} +\frac{a_2 (x-x_{\rm pole})^2}{x_{\rm pole}^2} +...\right]\,,\quad C=(\mu/2)^{-1/2}\,,\label{TFP}
\ee
$a_1 =-1/3$, $a_2 =2/9 +\mu x_{\rm pole}^2/6$, ...
The substitution of (\ref{TFP}) in Equation (\ref{weakTF}) allows for finding coefficients $a_n$, but does not allow for recovering dependence $x_{\rm pole}(\mu)$. To obtain a full solution of the problem, we need to solve Equation (\ref{weakTF}) with the boundary conditions (\ref{boundTF}) in the whole interval $1>x>x_{\rm pole}(\mu)$.  The numerical solution yields
\be
x_{\rm pole}(\mu)=r_{\rm pole} (\mu)/r_i = D(\mu)e^{-1/(8\mu)}\,,\quad \mu\to 0\,.\label{Dmux}
\ee

\begin{figure}[H]
\includegraphics[width=6.5cm,clip]{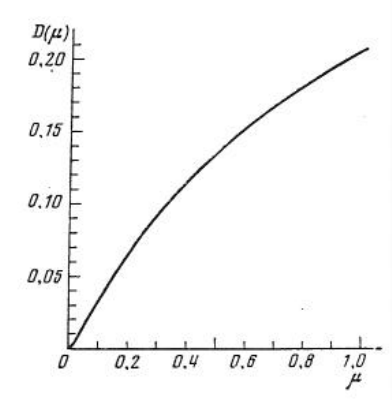}
\caption{Pre-exponential
factor $D(\mu)$ in Equation (\ref{Dmux}), cf. \cite{Eletskii:1977na}.}\label{Dmu}
\end{figure}

Pre-exponential factor $D(\mu)$ is shown in Figure \ref{Dmu}. For $Z_{\rm obs}\gg 1/(2e^2)$ with increasing $Z_{\rm obs}$ the pole moves towards the value  $1/m$.

We conclude that, in the many-particle problem, including the electron condensation but not including the  polarization of the vacuum, the falling to the center manifests itself in the presence of the pole at a  distance $r_{\rm pole}(\mu)$. Accordingly,  in the problem of the distribution of the charge at $r\ll 1/m$, there appeared a typical  size $r_{\rm pole}(\mu)$, which characterizes the electron condensation, where all of the states are occupied according to the Pauli principle.
Thus, we have found a relation between $Z_{\rm obs}$ and  $Z (r_0(\mu))$, for the size of the source  $r_0>r_{\rm pole}(\mu)$.
To match this exterior solution with the interior solution for $r<r_0$, we may use either model I or model II. It is important that $r_0$ should be larger than $r_{\rm pole}(\mu)$.

At this instance, we should remind about the existence of the Landau pole for \linebreak $r=r_{\rm L}\simeq e^{-3\pi/(2e^2)}/m$, which appears within the multi-particle problem of the polarization of the electron-positron vacuum near the Coulomb center, cf. \cite{LL4}. Comparison of the exponential factors  shows that, for $Z_{\rm obs}< 1/(2e^2)$, we have $r_{\rm L}>r_{\rm pole} (\mu)$ and, for $Z_{\rm obs}> 1/(2e^2)$, we have $r_{\rm L}<r_{\rm pole} (\mu)$. Thus, in the  case $Z_{\rm obs}< 1/(2e^2)$, with decreasing $r$, first the polarization of the vacuum becomes effective  and only at $r$ in a narrow  vicinity of  $r_{\rm L}$, where $Z(r)> 1/e^2$, the electron condensation becomes to be efficient. For $Z_{\rm obs}> 1/e^2$, the electron condensation first becomes effective  and only at $r$ in a narrow vicinity of  $r_{\rm pole}(\mu)>r_{\rm L}$ the polarization of the vacuum begins to contribute, see a detailed discussion below \mbox{in  Section \ref{DistrVacCh}.}

Note that the value $Z_{\rm obs}e^2$ plays a role of an effective coupling in description of semimetals and  effects under discussion might be relevant in this case, cf. \cite{Voskresensky2012}.

It is curious to note that the inclusion of gravitational field of the source into consideration modifies the QED problem of the distribution of the charge while taking the electron condensation into account, cf. \cite{Voskresensky:1982pf}. Solution (\ref{TFP}) is modified at $r$ approaching $r_{\rm pole}$. After a growth, solution continues up to $r\to 0$ as $V\to -Z_0 e^2/r$ with $Z_0 \sim Z_{\rm obs}^{2/3}/(eGm^2)^{1/3}$, where $G$ is the gravitational constant. Additionally,  the pole solution (\ref{TFP}) disappears in case of the electron condensation in presence of a strong uniform magnetic field, cf. \cite{Agasian:1983mf}.

At the end, note  \cite{Eletskii:1977na} that Equation (\ref{TFvacShell}) can be solved within the main logarithmic approximation \cite{LL4,LAKh1954b}, being broadly used in different problems of the quantum field theory, see a discussion below in Section \ref{PolarizationConst}. Introducing variables $\psi =\phi (x)/x$, $t=-\ln x$, $x=r/r_i$, in ultra-relativistic limit $|V|\gg m$, we obtain
\be
\phi_t^{\prime\prime} +\phi_t^{\prime}={\mu}\phi^3\,.
\ee

Assume $\phi =\sum_{n=1}^{\infty}\mu^n \phi_n$ with $\phi_n =C_n t^n +O(t^{n-1})$ for $t\to \infty$.
Subsequently, we get solution $C_n =2^{n+1}(2n)!/(n!)^2$ that finally yields $\psi (x\to 0)=C_n x^{-1}(-\ln x)^n+...$ A summation of these terms yields solution
\be
\psi (x)=2x^{-1}(1+8\mu \ln x)^{-1/2}\,,\label{spurious}
\ee
which has a spurious square-root singularity at $x\to x_0 =e^{-1/(8\mu)}$, whereas the exact solution has the pole. Thus, this example demonstrates the possible deficiencies of the main logarithmic approximation in cases when we deal with divergent series.

\section{Polarization of Vacuum}\label{Polarization}
\subsection{Polarization of Vacuum in Uniform Stationary Electric and Magnetic Fields}\label{PolarizationConst}

In the absence of external electromagnetic fields, electrons of the lower continuum have infinite energy
\be
{\cal{E}}_0 =\sum_{\vec{p}\sigma}\epsilon_{\vec{p}\sigma}^{0,-}\,,\label{InfConst}
\ee
where $\epsilon_{\vec{p}\sigma}^{0,-}=-\sqrt{m^2+\vec{p}^{\,2}}$ are negative-sign solutions of the dispersion relation of the free Dirac equation.
In pure QED, i.e., at ignorance of gravitational effects, infinite \mbox{constant (\ref{InfConst})} has no sense, being subtracted within   renormalization procedure.
In the presence of the electric and magnetic fields energy levels of the lower continuum, $\epsilon_{\vec{p}\sigma}^-$ are changed. \mbox{The difference}
\be
{\cal{E}}-{\cal{E}}_0 =\sum_{\vec{p}\sigma}\epsilon_{\vec{p}\sigma}^{-}-\sum_{\vec{p}\sigma}\epsilon_{\vec{p}\sigma}^{0,-}
\label{EE0}
\ee
has the physical meaning.

Heisenberg and Euler considered the polarization of the electron-positron vacuum in the static uniform stationary electric and magnetic fields \cite{Heisenberg1936}, cf. \cite{Weisskopf1936,LL4,Dunne2005}.  For the case of uniform purely magnetic field calculation is more transparent. Eigenvalues of the Dirac equation are
\be
\epsilon_{\vec{p}\sigma}^{\pm}=\pm\sqrt{m^2 +p_z^2 +|e|H(2n+1) +|e|H\sigma}\,\quad n=0,1,...,\quad \sigma =\pm 1\,.
\ee

The ground-state  corresponds to the "$-$'' sign solution.
To calculate the sum (\ref{EE0}), one uses that the number of states in the  interval $dp_z$ in the uniform magnetic field is given by
\be
\frac{|eH|}{(2\pi)^2} dp_z V_3\,,
\ee
cf. \cite{LL5}.
Taking into account the double degeneracy of levels with $n$, $\sigma =1$ and $n+1$, $\sigma =-1$ excluding ground state $n=0$, $\sigma =-1$, with $\epsilon_{\vec{p}\sigma}^{-}$ solution, one obtains
\be
{\cal{E}} =-\int_{-\infty}^{\infty}2\frac{|e|H}{(2\pi)^2}\sum_{n=1}^{\infty}
\sqrt{m^2 +p_z^2+2|e|Hn}\,dp_z V_3+\frac{|e|H}{(2\pi)^2}\int_{-\infty}^{\infty}\sqrt{m^2 +p_z^2}\,dp_z V_3\,.\label{EHtot}
\ee

The divergence of integrals is removed by the subtraction of ${\cal{E}}_0$. To do this renormalization, it is convenient to calculate a convergent derivative of the energy

\be
\frac{\partial^2 {\cal{E}}}{\partial (m^2)^2}=\frac{|e|H}{8\pi^2}\int_{0}^{\infty}e^{-m^2\eta}\left[\frac{2}{1-e^{-2|e|H\eta}}-1\right]d\eta V_3=\frac{|e|H}{8\pi^2}\int_{0}^{\infty} e^{-m^2\eta}\mbox{cth}(|e|H\eta)d\eta V_3\,.
\ee

After double integration and subtraction of the value  ${\cal{E}}_0 $, we obtain
\be
{\cal{E}}-{\cal{E}}_0 =\frac{V_3}{8\pi^2}\int_{0}^{\infty}\frac{e^{-m^2\eta}}{\eta^3}[\eta |e|H\mbox{cth}(\eta |e|H)-1]d\eta +C_1+C_2 m^2\,.\label{contrterm}
\ee
The contr-terms  $C_1$ and $C_2$ do not depend on $m^2$, but may depend on $H$.

In the case of uniform stationary fields $\vec{E}$ and $\vec{H}$, the Lagrangian density ${\cal{L}}=-{\cal{E}}$ can only be a function of
Lorentz invariants $\vec{E}^2 -\vec{H}^2$ and $\vec{E} \vec{H}$. Note here that, in the presence of the sources of the current, the Lagrangian density additionally depends on $j^\mu A_\mu$.

In the case under consideration employing arguments of dimensionality and parity in $\vec{H}$, one can write
\be
{\cal{L}}(H)={\cal{L}}_0(H)+{\cal{L}}^{\,\prime}(H)=-\frac{H^2}{8\pi}+m^4 f(H^2/m^4)\,.\label{L0Lprime}
\ee
The first term is  the ordinary  Lagrangian density in the magnetic field, whereas the second term is the contribution of the
polarization of the vacuum in the magnetic field. In Equation (\ref{L0Lprime}), there are no terms odd in $m^2$, so $C_2=0$. Using that $\mbox{cth} x =x^{-1}+x/3$ for $x\to 0$, we may see that the absence of $H^2$ term  ${\cal{L}}^{\,\prime}(H)$
corresponds to the choice
\be
C_1 =-\frac{H^2 e^2}{3\cdot 8\pi^2}\int_{0}^{\infty}e^{-\eta}{\eta}d\eta\,.\label{C1H}
\ee

In the case of uniform static magnetic and electric fields, function $f (H)$  in (\ref{L0Lprime}) should be replaced by
\be
f (H,E)=f(H^2-E^2,(\vec{E}\vec{H})^2)\,.\label{fHE}
\ee

At $H=0$, thereby $f(0,E)=f(-E^2,0)$. At $E=0$, $f(H,0)=f(H^2,0)$.
From here, we see that $f(0,E)=f(H=iE,0)$, i.e., the expression (\ref{L0Lprime})   for the case $H\neq 0$, $E=0$,
remains  valid after replacement $H\to iE$. Note that $f(-E^2,0)$ has a small imaginary part associated with a possibility of the tunneling of  a part of electrons, which initially occupied levels of the lower continuum, to the upper continuum. Created in a sufficient number, the electron-positron pairs change the spatial dependence of the  electric field. In a realistic treatment of the problem one should consider electron and positron condensates occurred near the plates of the capacitor, which produced initially uniform electric field.

In case of strong uniform electric and magnetic fields $|eE|/m^2\gg 1$ and $|eH|/m^2\gg 1$, with a logarithmic accuracy from Equations (\ref{contrterm}), (\ref{C1H}), one finds expressions for the dielectric and magnetic permittivities \cite{Heisenberg1936,LL4}:
\be
\varepsilon_{\rm HE} (E) =1-\frac{e^2}{3\pi}\ln (|eE|/m^2)+O(e^2)\,,\quad \mu_{\rm HE} (H) =1-\frac{e^2}{3\pi}\ln (|eH|/m^2)+O(e^2)\,.\label{epsmu}
\ee

The corresponding contributions to the energy of the lower continuum are
\be
 {\cal{E}}_E =\int d^3 x \frac{\varepsilon (\nabla V)(\nabla V)^2}{8\pi e^2}\,,\quad {\cal{E}}_H =\int d^3 x \frac{\mu (H)H^2}{8\pi e^2}
\,.\label{mueps}
\ee

\begin{figure}[H]
\includegraphics[width=7cm,clip]{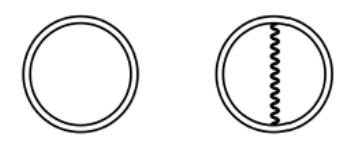}
\caption{The 1PI effective action shown up to two-loops. Double solid line shows the electron Green function dressed by the background field. Wavy line shows photon Green function.}\label{1PI}
\end{figure}

Note that expressions (\ref{epsmu}) are derived with the logarithmic accuracy, i.e., at the assumption that $\ln (|eE|/m^2)|\gg 1$ and $\ln (|eH|/m^2)|\gg 1$.
Thereby, they are also formally applicable for negative values of $\varepsilon$ and $\mu$ provided for the calculation of the vacuum energy in stationary uniform electric and magnetic fields one may employ the single-particle Dirac equation. At this assumption they are invalid only in a narrow region of fields, where $|\frac{e^2}{3\pi}\ln (|eE|/m^2)|\sim O(e^2)$ and $|\frac{e^2}{3\pi}\ln (|eH|/m^2)|\sim O(e^2)$. The result (\ref{epsmu}) also follows from the Dyson equation for the photon propagator that was calculated  at  one-loop, but with the electron Green functions that are dressed by the background field. In such an approximation, the radiative photon corrections to the electron Green function and vertices in the photon polarization operator are dropped.
Figure \ref{1PI} shows the effective action with one-particle irreducible (1PI) diagrams presented up to two-loops.
The same result (\ref{epsmu})  is also  recovered  within the so-called main logarithmic  resummation, when  $e^{2l}\ln^l (eE)$, $e^{2l}\ln^l (eH)$ terms in the Dyson equation for the photon Green function are summed up, whereas terms $e^{2l}\ln^{l-1} (eE)$, $e^{2l}\ln^{l-1} (eH)$  are disregarded, cf. \cite{LAKh1954b,LL4,Ritus1975,Ritus1998,Dunne2005,Dune2021}. The radiative photon corrections to the electron Green function continue to be disregarded. The difference between two approximations is only manifested  in the region where  $e^{4}\ln^2 (eE)\gsim e^2$, $e^{4}\ln^2 (eH)\gsim e^2$.  At the two-loop order, the term that is included in the effective action is given by the sandwich diagram (the one-particle-irreducible (1PI) contribution).  The resulting dielectric and magnetic permittivities up to correction terms are
 \begin{eqnarray}
 \varepsilon_{\rm 1PI} (E)\simeq \varepsilon_{\rm HE} (E), \quad \mu_{\rm 1PI} (H)\simeq \mu_{\rm HE} (H).\label{epsmu1}
 \end{eqnarray}

\begin{figure}[H]
\includegraphics[width=7cm,clip]{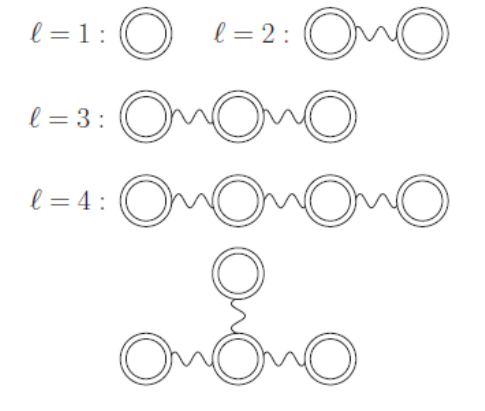}
\caption{The 1PR effective action shown up to  four loops summed up in \cite{Karbstein2020}. Double solid line shows electron Green function dressed by the background field. The wavy line shows photon \mbox{Green function.}}\label{1PR}
\end{figure}

Recently, Refs. \cite{Karbstein2016,Karbstein2017,Karbstein2020} studied the role of the one-particle reducible (1PR) loop diagrams. In this scheme, Figure \ref{1PR} shows the effective action up to four loops. These 1PR diagrams yield zero contribution in the case of constant fields \cite{Ritus1975}, since, in the case of purely constant classical fields, the four-current term is absent.
However, the argument for the vanishing of the current no longer holds as soon as the external field supports a slightest inhomogeneity somewhere in the space-time \cite{Karbstein2016}.

In the latter case, all possible 1PR loop diagrams, being included, can be constructed from the 1PI one-loop constant-field diagram.  The result of such a    resummation   of the diagrams in the strong-field limit yields \cite{Karbstein2020},
\vspace{-10pt}
\begin{eqnarray}
&\varepsilon_{\rm 1PR} (E) =1-\frac{e^2}{3\pi}\ln \frac{|eE|}{m^2}\left(1+\frac{1}{2}\frac{e^2 \ln (|eE|/m^2)}{3\pi(1-(e^2/(3\pi))\ln (|eE|/m^2)}\right) \left[1+O(1/\ln \frac{|eE|}{m^2}) \right]
\,,\label{epsmu2}\\
&\mu_{\rm 1PR} (H) =1-\frac{e^2}{3\pi}\ln \frac{|eH|}{m^2}\left(1+\frac{1}{2}\frac{e^2 \ln (|eH|/m^2)}{3\pi(1-(e^2/(3\pi))\ln (|eH|/m^2)}\right) \left[1+O(1/\ln \frac{|eH|}{m^2}) \right]\,.\nonumber
\end{eqnarray}

Note that, although, formally, these expressions are derived in the approximation $\ln (|eE|/m^2),\ln (|eH|/m^2)\gg 1$, as noticed in \cite{Karbstein2020},
they cannot be valid at least in the region where  $|1-(e^2/(3\pi)) \ln (|eE|/m^2)|,|1-(e^2/(3\pi))\ln (|eH|/m^2)|\lsim e^2$, due to the presence of the pole in expressions (\ref{epsmu2}). For example,  the dielectric permittivity $\varepsilon_{\rm 1PR} (E)\to -\infty$ for $(e^2/3\pi) \ln (|eE|/m^2)\to 1-\delta$ and $\varepsilon_{\rm 1PR} (E)\to +\infty$ for $(e^2/3\pi) \ln (|eE|/m^2)\to 1+\delta$ for $\delta\to 0$. Conversely, (\ref{epsmu}) and (\ref{epsmu1}) do not produce any non-physical singularities, yielding  zero, rather than the pole at $(e^2/3\pi) \ln (|eE|/m^2)\to 1$.  In the region where $\ln (|eE|/m^2),\ln (|eH|/m^2)\gg 3\pi/e^2$ expressions (\ref{epsmu}),  (\ref{epsmu2}) yield
\begin{eqnarray}
&\varepsilon_{\rm HE} (E) \to -\frac{e^2}{3\pi}\ln (|eE|/m^2)\,,\quad \mu_{\rm HE} (H) \to -\frac{e^2}{3\pi}\ln (|eH|/m^2)\,,\label{vareps}\\
&\varepsilon_{\rm 1PR} (E) \to -\frac{1}{2}\frac{e^2}{3\pi}\ln (|eE|/m^2)\,,\quad \mu_{\rm 1PR} (H) \to -\frac{1}{2}\frac{e^2}{3\pi}\ln (|eH|/m^2)\,.\nonumber
\end{eqnarray}

In the one-loop order, results (\ref{epsmu})--(\ref{epsmu2}) coincide. Beyond the one-loop approximation, various partial resummation schemes 
produce different results.

To proceed further, we will use  expression
\begin{eqnarray}
\varepsilon (E) =1-\nu \frac{e^2}{3\pi}\ln (|eE|/m^2)
\,.\label{epsmu3}
\end{eqnarray}

With $\nu =1$, we deal with the result \cite{Heisenberg1936,LL4,Dunne2005}, for $\frac{e^2}{3\pi}\ln (|eE|/m^2)<1$ being recovered within the main logarithmic approximation for the  1PI diagrams and, for \linebreak $\frac{e^2}{3\pi}\ln (|eE|/m^2)\ll 1$, being also recovered within the main logarithmic approximation applied for the  1PR diagrams. With $\nu$, being a very  smooth function of the tortoise variable $\ln (|eE|/m^2)$
varying from 1 at $|eE|
\sim m^2$
to $1/2$ for $\ln (|eE|/m^2)
\gg 3\pi/e^2$, we recover the asymptotic behavior that was derived in \cite{Karbstein2020} with the included 1PR loop diagrams.

At the end, we stress that  both main-logarithmic resummation schemes considered above
may be not valid for $(e^2/3\pi) \ln (|eE|/m^2), (e^2/3\pi) \ln (|eH|/m^2)\to \infty$, since the dropped sub-series of the diagrams may yield divergent contributions. We have demonstrated examples of such a kind in Section \ref{fall}, cf. \cite{Eletskii:1977na}. A summation of the 1PR diagrams leads to the appearance  of the pole in expressions $\varepsilon_{\rm 1PR} (E)$ and $\mu_{\rm 1PR} (H)$ for
$(e^2/3\pi) \ln (|eE|/m^2)=1, (e^2/3\pi) \ln (|eH|/m^2)= 1$.
 Moreover, recall that the expansion in the number of loops is a semiclassical series. The latter series is an asymptotic one, and  retaining too many terms may worsen the convergence of the series to the exact solution. Bearing this in mind, the result that is given by $\varepsilon_{\rm HE} (E), \mu_{\rm HE} (H)$  looks more physically motivated. Nevertheless, further on, we use Equation (\ref{epsmu3}) varying parameter  $\nu$ in the interval  $(1/2 , 1)$  to recover both asymptotics in Equation (\ref{vareps}).

\subsection{Noninteracting Photon, Electron, and Spin-Zero Boson  Propagators}

The Green function of the free photon is given by
\begin{eqnarray}
iD_{\mu\nu}^0(x-x^{\,\prime})={<0|\hat{T} \hat{A}^{\rm int}_\mu (x) \hat{A}^{\rm int}_\nu (x^{\,\prime})}|0>\,,\label{DAint0}
\end{eqnarray}
$\hat{T}$ is the ordinary time ordering, operators are in interaction picture, cf. \cite{LL4}.
The most general form is as follows,
\be
D_{\mu\nu}^0(x-x^{\,\prime})=g_{\mu\nu}D^0((x-x^{\,\prime})^2)-\partial_\mu\partial_\nu D^0_{(l)}((x-x^{\,\prime})^2)\,,\label{Dtl}
\ee
$g_{\mu\nu}$ is the metric tensor.
One usually uses the Feynmann gauge condition  $D^0_{(l)}=0$.

For $D^0_{xx}=-D^0$, we have
\be D^0(k^2)=\frac{4\pi}{(k^2+i0)}\,.
\ee

In the Feynmann gauge,
\be D^0_{\mu\nu}(k^2) =g_{\mu\nu}\frac{4\pi}{(k^2+i0)}\,.
\ee

The free propagator of spin $1/2$ electron is
\be
G_{ik}^{0}=-i <0| \hat{T}\hat{{\Psi}}_i^0 (x)\hat{\overline{\Psi}}_k^0 (x^{\,\prime})|0>\,,
\ee
where ${\overline{\Psi}}={\Psi}^{\dagger}\gamma_0$  and $\Psi_i^0 (x)$ satisfy the Dirac equation $(\gamma^\mu\hat{p}_\mu -m)\Psi_i^0 (x)=0$.

Thus, the Fourier transform is
\be
G^{0}(p)=\frac{1}{\gamma^\mu p_\mu -m}\,, \quad G^{0}(p)=\frac{\gamma^\mu p_\mu +m}{p^2-m^2}\,.
\ee

We may turn the contour in $p_0$  plane against clock arrow not touching poles and, then, we perform replacements $ip_0=p_4$, $ix_0=x_4$, $px=-\tilde{p}\tilde{x}=-(p_4x_4 +\vec{p}\vec{x})$,
$\tilde{p}=(\vec{p},p_4)$, $\tilde{x}=(\vec{x},x_4)$, $\int dp_0/i\to \int dp_4$. Let us present
\be
\frac{1}{\tilde{p}^2+m^2}=\int_0^\infty e^{-\alpha (\tilde{p}^2+m^2)}d\alpha\,,
\ee
\begin{eqnarray}
&G_{\rm b}^{0,\rm ch}(x)=\int \frac{d^4 p}{(2\pi)^4}\frac{e^{-ipx}}{p_0^2-\vec{p}^{\,2}-m^2+i\delta}
=-i\int_{-\infty}^{\infty}\frac{d^4\tilde{p}}{(2\pi)^4}\frac{e^{i\tilde{p}\tilde{x}}}{\tilde{p}^2+m^2}
\nonumber\\
&=-i\prod_{i=1}^{4} \int_{-\infty}^{\infty}\frac{d\tilde{p}}{2\pi}e^{i\tilde{p}_i\tilde{x}_i}\int_{0}^{\infty}d\alpha
e^{-\alpha (\tilde{p}_i^2 +m^2)}=-\frac{i}{16\pi^2}\int_0^{\infty}du e^{-m^2/u-\tilde{x}^2u/4}
\,,
\end{eqnarray}
$u=1/\alpha$. For $\tilde{x}m\ll 1$, we may put $m=0$ and find
\be
G_{\rm b}^{0,\rm ch}(x)=\frac{i}{4\pi^2x^2}\,,\quad \tilde{x}m\ll 1\,.
\ee

For $\tilde{x}m\gg 1$, we may use the pass method and present
$$-\frac{m^2}{u}-\frac{\tilde{x}^2 u}{4}\simeq -m\tilde{x}-
\frac{\tilde{x}^3 (u-u_m)^2}{8m}$$
and we find
\be
G_{\rm b}^{0,\rm ch}(x)=-i\sqrt{\frac{m}{32\pi^3\tilde{x}^3}}e^{-m\tilde{x}}\,, \quad \tilde{x}m\gg 1\,.\label{Gbchexp}
\ee

For Dirac electrons
\be
G^0(x)=\int\frac{d^4p}{(2\pi)^4}e^{-ipx}\frac{\gamma^\mu p_\mu +m}{p^2-m^2}=(m+
i\gamma^\mu\partial_\mu)G_{\rm b}^{0,\rm ch}(x)\,.
\ee

Thus, for $\tilde{x}m\ll 1$, we obtain $G^0(x)=\frac{\gamma^\mu x_\mu}{2\pi^2 x^4}$\,, the electron Green function is odd function of its coordinate argument. The power law increase of $G^0$ for $r\to 0$ reflects the fact that there is no scale of the length, which could  describe the free particle at $r\ll 1/m$. For $r\gg 1/m$, processes of polarization of the vacuum in the absence of external fields are suppressed as follows from Equation (\ref{Gbchexp}).

\subsection{Dyson Equation for Photon Propagator}

Taking the vacuum polarization diagrams in the first order perturbation theory in $e^2$ into account, the Dyson equation gets the form
\vspace{-12pt}
\begin{eqnarray}
&iD_{\mu\nu}(X_2-X_1)=iD_{\mu\nu}^0 (X_2-X_1)\\
& +\int d^4 X_3 d^4 X_4 iD_{\mu\lambda}^0 (X_2-X_3)\mbox{Tr}[
(-ie\gamma^\lambda) iG^{0}(X_4-X_3)(-ie\gamma^\rho) iG^{0}(X_3-X_4)]iD_{\rho\nu}^0 (X_4-X_1)\,.\nonumber
\end{eqnarray}

In the momentum representation, we obtain

\be
iD_{\mu\nu}(k)=iD_{\mu\nu}^0 (k)+iD_{\mu\lambda}^0 (k)\int \frac{d^4 p}{(2\pi)^4} \mbox{Tr}[
(-ie\gamma^\lambda) iG^{0}(p+k)(-ie\gamma^\rho) iG^{0}(p)]iD_{\rho\nu}^0 (k)(-1)\,.\label{Dimp}
\ee
The last factor $(-1)$ comes from the closed fermion loop.
The next terms in the full Dyson equation
are constructed analogously.

The sum of all irreducible  diagrams (which cannot be separated by a single photon line) is called the photon polarization operator, $-i\Pi_{\mu\nu}$. Thereby, in the lowest order $-i\Pi^{\lambda\rho}_0 =\mbox{Tr}[
(-ie\gamma^\lambda) iG^{0}(p+k)(-ie\gamma^\rho) iG^{0}(p)]$. In brief, notations
Dyson equation renders
\be
D=D^0+D^0\Pi D\,.\label{Dys}
\ee
In the lowest order in $e^2$ one has  $\Pi =\Pi_0$.

\subsection{Calculation of Photon Polarization Operator}

\subsubsection{Case of a Weak Static Electric Field. Renormalization of Charge}

To 
 remove divergencies in observables,
one employs renormalization procedure. Below, we  demonstrate this procedure on an example of renormalization of the charge. One assumes that, initially, the action enters the bare coupling $e_0^2$ rather than physical one, $e^2 =1/137$.  As we shall see, the polarization characteristics are divergent for $r\to 0$. At the same time, the $r\to 0$ limit is legitimate, because  QED is the theory with the local interaction. To proceed, one introduces the cut-value $r_0$, with performing the limit
 $r_0\to 0$  in final expressions. According to diagrammatic rules in the first non-vanishing order
\be
-i\Pi^{\mu\nu}_0 =\mbox{Tr}(-ie_0\gamma^\mu)iG^0(x)(-ie_0\gamma^\nu)iG^0 (-x)\,.\label{Pi0}
\ee

At $r> 1/m$,  in the case of  weak external fields, the effects of polarization of vacuum should be suppressed, since the electron Green function  and, thereby, the photon polarization operator decrease exponentially in Euclidean variables,  cf. Equation (\ref{Gbchexp}). Therefore, consider the opposite limit case $\tilde{x}\ll 1/m$ when the effects of the polarization of the vacuum can be significant. We recognize that at short distances there is no scale of length, except the Compton wave length. Thus, $G^0$ and $\Pi^0_{\mu\nu}$ should be power-law functions of $\tilde{x}$. We have
\be
-i\Pi^0_{\mu\nu}(x)=-e_0^2 \mbox{Tr}[\frac{\gamma_\mu \hat{x}\gamma_\nu \hat{x}}{4\pi^4 x^8}]=
-e_0^2 \frac{2x_\mu x_\nu -x^2 \delta_{\mu\nu}}{\pi^4 x^8}\,,
\ee
\be
-i\Pi^0_{00}(t,\vec{R}) =-e_0^2 \frac{t^2+\vec{R}^2}{\pi^4(t^2-\vec{R}^2)^4}\,.
\ee

In mixed  $\omega ,\vec{R}$ representation:
\be
\Pi^0_{00}(\omega =0,\vec{R})=\int d\tau \Pi^0_{00}(R)=
\frac{e_0^2}{4\pi^3 (\vec{R}^2)^{5/2}}\,.\label{MixedPi}
\ee

Using Equation
(\ref{Dtl}) with $D_{(l)}=0$, we have
\be
A^0(x) =\int d^4 x' D (x,x') j^0_{\rm ext}(x')\,.\label{ADj}
\ee

Multiplying Equation (\ref{Dys}) by $e_0^2 n_{{\rm ext}}(\vec{r})$ and integrating, we arrive at the Poisson equation for the static field $V(\vec{r})=e_0 A_0^{\rm n.ren}=eA_0^{\rm ren}$, being expressed  in terms of non-renormalized quantities,
\be
\Delta V (\vec{r}) =4\pi e_0^2 (-n_{\rm ext}(\vec{r})+4\pi\int K^0_{00}(\omega =0,\vec{R}) d^3 R \,V(\vec{r}+\vec{R}))\,,\label{Poisweak}
\ee
where in case of weak fields we took the polarization operator in the lowest order, i.e.,
$K^0_{00}(\omega =0, \vec{R})=\Pi^0_{00}(\omega =0, \vec{R})/e_0^2$.  $K^0_{00}(\omega =0, \vec{R})$ does not depend on $e_0^2$. As will be shown below, $K^0_{00}(\omega =0, \vec{R})$ diverges for $r_0\to 0$.

Now, our aim is  to  rewrite the Poisson Equation (\ref{Poisweak}) in the form
$$\Delta V=-4\pi e^2 n_{\rm ext}(\vec{r})\,.$$

To perform this procedure of renormalization of the charge, we continue to  consider the polarization of the vacuum in a weak field, i.e assuming $n_{\rm ext}$ to be small. Subsequently, we may use expansion
\be
V(\vec{r}+\vec{R})\simeq V(\vec{r})+\nabla V(\vec{r})\vec{R}+\frac{1}{2}\frac{\partial^2 V}{\partial R_i\partial_k R_k}R_i R_k+...\label{Vexp}
\ee

We may drop convergent terms in the expansion (\ref{Vexp}) irrelevant for the renormalization procedure.
The term $\int K^0_{00}(\omega =0,\vec{R})d^3 R V(\vec{r})$ should be put zero, since constant potential cannot produce polarization charges due to gauge invariance. The term $\int K^0_{00}(\omega =0,\vec{R})\vec{R} d^3 R \nabla V(\vec{r})=0$ due to isotropy of the vacuum in the weak field. \mbox{Hence, we obtain}
\be
\Delta V =-e^2_{\rm ren}{4\pi} n_{\rm ext}\,,\quad
e^2_{\rm ren} =e^2 =\frac{e_0^2}{1+\frac{4\pi}{6}{e_0^2}\int K^0_{00} (\omega =0,\vec{R}^2) \vec{R}^2 d^3 \vec{R}}\,
.\label{PiA1}
\ee

Finally, we derived a formal relation between the bare coupling constant $e_0^2$ and the physical one $e^2=1/137$. After this procedure is performed, we may say that all physical values  already depend only on  $e^2$.
 Thus, in the lowest approximation over $e^2_0$ using \mbox{Equation (\ref{MixedPi})} and relation between
$\Pi^0_{00}$ and $K^0_{00}$, we obtain
 \be e^2 =\frac{e_0^2}{1+\frac{e_0^2}{3\pi} \ln \frac{1}{m^2 {r}_0^2}}\,, \quad e_0^2 =\frac{e^2}{1-
 \frac{e^2}{3\pi } \ln \frac{1}{m^2 {r}_0^2}}\,, \quad {r}_0\to 0\,,\label{ee01}
 \ee

The formal solution of the first  equation  for any $e_0^2 >0$ yields $e^2 \to 0$, rather than $e^2 $. This is known as  "the problem of the zero charge'', (or ``Moscow zero''), cf. \cite{LL4}.
  Strictly speaking, such a consideration suffers from inconsistency, since the inverse relation given by the second equation
has so called Landau pole for
\be
r=r_{\rm L}=\frac{1}{m}e^{-3\pi/(2e^2)}\,.\label{LandauPole1}
\ee

From the second Equation (\ref{ee01}), for $r_0\to 0$,  follows the solution
\be
e_0^2 \to -\frac{3\pi}{\ln ((1/(m^2 {r}_0^2))}\left( 1+\frac{3\pi}{ e^2\ln (1/(m^2 {r}_0^2))}\right)\,,\label{eunren}
\ee
 corresponding to $e_0^2 <0$ and imaginary $e_0$.  A similar procedure could be performed in four-invariant form for the 4-potential $e_0A^\mu$, instead of $e_0 A_0$.

\subsubsection{Case of a Strong Static Electric field}

In the presence of a strong static electric field the electron polarization operator, even being considered  with the only one-loop diagram, should be calculated with full electron Green functions, $G$, instead of free ones \cite{Migdal1972,Migdal1978}. In this approximation, expression (\ref{Pi0}) is replaced by
\be
-i\Pi^{\mu\nu} =\mbox{Tr}(-ie_0\gamma^\mu)iG(x)(-ie_0\gamma^\nu)iG (-x)\,.\label{Pistr}
\ee
At this level, the Ward--Takahashi identity is only satisfied approximately. It can be fulfilled exactly  after taking the higher order diagrams into account.

Multiplying Equation (\ref{Dys}) by $e_0^2 n_{\rm ext}(\vec{r})$, we derive the Poisson equation for the static field $V(\vec{r})=e_0 A_0^{\rm n.ren}=eA_0^{\rm ren}$, expressed  in terms of non-renormalized quantities,
\be
\Delta V (\vec{r}) =-4\pi e_0^2 (n_{\rm ext}(\vec{r})-4\pi\int K_{00}(\omega =0, \vec{r}, \vec{R}) d^3 R V(\vec{r}+\vec{R}))\,,
\ee
where  $K_{00}(\omega =0, \vec{r}, \vec{R},e_0^2) =\Pi_{00}(\omega =0, \vec{r}, \vec{R},e_0^2) /e_0^2$. Being expressed in non-renormalized terms, both of these quantities depend on $e_0^2$. For $G\to G^0$, they transform to $K_{00}^0(\omega =0, \vec{r}, \vec{R}) =\Pi_{00}^0(\omega =0, \vec{r}, \vec{R},e_0^2) /e_0^2$.

We again use expansion (\ref{Vexp}).
The term $\int K_{00}(\omega =0,\vec{r}, \vec{R})d^3 R V(\vec{r})$ should be put to zero, since the constant potential cannot produce polarization charges due to the gauge invariance. The term $\int K_{00}(\omega =0, \vec{r},\vec{R})\vec{R} d^3 R =0$ due to the symmetry respectively replacement  $\vec{r}\leftrightarrow \vec{r}^{\,\prime}$. Accordingly, we obtain
\be
\Delta V=-4\pi e_0^2 (n_{\rm ext} +n_1)\,,\quad n_1 =\frac{1}{2}\int d^3 R K_{00} (\omega =0, \vec{r},\vec{R})R_iR_k \partial_i\partial_k V(\vec{r})+\delta n_1\,,
\ee
where we retained  the residual convergent term $\delta n_1$.

Let the field $E(\vec{r})$ be locally directed in the $z$ direction. Subsequently, we rewrite
\begin{eqnarray}
&\frac{1}{2}\int d^3 R K_{00} (\omega =0,\vec{r},\vec{R})R_iR_k \partial_i\partial_k V(\vec{r})=
\frac{1}{4}\int K_{00}(\omega =0,\vec{r},\vec{R})\rho^2 d^3 R \Delta V \\
&-\frac{1}{4}\int K_{00}(\omega =0,\vec{r},\vec{R}) \rho^2 d^3 R \partial_z^2 V
+\frac{1}{2}\int K_{00}(\omega =0,\vec{r},\vec{R}) z^2 d^3 R \partial_z^2 V\,,\nonumber
\end{eqnarray}
where $\rho^2 =x^2+y^2$.
The renormalization of the charge is performed by addition and subtraction to $n_1$ the term $$\frac{1}{4}\int K^0_{00}(\omega =0,\vec{R})\rho^2 d^3 R=\frac{1}{6}\int K^0_{00}(\omega =0,\vec{R})\vec{R}^2d^3 R,$$ where we used isotropy of the quantity $K^0_{00}(\omega =0,\vec{R}).$
\hspace{0.75cm}
Thus, we obtain
\begin{eqnarray}
&\Delta V =-e^2{4\pi}(n_{\rm ext} +n_1^{\rm ren})\,,\label{PiA11}\\
&n_1^{\rm ren} =\int (K_{00}(\omega =0,\vec{r},\vec{R})-K^0_{00}(\omega =0,\vec{R}))\frac{\rho^2}{4} d^3 R \Delta V \nonumber\\
&+
\int K_{00}(\omega =0,\vec{r},\vec{R}) (\frac{z^2}{2}-\frac{\rho^2}{4}) d^3 R \partial_z^2 V+\delta n_1\,,\nonumber
\end{eqnarray}
\hspace{-0.75cm}
where $e^2=1/137$, see Equation (\ref{PiA1}). From this moment, all of the functions  are expressed in terms of $e^2$.

Now, let us evaluate the electron Green function in a strong static electric field. For this, it is sufficient to use a semiclassical expression for the Green function in mixed space
$G(\omega, \vec{r},\vec{r}^{\,\prime})\propto
e^{iS (\vec{r})- iS (\vec{r}^{\,\prime})} $\,, with
$$ S(\vec{r})- S (\vec{r}^{\,\prime})\simeq \int_{\vec{r}}^{\vec{r}^{\,\prime}}p(l) dl
\simeq \int_{\vec{r}}^{\vec{r}^{\,\prime}}
(\omega -V(l))dl\,,
$$
where $V=V_0 +\nabla V \vec{R}+...$, $V_0$ is const. The quantity $p(l)$ can be estimated from the Klein--Gordon--Fock equation $\Delta \psi +((\omega -V)^2 -m^2)\psi=0$, since, in a strong field, spin effects can be neglected with a certain accuracy. Thus, we estimate $G(\omega, \vec{r},\vec{r}^{\,\prime})\propto
e^{i\omega^{\,\prime} |\vec{R}|-i eE \vec{R}^2 C}$\,, where $\omega^{\,\prime}=\omega -V_0$, $C\sim 1$ is a constant.
Thus, at $|\vec{R}|\ll 1/\sqrt{|eE|}$, $eE=-\nabla V$, the Green function  $G$, is reduced to
$G^0$, and with a logarithmic accuracy $\Pi_{00}(\omega =0)\simeq \Pi_{00}^0(\omega =0)$.
For $|\vec{R}|\gg 1/\sqrt{|eE|}$, the Green function $G$ rapidly oscillates and  with a logarithmic accuracy $\Pi_{00}(\omega =0)$ can be put zero. Thereby, from (\ref{PiA11}) with the logarithmic accuracy, we obtain
\be
n_1^{\rm ren} \simeq -\Delta V \frac{1}{4}\int_{\vec{R}^2>1/|eE|}K_{00}^0 \rho^2 d^3 R +\delta n_1\simeq -\Delta V\frac{1}{12\pi^2}\ln |eE|  +\delta n_1\,.
\ee

 Now, we should  take into account that $\int n_1 (r) d^3 r =0$ due to the conservation of the charge of the vacuum. Thereby, $n_1 =\mbox{div}\vec{P}$, where $\vec{P}$ is a polarization vector. Thus, with our logarithmic accuracy, we should replace
$$-\frac{\ln |eE|}{12\pi^2} \Delta V\to -\nabla \left(\frac{\ln |eE|}{12\pi^2} \nabla V\right)\,.$$

Accordingly, finally, we arrive at the Poisson equation
\be
\nabla (\varepsilon (E)\nabla  V)=-4\pi e^2 n_{\rm ext}\,,\label{PoisPolariz}
\ee
with
\be
\varepsilon (E)=1-\frac{e^2}{3\pi}\ln |eE|=
1-\frac{e^2}{3\pi}\ln \frac{Q(r)}{m^2r^2} \,.
\label{epsfirstlog}
\ee

For $Z=1$,  we have $\varepsilon (E)\simeq 1-\frac{e^2}{3\pi}\ln (1/(m^2r^2))+ O(e^4 \ln^2 (1/(m^2r^2)))$ with logarithmic accuracy, that reproduces known
Uehling law \cite{LL4}. Equation (\ref{epsfirstlog}) was derived with the inclusion of the one-loop diagram (although with full Green functions). We used approximation  $\frac{e^2}{3\pi}\ln (1/(m^2 r^2))\ll 1$. Otherwise, higher-loop order diagrams and vertex correction diagrams should be included.
 However, Ref. \cite{LL4} demonstrated that the given  expression might be  valid with a higher accuracy,
since, for $Z\ll 1/e^2$, it is also  recovered in the  main logarithmic approximation to 1PI diagrams in the action, which shows that
 $\varepsilon (E)\simeq 1-\frac{e^2}{3\pi}\ln (1/(m^2r^2))+ O(e^2, e^4 \ln (1/(m^2r^2)))$. Therefore, it might be also  valid for $e^2\ln (1/(m^2r^2))\gg 1$ but $e^4 \ln (1/(m^2r^2))\ll 1$, i.e., in a region, where $\varepsilon (E)<0$. Recall that the main logarithmic approximation means that  terms $\propto e^{2l}\ln^l (1/(m^2r^2))$ are summed up, but terms $\propto e^{2l}\ln^{l-1} (1/(m^2r^2))$  are dropped. As a precaution, we should emphasize that the sum of the sub-leading terms disregarded within the  main logarithmic approximation can be divergent.

We may also use another intuitive  argument in favor of a formal validity of this expression at $\varepsilon (E)<0$. For this, let us consider theory with $N\gg 1$ number of charged species with masses $\sim m$ and let the coupling is $e^2/N$, cf. \cite{KL78}. Afterwards, instead of \mbox{Equation (\ref{epsfirstlog})}, we immediately arrive at expression
\be
\varepsilon (E)=1-N\frac{e^2}{3\pi N}\ln |eE|+O(N \frac{e^4}{N^2}\ln^2 |eE|)=
1-\frac{e^2}{3\pi }\ln |eE|+O(1/N) \,,\label{epsfirstlog1}
\ee
being valid in the region, where $\varepsilon (E)>0$, as well as for $\varepsilon (E)<0$.
Note that, obviously, expressions (\ref{epsmu}) that are derived by Heisenberg and Euler for the cases of purely uniform \mbox{fields \cite{Heisenberg1936}} also continue to hold for slightly inhomogeneous fields provided
\be
|H/H^{\,\prime}|\gg R_H =1/\sqrt{|eH|}\,,\quad |E/E^{\,\prime}|\gg R_E =1/\sqrt{|eE|}\,,
\ee
where $R_H =1/\sqrt{|eH|}$ is the typical radius of the curvature of the charged particle trajectory in the magnetic field (Larmor radius) and $R_E =1/\sqrt{|eE|}$ is the typical radius of the curvature of the charged particle trajectory in the electric field. Thus, for the electric field of the form
\be
E=Q(r)/r^2\,,
\label{strE} \ee
criterion of applicability of approximation of a  uniform field coincides with inequality   $Q(r)\gg 1$ provided $rQ^{\,\prime}\ll 1$. Accordingly, the expression for the dielectric permittivity of the \mbox{vacuum (\ref{epsmu})} derived for the case of the uniform field coincides with (\ref{epsfirstlog})
\be\epsilon (E) =1-\frac{e^2}{3\pi}\ln (Q(r)/(r^2 m^2))+O(e^2)\,,\label{epsmuQ}\ee
with the  logarithmic accuracy and with the same accuracy we may write\linebreak interpolation expression
\be
\varepsilon (E) =1-\nu\frac{e^2}{3\pi}\ln \frac{(Q(r)+1)}{r^2 m^2}+O(e^2)\,.\label{epsmuQ1}
\ee
Here, we additionally inserted a smooth function  $\nu$  varying within the interval $(1/2,1)$. With $\nu =1/2$, we recover the asymptotic behavior that is found by a resummation of the sub-set of 1PR diagrams \cite{Karbstein2020}, as we have discussed above.

Once more, notice that we will use Equations (\ref{epsmu}), (\ref{epsfirstlog}), and (\ref{epsmuQ1}) for both
$\varepsilon (E)>0$ and $\varepsilon (E)<0$. There exist corrections to Equation  (\ref{epsmuQ1})
in the region, where $|\varepsilon (E)|\sim e^2$; however, as we have discussed,  there are no physical reasons to expect the presence of any singularities in this region. Therefore, it seems reasonable to use the same expression (\ref{epsmuQ1}) at all distances.

\subsection{Polarization of Vacuum and Electron Condensation}
In the presence of charge sources, the Lagrangian density is already not only a function of $\vec{E}^{\,2}$, as was  the case in the purely  uniform field, but it contains the term $n_{\rm ext} V$. The charge sources  always exist in a realistic problem. Indeed, the uniform electric field can only be constructed in a limited region of space, namely inside the capacitor with the length of plates $l\gg d$, where $d$ is the
distance between the plates. Outside the capacitor, the field decreases to zero.
The electron--positron pairs produced in the tunneling process inside the capacitor go to the plates. The electrons are localized near the positively charged plate and  positrons, near the negatively charged one.

Recall that the energy of the electron in a smooth field $V$ in the classical approximation is given by
\be
\epsilon =V\pm \sqrt{\vec{p}^{\,\,2} +m^2}\,,\label{ecl}
 \ee
cf. Figure \ref{ContinuaBoundaries}, demonstrating the boundaries of the upper and lower continua in the field $V<0$.
The upper sign solution corresponds to states that originate in the upper continuum, which can be occupied in an attractive field for electrons, $V<-2m$ in the case of a broad potential well, after the tunneling of electrons from the lower continuum. In the standard interpretation, see the discussion in Section \ref{Interpretation}, the lower sign solution corresponds to positrons after replacement $\epsilon\to -\epsilon$. Let us also study another interpretation when  the lower sign solution corresponds to electron states that originate in the lower continuum, being occupied by the electrons.  As we show below, this interpretation might be relevant in a specific case, when  $\varepsilon <0$ in some region and, thereby, the resulting potential  $V>0$.

The introduction of the electric field in the Dirac equation for electron corresponds to the replacement $\epsilon \to \epsilon - V$. Let us expand the potential $V(\vec{r}^{\,\prime})$ near a point $\vec{r}$:
\be
V(\vec{r}^{\,\prime})= V(\vec{r})-e\vec{E}(\vec{r})\vec{R}+...\,,\quad  \vec{R}=\vec{r}^{\,\prime}-\vec{r}\,.
\ee
Assuming $V(\vec{r})$ to be very smooth function of coordinates, we may only retain these two terms in the expansion.

It is easy to ascertain the consequences of the replacement $-e\vec{E}\vec{r}\to
V(\vec{r})-e\vec{E}\vec{r}$. The term $-\sum \int \psi^*e\vec{E}\vec{R}\psi d^3 R\,$ was already taken into account in the problem solved by Heisenberg and Euler in the case of purely uniform electric field.
 The expressions for the Lagrangian and the energy of the lower continuum in uniform fields are more easily calculated for the case of purely magnetic field as we have mentioned. We found \mbox{Equation (\ref{EHtot}),} where typical momenta $p_z$ contributing to  the sum are $p_z\sim \sqrt{|eH|}$. In case of  purely electric field the typical momenta contributing to the sum are $p_z\sim \sqrt{|eE|}$. Performing summation  in Equation (\ref{EHtot}) Refs. \cite{Heisenberg1936,LL4} derived expression
(\ref{L0Lprime}) and with the help of invariants recovered Equation (\ref{fHE}). After doing replacement $H\to iE$, $|eH|\to |eE|$ one arrived at expressions (\ref{epsmu}).

Now, see Equation (\ref{ecl}), in the expression for the energy, there appears an additional potential term
\be
\delta{\cal{E}}_V=\sum \int \psi^*V(\vec{r})\psi d^3 r =\mp \frac{|V^4|}{3\pi^2}\,,
\ee
since $\sum_{njm}|\psi_{nj m}|^2=\frac{|V^3|}{3\pi^2}>0$. The upper sign is for $V<-2m$ and the lower sign is for $V>0$ and we, for simplicity, assume $|V|\gg m$.

There is still a kinetic term in the energy, see Equation (\ref{ecl}),  which
we should  add while considering the condensation of electrons, corresponding to the region of momenta $|\vec{p}|\sim |V|\gg m$ rather than to $|\vec{p}|\sim \sqrt{|eE|}$, the latter term  we have included. At least in limit cases  $V^2\gg |eE|$ and $V^2\ll |eE|$, the mentioned contributions are not overlapped.
As a result,  the kinetic term is
\be
\delta {\cal{E}}_{\rm kin}(V)=\pm  \int^{|V|}_0  |\vec{p}|  \,\frac{2\cdot 4\pi \vec{p}^{\,2} d |\vec{p}|}{(2\pi)^3} d^3r \simeq  \pm
\int \frac{V^4}{4\pi^2}d^3 r\,.
\ee

The upper sign corresponds to the electron condensation on levels of the upper continuum that is occupied during the tunneling of electrons from the lower continuum in the field $V<0$. We have studied this case in Section \ref{RTF}.
The lower sign solution corresponds to the electron condensation on levels of the lower continuum, may be possible  for $V>0$, compare with  the first term in  Equation (\ref{EHtot}), which was summed up  in the case of the  magnetic field.

Finally, in the case of a  weakly inhomogeneous electric field we  obtain
\be
{\cal{E}}={\cal{E}}_E +\delta {\cal{E}}_V +\delta {\cal{E}}_{\rm kin}=-\int d^3 x \frac{\epsilon (\nabla V)(\nabla V)^2}{8\pi e^2}-\int n_{\rm ext} d^3 x
\mp\int \frac{V^4}{12\pi^2}d^3 x \,.\label{Enegative}
\ee

From the semiclassical derivation, one may see the difference between the condensation of electrons  on levels of upper continuum crossed the boundary $\epsilon =-m$, cf. \mbox{Equation (\ref{Enlim}),} and condensation on levels in the lower continuum
in a repulsive field. In the former case, vacant states with $\epsilon <-m$ are occupied only in the process of the  tunneling of electrons
from the lower continuum. In the upper continuum, the kinetic energy of electrons is positive
${\cal{E}}_{\rm kin}=+\sum\int \psi^* |\vec{p}|\psi d^3 x$, $|\vec{p}|>0$, whereas the kinetic energy of electrons occupying levels of the lower continuum
is negative, ${\cal{E}}_{\rm kin}=-\sum\int \psi^* |\vec{p}|\psi d^3 x$, $|\vec{p}|>0$, cf. the first term in  Equation (\ref{EHtot}), has been used  in the case of the uniform  magnetic field.

Variation of the energy yields the Poisson equation,
\be
\nabla (\varepsilon\nabla V)=4\pi e^2 (n_{\rm ext}- \theta(V^2+2mV)(V^2+2mV)^{3/2}/(3\pi^2))\,,\label{lowerupper}
\ee
cf. Equation (\ref{TFvacShell}), which  described the electron condensation in  the attractive potential of a supercharged nucleus at $\varepsilon \simeq 1$. Although we are interested in the case $|V|\gg m$, we recovered the dependence on $m$ in Equation (\ref{lowerupper}).
Now, for $\varepsilon >0$ and  $V<-2m$, we deal with the electron condensation on levels of the upper continuum crossed the boundary $\epsilon =-m$ with increasing $|V|$, as it follows from the standard interpretation of the levels, appearing from the upper continuum during an adiabatic increase of   $|V|$. Below, we will argue for a possibility of the condensation of electrons that originated in the lower continuum  in the problem of the screening of the positively charged source at ultrashort distances from it (at $r<r_{\rm L}$), $\varepsilon_{\rm ren} (r_{\rm L})= 0$,
$\varepsilon_{\rm ren}(r< r_{\rm L})<0$  and the  potential is repulsive due to that.

\section{Distribution of Charge at Super-Short Distances from the Coulomb Center}\label{DistrVacCh}

\subsection{Charge Distribution Near  the Charge  Source of Radius $r=r_0>r_{\rm L}$}\label{Polel}

\subsubsection{Electron Condensation is Not Included} Let us to be specific $n_{\rm ext}=Z_0\delta (\vec{r}-\vec{r}_0)$, $Z_0 >0$, which corresponds to the surface distribution of protons following model I,  and $r_0 > \tilde{r}_m $, see Equation (\ref{Qrmax}) below. First neglect a possibility of the electron condensation
and only include the polarization of the vacuum in consideration. We seek a solution of Equation (\ref{PoisPolariz})
in the form
\be
V=-Q_1(r)/r<0\,,\quad eE=-\nabla V =Q(r)/r^2>0\,.
\ee

Substituting it in (\ref{PoisPolariz}), we find solution
\be
Q(r)=\frac{C}{\varepsilon(r,Q(r))}\,,\quad \varepsilon (r,Q(r))=1-\frac{e^2}{3\pi}\ln \frac{Q(r)+1}{m^2r^2}\,,\quad C=const.
\label{Qr}
\ee
For $r\gsim 1/m$, we can set $\varepsilon (r,Q(r))\simeq 1$ and, thereby, we may put $C=Z_{\rm obs} e^2$.

The potential $V$ is easily recovered in the case of a smooth variation of the charge \mbox{$Q_1(r)$, when}
\be
Q(r)\simeq Q_1(r)\,.\label{QQ1}
\ee
This condition is fulfilled for $|Q_1^{\,\prime}|\ll |Q_1|/r$ that yields  $|\varepsilon (r)|\gg  e^2/(3\pi)$.

The solution of Equation (\ref{Qr}) has two branches, one corresponds to $\varepsilon (r,Q(r))>0$, other relates to  $\varepsilon (r,Q(r))<0$. We assume $Z=Z_{\rm obs}$ for $r\gsim 1/m$ and  find $Q(r)$ for decreasing $r$. Subsequetly, we obtain
\be
Q(r)=Z_{\rm obs} e^2/\varepsilon (r,Q(r))\,\label{Qrmax}
\ee
on the positive branch of $\varepsilon (r,Q(r))$.
Expression  (\ref{Qrmax}) has a kink at $r =\tilde{r}_m$, $\varepsilon (\tilde{r}_m)\sim e^2/(3\pi)$ and $Q(\tilde{r}_m)\sim 3\pi Z_{\rm obs}\gg 1$. Therefore, \mbox{Equation (\ref{Qrmax})} only has a meaning for  $r_0>\tilde{r}_m$. Only then can one find a relation between $Z_{\rm obs}$ and $Z_0$.
However, note that, actually, \mbox{Equation (\ref{Qrmax})} already becomes invalid at a slightly larger $r$ than $\tilde{r}_m$, when $\epsilon (r,Q(r))$ reaches values $\sim e^2$. At these distances, Equation (\ref{Qr}) for $\varepsilon$ becomes invalid
and  approximation (\ref{QQ1}), which  we have used, also fails.

 A comment is in order. Consider what would be, if we used Equations \mbox{(\ref{PoisPolariz}) and (\ref{Qr})} for $r<\tilde{r}_m$. Subsequently, we would get   $Q_1(r)=-Z_0 e^2/\varepsilon (Q_1)>0$, $\varepsilon (Q_1)<0$. This solution  becomes invalid in the vicinity of $\tilde{r}_m$, where  $-\varepsilon\sim e^2$ , now for $r<\tilde{r}_m$, and it cannot be smoothly matched with the solution we have derived for $r>\tilde{r}_m$.

\subsubsection{ Electron Condensation on Levels of Upper Continuum is Included}
In the region, where   $Q(r)> 1$, besides the vacuum polarization, cf. Equation (\ref{PoisPolariz}), we should include the electron condensation on levels of the upper continuum crossed the boundary  $\epsilon <-m$, cf. Equation (\ref{lowerupper}).  Thus, we have
\be
\nabla (\varepsilon (E)\nabla  V)=4\pi e^2 V^3/(3\pi^2)\,,\quad {\rm at}\quad r>r_0\,,\label{PoisPolarizCond}
\ee
$-V\gg m$.
The solution of this equation can be easily obtained in  the approximation (\ref{QQ1}). We have \cite{Migdal1978},
\be
Q^2 (r)=\frac{C^2}{\varepsilon^2 (r,Q(r))-2C^2}\,.
\ee

To be specific, consider the case $Q_{\rm obs}\ll 1$. Constant $C$ is determined from the condition $Q(r\gsim 1/m)\simeq Q_{\rm obs}=Z_{\rm obs}e^2 $, since $\varepsilon (r\gsim 1/m)\simeq 1$.
Thus, we obtain
\be
Q^2 (r)=\frac{Q_{\rm obs}^2}{\varepsilon^2 (r, Q(r))(2Q_{\rm obs}^2+1)    -2Q_{\rm obs}^2}\simeq \frac{Q_{\rm obs}^2}{\varepsilon^2 (r, Q(r))   -2Q_{\rm obs}^2}\,.\label{PolarEl}
\ee

This solution shows an apparent pole at $r=r_{\rm pole}^{\rm ap}$.
Near this  point, in the region where $\varepsilon (r, Q(r))-\sqrt{2}Q_{\rm obs}\lsim e^2/(3\pi)$, the condition (\ref{QQ1}) is no longer fulfilled and solution given by Equation (\ref{PolarEl}) loses its meaning.
Now, let  $r_{\rm pole}<r_0< r_{\rm pole}^{\rm ap}$.
To determine $Q(r)$ in immediate vicinity of the point  $ r_{\rm pole}$ (at $r_0$ approaching $r_{\rm pole}$) we, as before,
assume that $\varepsilon (r, Q(r))$ is a smooth function of coordinates but now  $Q(r)\gg Q_1 (r)$. Above we have found the pole  solution of the relativistic Thomas-Fermi equation for $\varepsilon (r, Q(r))=1$, cf. Equation (\ref{TFP})
and \cite{Eletskii:1977na}. Now with $\varepsilon (r, Q(r))\simeq const < 1$ assuming $Q(r)\gg Q_1 (r)$
we similarly get \cite{V1992},
\be
V=-\left(\frac{3\pi \eta\varepsilon (r, Q(r))}{2e^2}\right)^{1/2}\frac{1}{(r-r_{\rm pole})}\,,\quad 0<r-r_{\rm pole}\ll r_{\rm pole}\,.\label{Vepspole}
\ee

 The value $\tilde{r}_m$ is now irrelevant, because solution (\ref{Qrmax}) is modified due to inclusion of the  electron condensation. Solution (\ref{Vepspole}) with $\eta =1$ is valid for $\varepsilon (r, Q(r))\gg  e^2/(3\pi)$.
At very short distances from $r_{\rm pole}$, at which  $0<\varepsilon (r, Q(r))\lsim e^2/(3\pi)$, the condition that $\varepsilon (r, Q(r))$ varies smoothly with $r$ is  violated. In this region, we may present  $\varepsilon (r, Q(r))\simeq a(r-r_{\rm pole})$, for $a=const$ and then solution (\ref{Vepspole}) continues to be valid, but now for $\eta =1/8$.

Finally, we stress that solution (\ref{Vepspole}) corresponds to the charge distribution near the bare charge $Z_0$ for $r_0> r_{\rm pole}$. It looses the meaning for $r_0<r_{\rm pole}$. At  fixed $Z_{\rm obs}$  for $r\gsim 1/m$, the charge $Z_0 (r_0)$ that is related to this $Z_{\rm obs}$ is increased with decreasing $r_0$. Even for $Z_{\rm obs}\ll  1/e^2$, at tiny distances, $r\sim r_{\rm pole}$, the charge $Q(r)$ becomes very large, $Q(r)\gg 1$, and at these distances the electron condensation on levels of the upper continuum crossed the boundary  $\epsilon <-m$ comes into play. Our solution does not exist for $r_0 <r_{\rm pole}$,  $r_{\rm pole}= r_{\rm L}$, where $\varepsilon (r_{\rm L})= 0$.
The value of $r_{\rm pole}$ essentially depends on the value of $Z_{\rm obs}$. For $Q_{\rm obs}\gsim 1$ the value $r_{\rm pole}$ increases considerably, see  Figure \ref{Dmu} and Equation (\ref{Dmux}), being derived for $\varepsilon \simeq 1$.

\subsection{Charge Source of Radius $r= r_0 <r_{\rm L}$. Polarization of Vacuum and Electron Condensation on Levels in Lower Continuum}\label{DistrChsupershort}

Because QED is the theory with a local interaction,  the charge sources can be of arbitrary  sizes, including $r_0\to 0$. To attack  the zero-charge problem, let us reconsider the interpretation of the electron condensation in the field of the charged source of a very small size.

Because the Dirac equation in the spherically symmetric field does not change under simultaneous replacements
$\epsilon\to -\epsilon$ and $e\to -e$, i.e., $V\to -V$ and $\kappa \to -\kappa$, in the Coulomb field of a negative charge $Z_0<0$, there are electron levels (and in the field of a positive charge $Z_0>0$, there are positron levels), which originate in the lower continuum. With increasing $|Z_0|$, the energy of such level, $\epsilon_e$,  goes up and at a value $|Z_0|>137-170$ (depending on $r_0$), the level intersects the  boundary of the upper continuum $\epsilon_e =m$. According to the traditional interpretation, which we have used while considering $r_0>r_{\rm pole}$, the electron states with $\epsilon_e  >-m$, which appeared from the lower continuum already in a weak field of repulsion to the electron,  should be regarded as unphysical, and they should be reinterpreted as positron states with energies $\epsilon_{e^+}=-\epsilon_e$. As a consequence of such reinterpretation, for a nucleus with $-Z_0>1/e^2$, upon decreasing $r_0$, the lowest positron level reaches the energy $\epsilon_{e^+} =-m$. Subsequently, two positrons,  after  tunnelling from the lower continuum,  occupy  this empty level and two electrons move to infinity.
Similarly, positron states with $\epsilon_{e^+} >-m$ appeared from the lower continuum already in a weak field of attraction to electron (for $Z_0>0$) are regarded as unphysical, being interpreted as electron states with energies $\epsilon_e =-\epsilon_{e^+}$.
As we have demonstrated, such an interpretation  allows for solving the problem of the charge distribution only  for $r>r_0>r_{\rm pole}$, even while taking such multiparticle effects into account, such as  the polarization of the vacuum and (for $Z>0$) the electron  condensation on levels of the upper continuum crossed the boundary $\epsilon_e =-m$.

However, beyond the framework of a single-particle problem, there appears to be a possibility of another interpretation \cite{V1992,Voskresensky:1993uy}. Following this possibility, we may interpret the electron levels that originated in the lower continuum in the weak repulsive field (for $Z_0<0$), as levels have been occupied by electrons of the lower continuum, while taking into account that dielectric permittivity $\varepsilon (r)$ can be negative at small distances. Subsequently, no preliminary tunneling occurs from one continuum to another.
Near the positively charged center of radius $r_0<r_{\rm pole}$, the desired repulsive potential for the electrons appears, since the dielectric permittivity of the vacuum expressed in terms of the physical charge $e^2>0$ becomes negative at $r<r_{\rm pole}$. In terms of a not renormalized charge $\varepsilon_{\rm n.ren} (r\to r_0\to 0)\to 1$ but $e_0^2<0$ leading to the same result, $V(r)>0$, cf.
 (\ref{eunren}). Passage of the pole with decreasing $r$  becomes possible because of the phenomenon of electron condensation on levels originated in the lower continuum even in a weak field.

Above, dealing with the electron condensation on levels of the upper continuum, due to presence of the pole, we could not get a continues solution for all $r$. Now, dealing with  $\varepsilon <0$ at $r\to r_0\to 0$, we are able to find an appropriate solution connecting $Q(r>r_0\to 0)$ and $Q_{\rm obs}=Q(r\to \infty)$.

For $\varepsilon <0$ and $Z_0>0$, the resulting potential $V$ proves to be repulsive.
Thus, for a positively charged center, due to change of the sign of $\varepsilon$ there are electron levels coming from the lower continuum. Since the quantity $|Z_0/\varepsilon (r)|$ increases with increasing $r$, in a certain range of $r$, where $-Q(r)> 1$, in the bare potential there are many such levels.  To count them, one can use the relativistic  Thomas--Fermi approach, now  employing the electron density  $-V^3/(3\pi^2)$ for $V>0$. We have
\be
\nabla (\varepsilon (r)\nabla  V)=-\theta(V)4\pi e^2 V^3/(3\pi^2)+4\pi Z_0 e^2 \delta(\vec{r}-\vec{r}_0)\,,\label{PoisPolarizCond1}
\ee
 cf. \cite{V1992}, and Equation (\ref{lowerupper}) derived above.
Introducing tortoise coordinate, $\Xi =\ln (1/r^2 m^2)$, we obtain
\be
\frac{d(Q\epsilon)}{d\Xi}=-\frac{2e^2}{3\pi}Q_1^3 -2\pi r^3Q_0 \delta(\vec{r}-\vec{r}_0)\,,\quad Q=Q_1 +\frac{2dQ_1}{d\Xi}\,,
\ee
where $Q_0=Z_0 e^2>0$. With condition $Q\simeq Q_1$ (justified by the resulting distribution), \mbox{we have}
\be
\frac{du}{u^3}=-\frac{2e^2}{3\pi}\frac{d\Xi}{\varepsilon^3}\,,\quad u=\varepsilon Q\,,\quad r>r_0\,.\label{uQ}
\ee

Using explicit expression (\ref{epsmuQ1}) with approximately constant value $\nu$, and integrating further, we find
\be
Q^2 (r)=\frac{C^2}{\varepsilon^2+2C^2/\nu}=\frac{Q_0^2}{\varepsilon^2+2Q_0^2/\nu}\,.\label{Q0Q}
\ee

Choosing an appropriate sign of the solution corresponding to the repulsive potential for the electron due to $\varepsilon <0$ for $r<r_{\rm pole}$, we arrive at
\be
Q (r)=-\frac{Q_0}{\sqrt{\varepsilon^2+2Q_0^2/\nu}}\,.\label{solQ}
\ee

For $r\to r_0\to 0$, for any finite value of $Q_0 >0$ we obtain $Q(r)\simeq -Q_0/|\varepsilon|\to 0$. Thus, a test particle does not interact with the nucleus at ultrashort distances. Recall the asymptotic freedom property in the QCD for $r\to 0$. For $r\sim 1/m$, we have $\varepsilon\simeq 1$ and $Q(r) =Z_{\rm obs}e^2$. Thus, we obtain a relation between the bare and observed charges
\be
Z_{\rm obs}=-Z_0/(1+2(Z_0 e^2)^2/\nu)^{1/2}\,.
\ee

For $Z_0\ll 1/e^2$, we get $Z_{\rm obs}\simeq -Z_0$. The maximum possible value of $|Z(r)|$ is $|Z_{\rm max}|\simeq 1/(\sqrt{2/\nu}e^2)$, $\epsilon (r_{\rm max})=0$. All levels are occupied by electrons of the lower continuum. Thereby, in the region where $\varepsilon <0$, the vacuum remains stable.

It is important that, at distances $r\gg r_{\rm pole}$, the potential looks like an ordinary  Coulomb potential. Individual charges situated at these distances, each with  $Z_{\rm obs}\ll 1/e^2$, can be summed up to the total charge $Z>Z_{\rm cr}\sim 1/e^2$. At these distances $\varepsilon >0$ and it is close to unity for $r\gg r_{\rm pole}$, and there may appear the electron condensation on the levels in the upper continuum crossed the boundary $\epsilon =-m$. These levels become occupied  by electrons, after the  tunneling from the lower continuum, as we have demonstrated in Section \ref{RTF}. Thus  reconstruction of the interaction at $r<r_{\rm pole}$ does not affect any phenomena that can be observed experimentally occurring at much larger distances.

\textls[-15]{Note that solution (\ref{solQ}) is similar to the solution obtained within QCD in the \mbox{model \cite{Agasian:1983mf},}} which took a possibility of the quark condensation near the external color-charge source into account. The essential difference is in the dependencies of  $\varepsilon (r)$ in QCD and in  QED. In QCD within a logarithmic approximation $\varepsilon^{\rm QCD}(r)\simeq b_0 \ln(r_\Lambda^2/r^2)$ where $b_0$ and $r_\Lambda$ are some positive constants, i.e., $\varepsilon^{\rm QCD}(r\to 0)\to \infty$ and
$\varepsilon^{\rm QCD}(r\to \infty)\to -\infty$, whereas, within QED, we  employed that
$\varepsilon^{\rm QED}(r\to 0)\to -\infty$ and $\varepsilon^{\rm QED}(r\to 1/m)\to 1$.
In QCD, there appears to be condensation of quarks on levels that originate in the upper continuum and in the case under consideration in QED for $r_0 \to 0$, we included the electron condensation on levels that originate in the lower continuum.

Note that, in terms of not renormalized dielectric permittivity, Equation (\ref{PoisPolarizCond1}) renders
\be
\nabla (\varepsilon_{\rm n.ren}(r)\nabla  V)=-\theta(V)4\pi e_0^2 V^3/(3\pi^2)+4\pi e_0^2 Z_0 \delta(\vec{r}-\vec{r}_0)\,,\label{PoisPolarizCond11}
\ee
with $\varepsilon_{\rm n.ren}(r)=1-\nu\frac{e_0^2}{3\pi}\ln (r_0^2/r^2)$. Using that  $e_0^2$ is a function of  $e^2$,
we obtain $\varepsilon_{\rm n.ren}(r\to r_0)\to 1$ and $\varepsilon_{\rm n.ren}<0$ for $r> r_{\rm pole}$.
Thus, at small distances $r< r_{\rm pole}$, the non-renormalized dielectric permittivity is positive. Value $Z_0 e_0^2<0$  for $Z_0>0$, that corresponds to $V>0$, and $-e_0^2 V^3/(3\pi^2)$ is positive. Accordingly, near a negative external charge, there appear to be positive charges and, vise versa, near a positive external charge, negative charges arise, as expected in QED.

Additionally, recall that the Hamiltonian, where one replaced $p_\mu\to p_\mu -e_0 A^{\rm n.ren}_\mu$ should be Hermitian operator, as well as the same Hamiltonian that is expressed in terms of the renormalized charge, where one uses the replacement $p_\mu\to p_\mu -e A^{\rm ren}_\mu$. Within the ordinary second quantization scheme, one expands  $\hat{A}_\mu$ in series of plane waves, where the creation and annihilation operators appear, considering $A_\mu$ as the real quantity. Because $e_0$ is imaginary, $A^{\rm n.ren}_\mu$ should be considered as purely imaginary quantity.  Now, we should perform expansion for $e_0 A^{\rm n.ren}_\mu$, being  real quantity. The energy  is  reduced  to the energy of stable oscillators only after performing renormalization, i.e., being expressed in terms of $e A^{\rm ren}_\mu$.

\subsection{Distribution of Charge of Electron}

Up to now, we considered the charge distribution near the external charge source, which was assumed to be infinitely massive. For  description of the electron  mass distribution, $m(r)$, one needs to study   Dyson equation for the electron Green function, cf. \cite{LAKh1954b}.  At distances of our interest  $|V|\gg m(r)$ and the dependence of  $m(r)$ does not influence the charge distribution in the logarithmic approximation that we have used. Equation for the mass is given by \cite{LAKh1954a},
\be
\frac{d m(\Xi)}{d\Xi}=-\frac{3e_0^2}{4\pi}d_t (\Xi) m(\Xi)\,,\label{mxi}
\ee
where $d_t$ is the so called  $d$-function of the photon and $\Xi =\ln (1/(r^2m^2))$ is the tortoise coordinate introduced above.

A clarification is in order. As is known, the presence of a zero in the expression for the dielectric permittivity $\tilde{\varepsilon}(\Xi)$ defined via the photon $d$-function,
\be
e_0^2 d_t =e^2(\Xi)\,,\quad \tilde{\varepsilon}(\Xi)=e^2/e^2(\Xi)\,,
\ee
according to the K\"allen--Lehmann expansion, would correspond  either to the violation of the causality or to the  instability of the vacuum \cite{Kirzhnits85}. However, note that, in our case, the quantity $\tilde{\epsilon}(\Xi)$ does not have  zero,
\be
\tilde{\varepsilon}(\Xi)=\frac{e^2}{e^2(\Xi)}=(\varepsilon^2 (\Xi)+2e^4/\nu)^{1/2}\,,
\ee
as follows from (\ref{Q0Q}) for $Q_0=Z_0 e^2$, $Z_0=1$. Thus, the quantity
$\tilde{\epsilon}(\Xi)$ does not coincide with $\varepsilon (\Xi)$. The latter quantity may vanish and it can even be negative, whereas the "true'' value $\tilde{\varepsilon}(\Xi)>0$.

Integrating (\ref{mxi}), we  obtain \cite{V1992},
\be
m(\Xi)=m\left(\frac{\varepsilon (\Xi)+(\varepsilon^2 (\Xi)+2e^4/\nu)^{1/2}}{1+(1+2e^4/\nu)^{1/2}}\right)^{9/4}\,,
\ee
where $m$ is the observed electron mass. Thus $m(\Xi\to \infty)\to 0$ and $m(\Xi\to 1)\to m$, i.e., in this case, the entire electron mass is of purely electromagnetic origin.

Concluding, we presented some arguments for the logical consistency of QED.

\section{Conclusions}\label{Conclusion}

 Most actively, the problem of a spontaneous production of positrons from the QED vacuum in strong fields has been  attacked in theoretical works in Moscow (in the group of V. S. Popov in 1970s) and in Frankfurt (in the group of W. Greiner in 70s and 80s of the previous century). The experiments performed at GSI Darmstadt in 1980s had turned out puzzling line structures in the energy spectra. These results were not confirmed by the subsequent experiments performed in the 1990s. Questions regarding the experimental confirmation  of existence of the spontaneous positron production in low-energy heavy-ion collisions remained open. Now, interest in this problem is renewed  \cite{Popov:2020xmd}, in connection with the possibility to perform new experiments  at the upcoming accelerator facilities in Germany, Russia, and \mbox{China \cite{Gumberidze2009,Ter2015,Ma2017}.}
 The study of many-particle effects in description of the QED vacuum in strong fields is of of principal  interest.
The problem of the zero-charge remains one of the most important fundamental problems of QED already about 70 years. In the given paper, these problems were studied within a common relativistic semiclassical approach that was developed in the reviewed papers.

 In the given paper, first, the   problems of the falling to the Coulomb center for the charged spinless boson  and for the fermion  were  considered within the  single-particle picture. Subsequently, focus was concentrated on a case of the spontaneous positron production in the field of a finite supercritical nucleus with the charge $Z>Z_{\rm cr}\simeq (170-173)$. The behavior of deep electron levels that crossed the boundary of the lower continuum  and the probability of the spontaneous positron production were studied. Subsequently, similar effects were  considered in application to the low-energy  collisions of heavy ions, when, for a short time, the electron level of the quasi-molecule crosses the boundary of the lower continuum $\epsilon =-m$. Next, the phenomenon of the electron condensation on levels of the upper continuum crossed the boundary of the lower continuum  in the field of a supercharged nucleus with  $Z\gg Z_{\rm cr}$ was studied. Subsequently, focus was concentrated on  many-particle problems of the polarization of the QED vacuum and the electron condensation at ultra-short distances from the  source of the charge. Arguments were presented for the important difference of the cases, when the size of the source is larger than the pole size $r_{\rm pole}=r_{\rm L}$, at which the dielectric permittivity of the vacuum reaches  zero, and smaller $r_{\rm pole}$. Subsequently, distributions of the charge and mass of the electron were considered and arguments were given in favor of the logical consistency of QED. Additionally, I believe that at least some of the results reviewed in this paper can find applications in the description of semi-metals and  stack of layers of graphene.
\vspace{6pt}

\reftitle{References}


\end{document}